\documentclass[11pt]{article}
\usepackage[margin=1in]{geometry}

\usepackage{amsopn}
\usepackage[doi=false,url=false,backref=true,,sorting=nyt,sortcites]{biblatex}
\usepackage[colorlinks=True]{hyperref}
\usepackage{xcolor}
\definecolor{cadmiumgreen}{rgb}{0.0, 0.42, 0.24}
\definecolor{hotpink}{rgb}{1.0, 0.41, 0.71}
\definecolor{cgreen}{RGB}{0, 180, 100}
\hypersetup{
    colorlinks,
    linkcolor=cgreen,
    citecolor=cgreen,
    urlcolor=blue,
    linktocpage
}

\usepackage{tcolorbox}
\tcbset{colback=white,boxrule=0.5pt,arc=2pt,left=4pt,right=4pt,top=4pt,bottom=4pt}

\usepackage{authblk}
\author[1,2,*]{Simon Welker}
\author[3]{Lorenz Kuger}
\author[4,5]{Tim Roith}
\author[7,8]{Berthy Feng}
\author[3,6]{Martin Burger}
\author[1]{Timo Gerkmann}
\author[2,9,10]{Henry Chapman}
\affil[1]{Department of Informatics, University of Hamburg, Bundesstr. 56b, 20146 Hamburg, Germany}
\affil[2]{Center for Free-Electron Laser Science CFEL, Deutsches Elektronen-Synchrotron DESY, Notkestr. 85, 22607 Hamburg, Germany}
\affil[3]{Department of Mathematics, Bundesstr. 55, University of Hamburg, 20146 Hamburg, Germany}
\affil[4]{CIT School, Technical University of Munich, Garching bei München, Germany}
\affil[5]{Munich Center for Machine Learning, München, Germany}

\affil[6]{Helmholtz Imaging, Deutsches Elektronen-Synchrotron DESY, Notkestr. 85,  22607 Hamburg, Germany}
\affil[7]{Massachusetts Institute of Technology (MIT), Cambridge, MA 02139, USA}
\affil[8]{The NSF AI Institute for Artificial Intelligence and Fundamental Interactions, Cambridge, MA 02139, USA}
\affil[9]{The Hamburg Center for Ultrafast Imaging, Luruper Chaussee 149, 22761 Hamburg, Germany}
\affil[10]{Department of Physics, University of Hamburg, Notkestr. 9-11, 22607 Hamburg, Germany}
\affil[*]{\small Corresponding author -- \href{mailto:simon.welker@uni-hamburg.de}{simon.welker@uni-hamburg.de}}
\date{\today}
\title{Position-Blind Ptychography: Viability of image
reconstruction via data-driven variational inference}

\usepackage{amsfonts}
\usepackage{epstopdf}
\usepackage{graphicx}%
\usepackage{subcaption}
\usepackage{multirow}%
\usepackage{amsmath,amssymb,amsfonts,amsthm}%
\usepackage{mathtools}%
\usepackage{mathrsfs}%
\usepackage[title]{appendix}%
\usepackage{xcolor}%
\usepackage{textcomp}%
\usepackage{manyfoot}%
\usepackage{booktabs}%
\usepackage{algorithm}%
\usepackage{algpseudocode}%
\usepackage{listings}%
\usepackage{csquotes}%
\usepackage{hyperref}
\usepackage[capitalise]{cleveref}%
\usepackage{hyperref}
\usepackage{xfrac}
\usepackage{microtype}
\newcommand{\tablefontsize}{\scriptsize}

\usepackage{adjustbox}%
\usepackage[nohyperlinks,nolist]{acronym}

\newtheorem{theorem}{Theorem}[section]
\newtheorem{lemma}[theorem]{Lemma}

\usepackage{xhfill}

\DeclareMathOperator*{\argmin}{arg\,min}
\DeclareMathOperator*{\argmax}{arg\,max}

\DeclareMathOperator{\KL}{KL}
\newcommand{\KLdiv}[2]{\ensuremath{\KL(#1\,||\,#2)}}
\renewcommand{\epsilon}{\varepsilon}

\newcommand{\abs}[1]{\left\vert #1 \right\vert}

\newcommand{\bbR}{\mathbb{R}}

\newcommand{\bbE}{\mathbb{E}}

\newcommand{\calH}{\mathcal{H}}
\newcommand{\calL}{\mathcal{L}}
\newcommand{\calR}{\mathcal{R}}

\newcommand{\calX}{\mathcal{X}}
\newcommand{\calY}{\mathcal{Y}}
\newcommand{\bfr}{\mathbf{r}}

\newcommand{\rmd}{\,\mathrm{d}}
\newcommand{\ptheta}{\hat{p}_\theta}
\newcommand{\itc}{l}

\newtheorem{remark}{Remark}[section]

\numberwithin{equation}{section}

\newcommand{\B}[1]{\textbf{#1}}
\newcommand{\U}[1]{\underline{#1}}

\newcommand{\shiftdomain}{\mathsf{R}}
\newcommand{\fshift}{\Upsilon}

\usepackage{xcolor}

\newcommand{\rev}{}
\newcommand{\revtwo}{\color{blue!80!black}\relax}
\newcommand{\nc}{\normalcolor\relax}
\def\!r{\rev}
\def\@n{\nc}

\newcommand{\norm}[1]{\left\lVert#1\right\rVert}
\newcommand{\sbar}{\,|\,}

\DeclareMathOperator{\diag}{diag}
\DeclareMathOperator{\DFT}{DFT}

\newcommand{\bSDE}{b_\theta^{\mathrm{SDE}}}
\newcommand{\scoremodel}{s_\theta}
\newcommand{\batchsize}{B}

\newcommand{\img}{x}
\newcommand{\probe}{p}
\newcommand{\aperture}{a}
\newcommand{\fwd}{\mathcal{A}}
\newcommand{\fwdprop}{\mathcal{P}}
\newcommand{\propagator}{\mathcal{F}}
\newcommand{\shifter}{S}
\newcommand{\padder}{\textsc{Pad}}
\newcommand{\cropper}{\textsc{Crop}}

\newcommand{\meas}{y}
\newcommand{\imeas}{k}
\newcommand{\nmeas}{K}

\newcommand{\coordx}{w}
\newcommand{\coordy}{h}
\newcommand{\shiftx}{\Delta\coordx}
\newcommand{\shifty}{\Delta\coordy}
\newcommand{\imgsizex}{W}
\newcommand{\imgsizey}{H}
\newcommand{\probesizex}{W_\probe}
\newcommand{\probesizey}{H_\probe}
\newcommand{\apdiam}{d_\text{ap}}
\newcommand{\maskblocksize}{b}
\newcommand{\pos}{r}

\newcommand{\lambdaRD}{\lambda_{\text{RD}}}

\newcommand{\noise}{\varepsilon}
\newcommand{\noisestd}{\sigma_\noise}
\newcommand{\nphot}{n_{\text{phot}}}  % Expected number of photons in Poisson model

\newcommand{\Nopt}{N}  % Number of outer optimization loop steps

\newcommand{\estimg}{\hat{\img}}
\DeclareMathOperator{\PSNR}{PSNR}
\DeclareMathOperator{\SSIM}{SSIM}
\DeclareMathOperator{\aPSNR}{aPSNR}
\DeclareMathOperator{\aSSIM}{aSSIM}
\DeclareMathOperator{\cRMS}{cRMS}
\newcommand{\cRMScorr}{\gamma}  % correction scalar used in cRMS
\newcommand{\cconj}[1]{{#1}^{*}}

\newtheorem{ass}[theorem]{Assumption}
\crefname{ass}{Assumption}{Assumptions}

\newcommand{\ditto}[1][.4pt]{\xrfill{#1}\ "\ \xrfill{#1}}

\AtEveryBibitem{%
  \iffieldequalstr{keywords}{rev}{\rev }{}{}%
}

\renewcommand{\rev}{}
\renewcommand{\revtwo}{}
\begin{document}

\begin{acronym}
\acro{XFEL}{X-ray free electron laser}
\acro{SPI}{single-particle diffractive imaging}
\acro{TV}{total variation}
\acro{SDE}{stochastic differential equation}
\acro{VI}{variational inference}
\acro{VP-SDE}{variance preserving stochastic differential equation}
\acro{VE-SDE}{variance exploding stochastic differential equation}
\acro{RED}{regularization by denoising}
\acro{MAP}{maximum a posteriori}
\acro{ODE}{ordinary differential equation}
\acro{ELBO}{evidence lower bound}
\acro{DFT}{discrete Fourier transform}
%\acro{SP}{surrogate prior}
\acro{SSP}{surrogate score-based prior}
\acro{DPS}{diffusion posterior sampling}
\acro{EM}{expectation maximization}
\acro{PnP}{Plug-and-Play}
\acro{SNR}{signal-to-noise ratio}
\acro{AM}{alternating minimization scheme}
\acro{NLL}{negative log-likelihood}
\end{acronym}

% Refs for Adam in Ptychography:
% * Jiang et al. https://opg.optica.org/boe/fulltext.cfm?uri=boe-9-7-3306&id=392827
%   -> this one mentions L1 loss instead of L2 loss and discusses that " ‘L1 intensity’ and ‘L2 exit wave’ give the best results." May relate to our observations
% * Guzzi et al. https://www.mdpi.com/2410-3896/6/4/36

\maketitle

% \listoftodos

\begin{abstract}%
In this work, we present and investigate the novel blind inverse problem of \emph{position-blind ptychography}, i.e., ptychographic phase retrieval without any knowledge of scan positions, which then must be recovered jointly with the image. The motivation for this problem comes from single-particle diffractive X-ray imaging, where particles in random orientations are illuminated and a set of diffraction patterns is collected. If one uses a highly focused X-ray beam, the measurements would also become sensitive to the beam positions relative to each particle and therefore ptychographic, but these positions are also unknown. We investigate the viability of image reconstruction in a simulated, simplified 2-D variant of this difficult problem, using variational inference with modern data-driven image priors in the form of score-based diffusion models. We find that, with the right illumination structure and a strong prior, one can achieve reliable and successful image reconstructions even under measurement noise, in all except the most difficult evaluated imaging scenario.

\par\vskip\baselineskip\noindent
\end{abstract}
\section{Introduction}
In the past two decades, there has been a strong push toward imaging ever smaller specimens such as nanoparticles, virus particles or even single proteins at \acp{XFEL}, and significant developments have been made on the experimental \cite{Bogan:2008,Seibert:2011,Clark:2013,Gorkhover:2018,Ayyer:2020,Colombo:2023} and algorithmic \cite{lohReconstructionAlgorithmSingleparticle2009,ayyerDragonflyImplementationExpand2016a,tegzeComparisonEMCCM2021,mobleyTheoreticalTechniquesRecovery2024,wollterPhaseRetrievalOrientation2024} fronts to realize these imaging modalities. To obtain measurable signals, intense femtosecond-duration X-ray pulses are used, which destroy the sample but only after the pulse has traversed the sample. \ac{SPI} therefore usually combines individual measurements from a stream of reproducible objects, each recorded in a random and unknown orientation. This approach promises benefits such as time-resolved, in-situ imaging of macromolecules such as proteins, including those that are not amenable to forming large crystals \cite{neutzePotentialBiomolecularImaging2000}.

At the same time, the coherent diffractive imaging method of \emph{ptychography} has shown remarkable success in microscopy with electron beams, optical light, and X-rays \cite{Rodenburg2019Ptychography,Pfeiffer2017Xrayptychography}. Its advantage lies in the use of structured measurement redundancy by illuminating parts of an object in a scanning fashion and capturing multiple local diffraction patterns that can be merged into a single image with specialized algorithms \cite{Rodenburg2019Ptychography}. This helps to avoid the need for prior information on the object under investigation and can achieve diffraction-limited resolution, even when various sources of experimental errors are present or the structure of the illuminating beam is unknown \cite{Rodenburg2019Ptychography}.

Recent innovations in X-ray optics can now achieve a beam focus with a spot size below three nanometers \cite{Bajt:2018,Dresselhaus2024Focusing}, which is well within the size range of single biological macromolecules. Illuminating a single particle or nanocrystal with such a small beam would result in a ptychographic measurement, where only a part of the object is strongly illuminated for each diffraction pattern as the illuminating beam decays off the main beam spot. However, a significant advantage of the ptychographic measurement, the knowledge of the scan positions, is fully lost when imaging in the destructive regime of XFEL pulses. This leads to a blind inverse problem, where both the scan positions and the image of the object have to be recovered jointly. Well-established techniques \cite{maiden2012annealing,zhang2013translation} and newer developments \cite{dwivedi2018position,liu2024position} exist for the case of correcting local position errors, which is in a sense a semi-blind problem. For instance, Zhang et al. \cite{zhang2013translation} find that their method, based on serial cross-correlation of objects in a modification of the ePIE algorithm \cite{maiden2009improvedpty}, ceases to work well when the initial position error exceeds 20 pixels. The full position-blind problem has, to the best of our knowledge, not been investigated in prior works and presents a reconstruction task of high difficulty. This scenario is the subject of the present work.

Here, we perform a computational study on the viability of a simplified type of position-blind ptychographic imaging for small specimens such as single macromolecules. For simplicity and computational efficiency, we assume that the specimen is a thin sheet 2-D object of finite extent within the plane, and that the illuminating beam (also called the \emph{probe}) is concentrated in a region roughly of the size of the specimen or smaller. \revtwo Our focus with this work is to establish that this inverse problem, which is highly ill-posed and difficult even in a simplified setting, can be solved at all. The full 3-dimensional problem is therefore out of scope for this work. However, we provide guidance on future extensions to more realistic scenarios in \cref{sec:towards-realistic-spi}. \nc Due to the increased difficulty caused by the loss of position information, we incorporate prior knowledge about the imaged object via diffusion models in order to facilitate the reconstruction process.

Generative diffusion models have strongly impacted the field of machine learning in the past few years, with widespread applications from unconditional and conditional image and audio generation \cite{ho2020ddpm,Song2021Score,rombach2022high,liu2023audioldm} to data-driven approaches for solving inverse problems \cite{Chung2023Diffusion,kawar2022ddrm,zhu2023denoising,Mardani2024Variational}.
\rev
Multiple prior works have also investigated the use of (both diffusion-based and not diffusion-based) generative models for ptychography specifically \cite{shamshad_deep_2019,shamshad_adaptive_2019,Seifert2024GenerativePty,lee2025threedimpty-diff}\revtwo, but unlike our present work, these typically do not treat parameter-blind (probe- or position-blind) variants of the ptychographic problem, and do not follow a Bayesian variational approach.
\nc
In this work, we use score-based generative models, a subclass of diffusion models formulated via a continuous-time diffusion process, as data-driven priors, which have recently begun to be employed in real imaging problems, for example in a \ac{PnP}-based method for (non-blind) ptychography \cite{Denker2025PlugandPlay}.
We compare these data-driven priors against the use of either no prior information or a simple model-based \ac{TV} prior. The comparison is carried out using two algorithmic frameworks for variational inference \cite{Feng2024Variational,Mardani2024Variational} suited for model-based or diffusion-based priors, as well as simpler optimization-based procedures. \rev In both classes of methods, we make use of auto-differentiation, which has been successfully used in multiple works on ptychography \cite{10.1093/mam/ozaf070,Seifert2021AutodiffPty,kandel2019autodiffpty,ghosh2018autodiffpty,nashed2017distrautodiffpty}. \nc

\rev
Our key contributions in this work are as follows: \B{(1)} we introduce and discuss the novel and challenging blind inverse problem of \emph{position-blind ptychography}\revtwo; \B{(2)} we model this blind inverse problem as a joint posterior sampling problem under a Bayesian variational framework, deriving an alternating variational optimization scheme for successful reconstructions\nc; \B{(\revtwo3\nc)} we derive an extension of the modern generative \emph{\acl{SSP}} (\acs{SSP}) variational reconstruction method \cite{Feng2024Variational} to the parameter-blind setting; \B{(\revtwo4\nc)} we run extensive numerical simulations, employing and comparing classic model-based and modern data-driven reconstruction methods in multiple variants of the position-blind ptychography problem; \B{(\revtwo5\nc)} we offer empirical insights into the inherent difficulties associated with position-blind ptychography, as well as possible remedies inspired by classic optimization ideas and recent phaseless imaging literature; % We focus on application-level novelty
\B{(\revtwo6\nc)} we make our code available to the research community at \url{https://github.com/sp-uhh/PositionBlindPtycho}.
\nc

\section{Background on position-blind ptychography}
In this section, we explain how a position-blind ptychographic reconstruction problem may arise in a potential \ac{SPI} setup\revtwo, and detail our modeling and underlying assumptions for the toy problem variant we investigate here.\nc

%\subsection{Imaging setup}
\subsection{Phase Retrieval}
In phase retrieval imaging problems, one seeks to recover the complex-valued image $\img \in \calX \coloneqq \mathbb{C}^d$ from (noisy) intensity values $\meas$. This typically poses reconstruction problems of the form 
$$\meas = \abs{\propagator \img}^2 + \noise,$$
where $\propagator$ is a linear operator describing the light propagation in the measurement process and $\epsilon$ is measurement noise. For far-field data in applications like \ac{SPI}, X-ray crystallography and ptychography, $\propagator$ is typically the Fourier transform, whereas for near-field measurements $\propagator$ is the Fresnel integral operator with an experiment-dependent defocus value. The image reconstruction task is ill-posed since the phase problem is non-linear and subject to several sources of measurement errors in practice. Robust and efficient reconstruction algorithms are subject to ongoing research within the mathematical and machine learning literature \cite{Fienup1982Phase,Shechtman2015Phase,Cherukara2018Realtime,Fannjiang2020Numerics,Dong2023Phase,Elser-book}. 

\subsection{Ptychography}
Using modern X-ray sources at synchrotron facilities, coherent diffraction imaging (CDI) aims to solve the phase retrieval problem by reconstructing an image from the diffraction patterns $\meas$ generated from a highly coherent X-ray beam illuminating the sample $\img$. \emph{Ptychography} \cite{Pfeiffer2017Xrayptychography,Rodenburg2019Ptychography} is a special case of CDI allowing the reconstruction of high-resolution images from a collection of (far-field or near-field) diffraction patterns. It uses measurement redundancy by illuminating a sample multiple times at different positions, generating a set of diffraction patterns $\meas_\imeas$ with $\imeas = 1,\dots, \nmeas$. Each pattern $\meas_\imeas$ is collected by strongly illuminating only a part of the object under investigation (real-space ptychography) or part of the diffraction space (Fourier ptychography) \cite{Rodenburg2019Ptychography}. In real-space ptychography, the scan positions $\pos_\imeas$ are placed so that the illuminated parts overlap, and the resulting redundancy in the data mitigates the ill-posedness of the reconstruction problem.

Ptychography can also be interpreted as a specific type of \emph{coded illumination} or \emph{coded aperture} method, making links to other imaging disciplines \cite{Dong2023Phase,Allain2025Phasebook}, or as a variant of the \emph{short-time Fourier transform phase retrieval} problem, with links to signal and audio processing \cite{gerkmann2015phase,Shechtman2015Phase,gerkmann2012phase}.

Ptychography has been remarkably successful at retrieving high-resolution images of the object under investigation \cite{Rodenburg2019Ptychography}. Moreover, it allows joint reconstruction of the object and \revtwo the illuminating probe \cite{Rodenburg2019Ptychography,chang2019blind,chang2019advanced}\nc, an advantageous property since the latter is often only known approximately in phase retrieval tasks. It was also shown to be robust to various sources of measurement errors \revtwo and noise \cite{Rodenburg2019Ptychography,chang2019advanced}\nc. An important measurement error arises from imperfectly known scan positions. Such errors usually occur due to experimental factors such as imperfect scanning stages and thermal noise. Several prior works \cite{maiden2012annealing,zhang2013translation,dwivedi2018position,liu2024position} have proposed correction methods that have shown to work well as long as the initial estimates of the positions lie in a local vicinity of the true positions\revtwo{} but, to our knowledge, no published works so far model and treat the problem of fully position-blind ptychography.\nc

% \todo{SW or TR: add citations R2 suggested: \cite{Seifert2021AutodiffPty, Seifert2024GenerativePty, Allain2025Phasebook}}

% \begin{figure}[t]
%     \centering
%     \hfill%
%     \begin{subfigure}[c]{0.49\textwidth}
%         \includegraphics[width=\textwidth]{example-image-a}
%         \subcaption{Ptychographic imaging setup}
%         \label{sfig:ptycho-sketch}
%     \end{subfigure}
%     \hfill%
%     \begin{subfigure}[c]{0.49\textwidth}
%         \includegraphics[width=\textwidth]{fig/lysozyme-FEL-pty.png}
%         \subcaption{Ptychographic SPI setup}
%         \label{sfig:spi-sketch}
%     \end{subfigure}
%     \hfill%
%     \caption{We assume we have measured multiple diffraction patterns from an SPI setup. The particles move through the focus in an uncontrolled manner, so the positions of the ptychographic measurements are unknown.}
% \end{figure}
\begin{figure}[t]
    \centering
    \includegraphics[width=.8\textwidth]{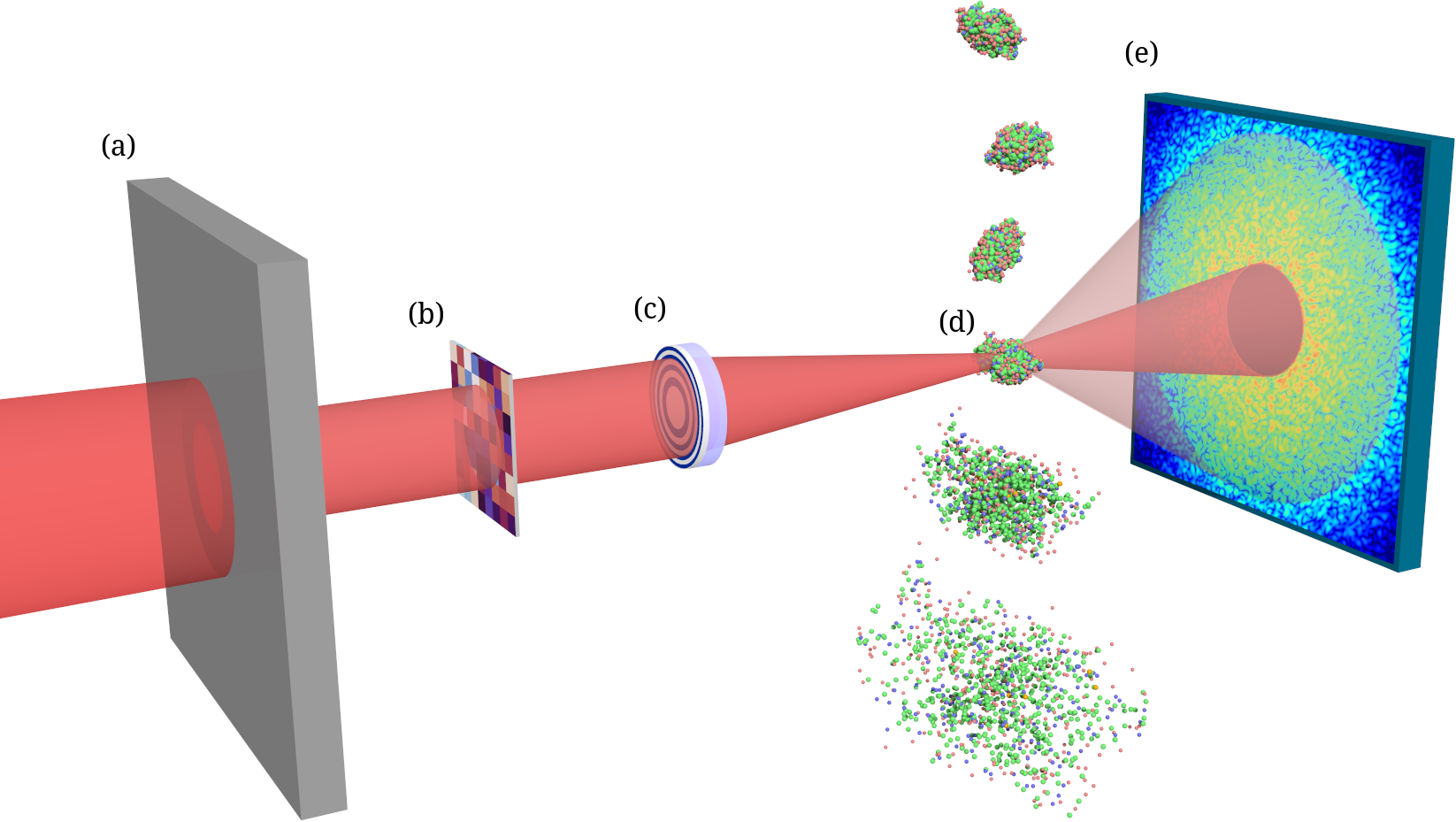}
    \caption{The ptychographic \acf{SPI} setup. Components from left to right: \textbf{(a)} a beam aperture, \textbf{(b)} an optional random phase mask, \textbf{(c)} a focusing optic, here illustrated as a Fresnel zone plate, \textbf{(d)} the interaction region at the beam focus, and \textbf{(e)} a detector. The photon beam is illustrated in transparent red. The particles move through the interaction region in an uncontrolled manner, and each particle generates a single diffraction pattern before disintegrating (\emph{diffraction before destruction} \cite{neutzePotentialBiomolecularImaging2000,chapman2014diffraction}). This makes the particle position and orientation relative to the beam unknown in every measurement.}
    \label{fig:pty-spi-sketch}
\end{figure}

\rev
\begin{remark}\label{rem:wellposed}
The ill-posedness of the ptychographic phase retrieval problem has been studied in multiple works, see \cite{grohs2020phase,chen2018coded,fannjiang2020blind,forstner2020well,grohs2021stable}. For example, \cite{grohs2020phase} proves that the solution is unique (up to a global phase factor) assuming the short-time Fourier transform of the probe function does not vanish, and gives sufficient conditions on the probe function to verify this. Results of this type, however, rely on knowledge of the measurement positions. In particular, well-posedness can be ensured given that there is enough overlap between measurements at different positions. In the position-blind setting, a similar assumptions seems problematic due to the stochastic nature of the measurement process and the scan positions. An open problem is to identify a condition that holds with high-probability under which uniqueness can be guaranteed. We later empirically support the high difficulty of the position-blind scenario by showing the high amount of non-convexity of the position recovery loss landscape, even in the idealized setting where the target object is already known, see \cref{fig:probe-comp-shift-difference-norm}.

% \textcolor{red}{possible further sources:
% \url{https://link.springer.com/article/10.1007/s43670-023-00079-1} and\url{https://ems.press/content/serial-article-files/32675}
% }
\end{remark}
\nc

\subsection{Single-Particle Imaging and Ptychography}
In this work, we investigate the extreme case of completely unknown scan positions. The underlying motivation comes from the methodology of \acf{SPI} \cite{neutzePotentialBiomolecularImaging2000}. 
Here, the imaged objects are micro- to nanometer-sized and do not reside on a well-controllable scanning stage, but are, e.g., in free flight within a \emph{molecular beam} injected into the experimental chamber and exposed with a very short-duration X-ray pulse; see \cref{fig:pty-spi-sketch}. 
Each object can only be exposed once, but it is assumed that all particles are identical in structure such that the reconstructed image $\img$ represents a single object. The SPI setup potentially enables time-resolved investigations of biochemical processes, but is experimentally limited by low probabilities that the randomly injected particles are hit by an X-ray pulse. 
The measurements further suffer from extremely low signal-to-noise ratio (SNR), since only a handful of photons are captured on the detector for each successful hit. We refer the interested reader to a review discussing these ideas and problems in more detail \cite{chapmanXRayFreeElectronLasers2019}. 
A way to increase the SNR is to use a highly focused beam with a focus size below 10~nm, which has recently become possible through advances in X-ray optics \cite{Dresselhaus2024Focusing,yamadaExtremeFocusingHard2024,Zakharova:2025}. 
When these beams have a focus size smaller than the object under investigation, the measurements become ptychographic: each diffraction pattern encodes only a part of the object. While this increases the photon flux through the sample and hence the measurement SNR by allowing more photons to diffract, it also introduces severe measurement uncertainty since the \enquote{scan} positions $\pos_\imeas$ are unknown and must be recovered jointly with the object $\img$. Since the measurement operator\revtwo{} $\fwd$\nc{} depends on the scan positions $\pos_\imeas$, the reconstruction problem is an instance of a \emph{blind} inverse problem.
\revtwo
One key difficulty in this novel blind problem is that the diffraction pattern measurements $y_\imeas$ are highly nonlinearly transformed and lossy frequency-space representations of the object image $\img$, which \emph{a priori} contain no spatial information at all and cannot directly be inverted to image estimates. This forbids the use of, e.g., registration-based methods that align estimates of image patches to retrieve some approximate positions $\hat{\pos}_\imeas$: no image exists before reconstruction that could be used to estimate positions, but without some at least partially correct estimate of the positions, no partially correct image can be reconstructed. This 
%complex \emph{chicken-and-egg}-style 
interdependence between image and measurement positions also motivates our later choice of an alternating minimization scheme, see \cref{sec:generalized-vi-for-blind-problems}.
\nc

\rev
\subsection{Position-blind ptychography}\label{sec:pos-blind-ptycho-and-assumptions}
In this work, we will treat a simplified 2D far-field real-space position-blind ptychography problem. For clarity, we preface this section with the following simplifying assumptions we make for computational viability in our simulated setting:

\rev
\begin{ass}\label{ass:knownprobe}
The illuminating wavefront (the probe function \revtwo{}$\probe_k$ in measurement $k$\nc{}) is fully known and\revtwo{} the same in all measurements, i.e., $\exists\, p\in\mathbb{C}^{\probesizey \times \probesizex}: \probe_\imeas = \probe,\, \forall \imeas \in \{1, \ldots, \nmeas\}$.
\end{ass}

\begin{ass}\label{ass:measnoiselvl-known}
The measurement noise level is known and \revtwo{}the same in all measurements\revtwo, i.e.,  in the case of Gaussian measurement noise $\exists\, \sigma> 0: \noise_\imeas \sim \mathcal{N}(0, \sigma^2 I), \, \forall \imeas \in \{1, \ldots, \nmeas\}$.
\end{ass}

\begin{ass}\label{ass:2d}
The imaged object $\img$ is treated as infinitely thin sheet such that it can be treated as a 2-dimensional complex transmission function, \revtwo i.e., $\img \in \mathbb{C}^{H \times W}$.
\end{ass}

\begin{ass}\label{ass:orientation}
The object $\img$ is
%\revtwo{}constant and\nc{}
always oriented the same in every measurement, i.e., \revtwo{}denoting by $\{\varphi_k\}_{\imeas=1,\ldots,K}$ the rotation angles for each measurement, we assume $\exists \varphi: \varphi = \varphi_k, \forall k\in\{1,\ldots, K\}$. In particular this means there are no unknown rotational parameters we need to recover. \nc{}  
\end{ass}

\begin{ass}\label{ass:probe-centered-on-object}
The probe's center is positioned somewhere on the object in every measurement, hence, each measurement is significantly influenced by some portion of the object (no measurements of empty space). \revtwo{} This means that $\pos_\imeas = (\shifty_\imeas, \shiftx_\imeas)\in \shiftdomain$ for all $k\in\{1,\ldots,K\}$, where
\begin{align*}
\shiftdomain :=
\left\{\pos =(\shifty,\shiftx):
0\leq \shifty \leq H, 0\leq \shiftx\leq W\right\}.
\end{align*}
i.e., at least $25\%$ of pixels in the probe array overlap with the object array.
% \todo[inline]{$r_k$ given by ... to lie on object domain in terms of bounds, yields a specific object-probe overlap constraint of $\geq 25\%$ overlap on the probe array (no matter what size the probe)}
\end{ass}
\nc
\rev
These assumptions together lead to a simplified toy problem variant of the full 3-D problem, as one would have to treat many additional unknowns (e.g., the probe function, sample rotations) in a realistic \ac{SPI} setting. However, due to the difficulty that our simplified problem setup already exhibits, and due to the need for parameter tuning in more complex settings, we leave relaxations of the above assumptions for future work. Nonetheless, we note that the reconstruction methods we present herein can be extended to this end, and provide concrete ideas for such future extensions in \cref{sec:towards-realistic-spi}.
%\nc

%\todo[inline]{SW, TR: check that pad, crop are correct for our modeling?}
We now turn our attention to the concrete forward model we choose.
Let $\img \in \mathbb{C}^{\imgsizey\times\imgsizex}$ be an image of the complex object transmission function which we want to recover,
\begin{equation}\label{eq:refractive-index}
\img[\coordy,\coordx] = \exp(\mathrm{i}kt\cdot (n(\coordy/\imgsizey, \coordx/\imgsizex)-1)),\quad n(h',w') = 1-\delta(\coordy',\coordx')+\mathrm{i}\beta(\coordy',\coordx'),
\end{equation}
where $n(\coordy',\coordx')$ is a complex-valued function describing the complex refractive index at normalized coordinates $\coordy',\coordx'$, with $\delta$ describing the local phase shift and $\beta$ describing the local absorption induced by the object material. For simplicity we assume a unit wavenumber $k=1$ and a unit thickness $t=1$. We further assume that the image is square, i.e., $H=W$. Let $\meas_\imeas$ be the $\imeas$-th measurement (diffraction pattern) at scan position $\pos_\imeas$, $\imeas = 1,\dots, \nmeas$. We model $\img$ and $\meas_\imeas$ to be related by the following \revtwo sub-\nc{}differentiable forward operator\revtwo{} $\fwd_\imeas$ and noise instance $\noise_\imeas$:\nc
\begin{alignat}{3}
    \meas_\imeas &= \revtwo \fwd_\imeas(x) + \noise_\imeas = \nc \abs{\revtwo\fwdprop\nc(\pos_\imeas,\img)}^2 + \noise_\imeas, \, &&\revtwo{} \fwd_\imeas&&\revtwo{}:\mathbb{C}^{H\times W}\to\mathbb{R}^{H_p\times W_p},\nc\label{eq:forward-model} \\
    \revtwo\fwdprop\nc(\pos_\imeas,\img) &= \propagator(\probe \odot \textsc{Crop}(\shifter(\pos_\imeas, \img)))\,, \,
    &&\revtwo{} \fwdprop&&\revtwo{}:\mathbb{R}^2\times\mathbb{C}^{H\times W}\to\mathbb{C}^{H_p\times W_p},\nc
    \label{eq:forward-model-linear-part}\\
    \revtwo\tilde{\shifter}(\pos_\imeas, \img)[\coordy,\coordx] &= \revtwo%\DFT^{-1} \left\{
    \DFT(\padder(\img))[\coordy,\coordx] \cdot \exp\left(\mathrm{i} \fshift(\pos_\imeas, \coordy, \coordx) \right) %\right\}
    \,, \\
    \revtwo \shifter(\pos_\imeas, \img) &= \revtwo \DFT^{-1}(\tilde{\shifter}(\pos_\imeas, \img)), \label{eq:forward-model-shift-operator}
    \hspace{4em}\mathrlap{\revtwo{} \tilde{\shifter}, \shifter: \revtwo{}
    \mathbb{R}^2\times\mathbb{C}^{H\times W}\to\mathbb{C}^{(H + H_p)\times (W + W_p)},}
    % Cutout operator: in the simplest case, 1 for region centered around $r_k$ where probe is nonzero, 0 everywhere else. Somewhat annoyingly however, this 'centering' in S_{r_k} should probably involve subpixel shifts when r_k is non-integer (in units of object pixels)
    % TODO replace the $\Delta x$, $\Delta y$ by some function of r
\end{alignat}
\noindent\revtwo{}where $\fshift$ is the Fourier-space phase shift corresponding to the real-space shift parameter $\pos_\imeas$ as
\begin{align}
\revtwo\fshift(\pos_\imeas, \coordy, \coordx) &= \fshift((\Delta\coordy_\imeas, \Delta\coordx_\imeas), \coordy, \coordx) \\
        &= \revtwo\shifty_\imeas \left(\frac{2\pi \coordy}{\imgsizey+\probesizey} - \pi \right) + \shiftx_\imeas \left( \frac{2\pi \coordx}{\imgsizex+\probesizex} - \pi \right)\,, \notag
\end{align}
\nc
\revtwo{}$\fwd_\imeas$ is the forward operator of the $\imeas$-th measurement and \nc$\noise_\imeas$ is the measurement noise of the $\imeas$-th measurement \revtwo (see \cref{sec:noise-model} for details on the used noise models), $\fwdprop$ models the full complex wavefront resulting from the probe-object interaction and propagation, and the operator $\propagator$ models \nc the wavefront propagation to the detector.
$\DFT$ denotes the \acl{DFT}, the probe array $\probe \in \mathbb{C}^{\probesizey \times \probesizex}$ models the complex-valued wavefield of the probe in the plane of the object and is applied through element-wise multiplication $\odot$, and $\shifter(\pos_\imeas, \cdot)$ is a shift operator depending on a two-dimensional shift $\pos_\imeas = (\shifty_\imeas, \shiftx_\imeas)$ which we refer to as the \emph{scan position} in the following. The $\cropper$ operator crops its input to the same array size as the probe $\probe$ before element-wise multiplication\revtwo, starting from the top left input pixel\nc. The $\padder$ operator pads its input with one full probe array size of empty space entries \revtwo in a centered fashion.
Mathematically, the cropping and padding operations are defined as
\begin{align*}
\textsc{Crop}&:\mathbb{C}^{(H + H_p) \times (W + W_p)} \to\mathbb{C}^{H_p\times W_p},\quad
\textsc{Crop}(z)[\coordy,\coordx]
\revtwo= z[h,w],\\
\textsc{Pad}&:\mathbb{C}^{H\times W} \to\mathbb{C}^{(H + H_p) \times (W + W_p)},\\
\textsc{Pad}(\img)[\coordy,\coordx]
&= 
\begin{cases}
x[h - \frac{H_p}{2}, w - \frac{W_p}{2}] &\text{if } \frac{H_p}{2} < h \leq H + \frac{H_p}{2}, \frac{W_p}{2} < w \leq W + \frac{W_p}{2},\\
1 + 0\mathrm{i} &\text{else},
\end{cases}
\end{align*}
where, for notational simplicity, we assumed that $\imgsizey, \imgsizex, \probesizey, \probesizex$ are even integers.
\nc{}%
%; see \cref{sec:scan-positions}.

\revtwo The \ac{DFT} we use in the shift operator $\shifter$ implicitly treats the input signal as periodic. We use this fact to make it impossible for estimated positions to escape into empty space, as $\pos_\imeas + [a(\imgsizey+\probesizey),b(\imgsizex+\probesizex)]$ is equivalent to $\pos_\imeas$ for all $a,b\in\mathbb{Z}$ under our DFT-based definition of $\shifter$. Since the target image itself is however a single particle and thus aperiodic, we use the padding operator $\padder$, which pads $\img$ with one full probe array size of free space ($1+0\mathrm{i}$) entries. The operator $\shifter(\pos_\imeas, \cdot)$ then effectively models shifting of an aperiodic object inside an indefinitely repeating reconstruction box.
\nc

\revtwo Since the full measurement consists of all $\imeas$ individual measurements taken, we may define an overall forward operator $\fwd$ as generating the tuple of all individual measurements,
\begin{align}
    \fwd(\img) &= \left(
        \fwd_1(\img) + \varepsilon_1,
        \ \ldots,\ 
        \fwd_\nmeas(\img) + \varepsilon_\nmeas
    \right)\in\mathbb{R}^m,
\end{align}
%
%where $m=\underbrace{(H_p\times W_p)\times\ldots\times (H_p\times W_p)}_{\nmeas\ \text{times}}$.
% \nc
%
% %\subsubsection{Propagation model}
%
% \revtwo
%We have devised $\fwd_\imeas$ to be (sub-)differentiable with respect to both $\img$ and $\pos_\imeas$, in order to be able to optimize both quantities using gradient descent.
where $m = k\times(H_p\times W_p)$. Note, that each operator $\fwd_\imeas$ is sub-differentiable, with respect to both $\img$ and $\pos_\imeas$, instead of differentiable due to the involvement of the absolute value operation $|\cdot| : \mathbb{C}\to\mathbb{R}_{\geq 0}$, the presence of which is an unavoidable property of the physics underlying the phase retrieval problem. This is, however, enough for the practical implementation using automatic differentiation and the Adam optimizer \cite{kingma2014adam} in our experimental sections.
%Since we will devise our variational solution methods based on automatic differentiation, the resulting sub-differentiality is however unproblematic due to ... \todo[inline]{SW4TR: can you revise the above and finish the paragraph here? :)}
\nc

Since we assume a \emph{far-field ptychography} problem, we set $\mathcal{F}$ to be the (discrete) Fourier transform. It may be interesting to investigate the case of Fresnel propagation (near-field ptychography), where position recovery could be simpler due to some real-space information being encoded in the diffraction patterns. However, we focus on the far-field case here, which should be more challenging due to a complete lack of position information in the measurements.
\section{Reconstruction and Sampling Methods}\label{sec:recmethods}

In the following, we describe the theory behind the investigated methods for image reconstruction and sampling with and without score-based priors.

\subsection{Score-Based Priors for Imaging Inverse Problems}

%\subsubsection{(Blind) Inverse Problems}
In imaging inverse problems, we aim to recover a $d$-dimensional image $\img \in \calX$ from a measurement $\meas\in  \calY \coloneqq\bbR^m$, where the two are related via
\begin{equation}\label{eq:inverse-problem}
    \meas = \fwd(\img) + \noise.
\end{equation}
For now, $\fwd$ is a general forward model describing the measurement acquisition and $\noise$ is measurement noise. Typically the inverse of $\fwd$ is discontinuous, making the solution of this problem ill-posed. The reconstruction of $\img$ from $\meas$ therefore requires some form of regularization \cite{Benning2018Modern}. In order to quantify uncertainty in the reconstruction, it is further necessary not only to reconstruct a single solution to \eqref{eq:inverse-problem}, but we have to make statements about the statistical properties of $\img$. The distribution of possible solutions can be described using Bayesian inference \cite{Stuart2010Inverse}. In the Bayesian formulation, we recover the posterior distribution of the image $\img$, the density of which is given by 
\begin{equation}\label{eq:bayes-non-blind}
    p(\img \sbar \meas) = \frac{p(\meas \sbar \img) p(\img)}{Z}.
\end{equation}
Here $p(\img)$ is the density function of the prior distribution that is imposed on the space of possible images, which acts as the regularization of the problem \cite{Stuart2010Inverse}. The term $p(y\sbar x)$ is the measurement likelihood implied by \eqref{eq:inverse-problem}. A standard point estimate for this posterior which can serve as an exemplary solution to the inverse problem is the \ac{MAP} estimate $\img^{\mathrm{MAP}} = \argmax_\img p(\img\sbar \meas)$, but the posterior also allows to compute more advanced statistical properties like moments or confidence sets. In order to carry out these computations, we need to draw samples from the posterior. The density function $p(\img\sbar \meas)$, however, is generally intractable since its normalizing constant, the model evidence $Z = p(\meas) = \int p(\meas\sbar \img) p(\img) \rmd\img$, is a high-dimensional integral and as such unknown. Algorithms therefore typically fall back to approximating the posterior by a simpler distribution with tractable density that is easy to sample from, or sample from the posterior using methods that do not require knowing $Z$, e.g., Markov chain Monte Carlo (MCMC) algorithms \cite{Kaipio2005Statistical}.

Inverse problems are termed \emph{blind} if the forward model depends on an unknown parameter $\bfr \in \calR \subset \bbR^K$. Instead of \eqref{eq:inverse-problem}, the measurement in a blind problem is given by \begin{equation}
y = \fwd(\bfr,\img)+\noise \,.
\end{equation}
In the Bayesian setting, the model parameters can be treated in a similar way as the unknown: The likelihood term is now $p(y\sbar x,\bfr)$ and the posterior of both image and model parameter is given by
\begin{equation}\label{eq:bayes-blind}
    p(\img,\bfr\sbar \meas) = \frac{p(\meas\sbar \img,\bfr)p(\img,\bfr)}{Z}.
\end{equation}
Depending on the application, it can often be assumed that $\img,\bfr$ are independent under the prior, so that $p(\img,\bfr) = p(\img)p(\bfr)$ with respective prior distributions $p(\img)$ and $p(\bfr)$. Any inference task is now carried out with respect to the joint posterior $p(\img,\bfr\sbar \meas)$.

\subsubsection{Learning Priors with Diffusion Models}\label{sec:learnedpriors}
Score-based diffusion models are a popular method for generative modeling due to their ability to learn complex distributions of training image datasets. In recent research, their adaptability to conditional/posterior distributions in inverse problems has been showcased with promising results \cite{Daras2024Survey}. The main concept behind the unconditional model is to transform an unknown distribution $p_0(x)$ of images to a normal distribution via a diffusion process. To draw new samples, one samples from the normal distribution and simulates an associated backward diffusion. The forward diffusion process is given by the \ac{SDE}
\begin{equation}\label{eq:diffusion-sde}
\rmd x_t = f(x_t,t) \rmd t + g(t) \rmd W_t,\qquad t \in [0,T],
\end{equation}
with $x_0 \sim p_0$, where $W_t$ denotes standard Brownian motion. We will denote $p_t$ for the distribution of $x_t$ in \eqref{eq:diffusion-sde} at time $t$. The drift $f(x,t)$, the diffusion coefficient $g(t)$ and the final time $T$ are chosen such that $p_t$ approximately equals an analytically tractable distribution $\pi$ at the final time $t=T$, i.e. $p_T \approx \pi$.

In order to sample from $p_0$, score-based models simulate the time-reversed \ac{SDE}
\begin{equation}\label{eq:reverse-sde}
\rmd x_t = \left( f(x_t,t) - g(t)^2 \nabla_x \log p_t(x_t) \right) \rmd t + g(t) \rmd \bar W_t,
\end{equation}
where $\bar W_t$ is a time-reversed Brownian motion. If the reverse \ac{SDE} is initialized at time $t=T$ as $x_T \sim p_T$, then under mild conditions on the coefficients $f,g$, the process at time $t=0$ obeys $x_0 \sim p_0$ \cite{Anderson1982ReverseTime}. We will compare results for two standard SDE choices. The first one is the \ac{VP-SDE} with $f(x_t,t) = -\frac{\beta(t)}{2}x$ and $g(t) = \sqrt{\beta(t)}$, with $\beta(t)$ linear and monotonically increasing so that $p_t$ converges to the standard normal $\pi = \mathcal{N}(0,I)$. The second one is the variance-exploding variant (\acs{VE-SDE}\acused{VE-SDE}) with $f\equiv 0$ and $g(t)$ a monotonically increasing schedule, with $g(T)$ large enough such that $p_T \approx \mathcal{N}(0,g(T)^2 I)$. Samples from $p_0$ can thus be generated by sampling from the tractable distribution $x_T \sim \pi \approx p_T$ and simulating \eqref{eq:reverse-sde}. 

The practical difficulty of this approach lies in accurately approximating the score function $\nabla_x \log p_t(x_t)$ in \eqref{eq:reverse-sde}. This is possible if we already have access to sufficient training data drawn from $p_0$, since we can then train a score model $s_\theta(x_t,t) \approx \nabla_x \log p_t(x_t) $ which approximates the true score \cite{Hyvarinen2005Estimation,Song2021Maximum}. The score model is parametrized by a network $\theta$ and trained using \emph{denoising score matching}, see e.g. \cite{vincent2011connection}:
\begin{equation}\label{eq:dsm}
    \argmin_\theta \int_0^T \lambda(t) \bbE_{(x_0,x_t)\sim p(x_0,x_t)} \left[ \norm{s_\theta(x_t,t) - \nabla_x \log p(x_t\sbar x_0)}^2 \right] \rmd t,
\end{equation} 
where $\lambda(t)$ is a weighting factor balancing the approximation quality at different time steps. Solving \eqref{eq:dsm} requires estimating the expectation with respect to the joint distribution $p(x_0,x_t)$. If the drift coefficient $f$ in \eqref{eq:diffusion-sde} is affine linear, the forward in time conditional $p(x_t\sbar x_0)$ is a normal distribution with a known closed-form mean and variance. Sample pairs $(x_0,x_t)$ from $p(x_0,x_t) = p(x_0)p(x_t\sbar x_0)$ can hence be easily generated by drawing $x_0$ from the training data $p(x_0)$ and generating $x_t$ efficiently by calculating the closed-form mean and variance expressions and adding sampled Gaussian noise.

Once the score model is trained, new samples from $p_0$ can be generated by replacing $\nabla_x \log p_t(x_t)$ by $s_\theta(x,t)$ in \eqref{eq:reverse-sde} and then simulating the reverse \ac{SDE} by discretizing it using, e.g., a standard Euler--Maruyama scheme \cite{Song2021Score} and an initialization $x_T \sim \pi$. 

\subsubsection{Sampling from a Bayesian Posterior}
For unconditional sampling from $p_0$, we can employ the training data to estimate the expectation in \eqref{eq:dsm}. Suppose we aim to sample from a posterior distribution instead, where the initial distribution $p_0(x)$ in the \ac{SDE} would be of the form $p(x\sbar y)$ \eqref{eq:bayes-non-blind}. In that case, sampling is no longer possible since there are no representative samples from the posterior to begin with. To circumvent this problem, most methods train a score model $s_\theta(x,t)$ on the image prior distribution $p(x)$. Crucially, since the prior distribution $p(x)$ is not a function of $y$, it allows the score model to be pre-trained offline before making any measurement.

Focusing for the moment on the non-blind setting \eqref{eq:bayes-non-blind}, conditional sampling thus requires adjusting for a diffused likelihood term: If we want to simulate \eqref{eq:reverse-sde}, where the target at time $t=0$ is the posterior $p(\img\sbar \meas)$, the score is given by
\begin{align*}
    \nabla_x \log p_t(x_t\sbar \meas) = \nabla_x \log p_t(\meas\sbar x_t) + \nabla_x \log p_t(x_t)%\\
    \approx \nabla_x \log p_t(\meas\sbar x_t) + s_\theta(x_t,t)
\end{align*}
While the prior score $\nabla_x \log p_t$ can be efficiently approximated by the score model $s_\theta$, the term $\nabla_x \log p_t(\meas\sbar x_t)$ is generally intractable.

Some works have developed schemes to approximate $\nabla_x \log p_t(\meas\sbar x_t)$, e.g., by building approximations based on the chain rule $p_t(y\sbar x_t) = \bbE_{x_0}[p(\meas\sbar x_0)p(x_0\sbar x_t)]$. One instance of these methods is \ac{DPS} \cite{Chung2023Diffusion}, and we refer to \cite{Daras2024Survey} for an overview of several other such algorithms.

Other works avoid approximating the intractable likelihood score by not simulating the reverse \ac{SDE} \eqref{eq:reverse-sde} for the posterior at all. Instead, one can try to take a \ac{VI} approach, which has been done for the RED-Diff method \cite{Mardani2024Variational} and the works on principled score-based priors in \cite{Feng2023Score,Feng2024Variational}. Consider the posterior $p(x\sbar y)\propto p(\meas\sbar \img) p(x)$. One can use a learned approximation ${\ptheta}(x)$ for the prior term $p(x)$, where ${\ptheta}(x)$ is implicitly defined through a learned score-based prior $s_\theta(x, t)$. This defines an approximated, but still intractable posterior ${\ptheta}(\img\sbar y) \propto p(\meas\sbar \img) {\ptheta}(x)$, which can be modeled via \ac{VI} by a tractable parametric distribution $q_\phi$. To that end, one solves the optimization problem
\begin{equation}\label{eq:variational-inference-minkl}
    \phi^\ast \coloneqq \argmin_\phi \left\{\KLdiv{q_\phi}{{\ptheta}(\cdot\sbar y)}\right\}.
\end{equation}
Depending on the chosen parametric class, the resulting approximate posterior $q_\phi$ can allow for direct sampling and density evaluations. The complexity of recovering $\phi^\ast$ is controlled by its dimensionality and the chosen parametric family of distributions. For instance, $\phi$ could consist of the mean and covariance parameters of a simple Gaussian or Gaussian mixture \cite{Blei2017Variational}, or, allowing for more expressivity, $\phi$ could be the parameters of a neural network encoding a normalizing flow model \cite{Sun2021Deep}. Rewriting \eqref{eq:variational-inference-minkl}, those methods seek to recover
\begin{align}\label{eq:variational-inference-loss2}
    \phi^\ast &= \argmin_\phi \left\{\bbE_{x \sim q_\phi} \left[ - \log {\ptheta} (x,y) + \log q_\phi(x) \right]\right\}\notag\\
    &= \argmin_\phi \left\{\bbE_{x \sim q_\phi} \left[ - \log p(y\sbar x) - \log {\ptheta}(x)\right] - \calH (q_\phi)\right\},
\end{align}
where $\calH$ denotes the entropy functional $\calH(q) \coloneqq - \bbE_{\img \sim q} [\log q(\img)]$. 

\subsubsection{RED-Diff}
A simple variational distribution is a Gaussian with mean $\mu \in \mathbb R^d$ and isotropic covariance with a scalar $\sigma > 0$, i.e., $q_\phi = \mathcal{N}(\mu, \sigma^2 I)$, with $\phi = (\mu,\sigma)$. In \cite{Mardani2024Variational}, the authors prove that the \ac{VI} objective \eqref{eq:variational-inference-loss2} can then be written as
\begin{equation}\label{eq:red-diff-loss-unchanged}
    \argmin_{\phi} \left\{ -\bbE_{x \sim q_\phi}\left[ \log p(y\sbar x) \right] + \int_0^T \omega(t) \bbE_{x\sim q_t(\cdot\sbar y)} \left[ \norm{\nabla_x \log q_t(x\sbar y) - \nabla_{x} \log p_t (x)}_2^2 \right]\rmd t \right\},
\end{equation}
where $\omega(t)$ is a suitable weight, and $q_t = \mathcal{N}(\alpha(t) \mu, (\alpha(t)^2\sigma^2 + \sigma(t)^2)I)$ is the distribution that arises from simulating the forward \ac{SDE} \eqref{eq:diffusion-sde} with initial condition $q_\phi$. The functions $\sigma(t), \alpha(t)$ are defined as $\sigma(t) = 1 - \exp(-\int_0^t \beta(s) \rmd s)$ and $\alpha(t) = \sqrt{1-\sigma(t)^2}$ (for \ac{VP-SDE}) or $\sigma(t) = g(t)$ and $\alpha(t) \equiv 1$ (for \ac{VE-SDE}), respectively. Since the diffused variational density $q_t$ is available in closed form and $\nabla_x \log p_t(x)$ can be replaced by the trained score model $s_\theta(x,t)$, the terms under the integral can be evaluated efficiently. By modifying the weight $\omega(t)$ in the integral, the authors arrive at a loss function that bears similarities to the \ac{RED} approach to \ac{MAP} estimation in inverse problems \cite{Romano2017Little}, despite it not being equal to the original \ac{VI} loss anymore. Additionally, the authors assume for their numerical experiments that $\sigma \approx 0$, essentially fitting a point mass to the posterior and deviating from the Bayesian motivation of the \ac{VI} approach. Despite these modifications, the method is reasonably fast and the reconstructed images of promising quality.

\subsubsection{Variational inference with principled score-based priors}\label{sec:vi-principled-priors}
In two other works \cite{Feng2023Score,Feng2024Variational}, the \ac{VI} loss is optimized without reweighting in time. Evaluating the objective (or its gradients) in \eqref{eq:variational-inference-loss2} requires evaluating the prior log-density $\log {\ptheta}(x)$ for unseen data $x$. The score model $s_\theta(x_t,t)$ is typically unstable around $t\approx 0$, but as previously shown by Song et al. \cite{Song2021Score}, log-density values can instead be obtained by solving the initial value problem for the forward probability flow \ac{ODE} 
\begin{equation}\label{eq:probability-flow-ode}
    \frac{\mathrm d x_t}{\mathrm d t} = \underbrace{f(x_t, t) - \frac{1}{2} g(t)^2 \nabla_x \log p_t(x_t,t)}_{\rev =: v(x_t,t)}, \qquad x_0 = x.
\end{equation}
This generates the same dynamics of the distribution $p_t$ as \eqref{eq:diffusion-sde}, since the \rev continuity equation of \eqref{eq:probability-flow-ode} coincides with the Fokker--Planck equation of \eqref{eq:diffusion-sde}. For velocity fields $v\in C^2$, \cite[Cor. 10.4]{wald2025flow} proves the \emph{instantaneous change of variables} formula
\begin{align*}
\frac{\mathrm d}{\mathrm d t} p_t = - p_t(x_t) \nabla \cdot v(x_t,t) \quad\Leftrightarrow\quad
\frac{\mathrm d}{\mathrm d t} \log p_t = -\nabla \cdot v(x_t,t)
\end{align*}
see also \cite{chen2018neural}. We can thus write $\log p_0$ as the solution of an initial-value problem:
\begin{align*}
\log p_0(x_0) = 
\log p_T(x_T) + \int_0^T \nabla\cdot v(x_t,t) \mathrm d t
\end{align*}
\nc
%, since the Liouville equation of \eqref{eq:probability-flow-ode} coincides with the Fokker--Planck equation of \eqref{eq:diffusion-sde}.
Replacing the score in \eqref{eq:probability-flow-ode} by a score model $s_\theta$ \rev induces a learned velocity field $v_\theta$ which can be used to solve the ODE numerically. This results in the approximation $\log {\ptheta}^{\mathrm{ODE}} \approx \log {p}$.\nc{}
The authors of \cite{Feng2023Score} thus proposed to use $\log {\ptheta}^{\mathrm{ODE}}$ to evaluate the objective \eqref{eq:variational-inference-loss2}. As the numerical results of \cite{Feng2023Score} showed, the resulting \ac{VI} approach has a high computational cost, but generates very accurate approximations of the true posterior.\!r A follow-up paper \cite{Feng2024Variational} resolves the costly evaluation of $\log {\ptheta}^{\mathrm{ODE}}$ by instead employing a surrogate, that is reminiscent of the \ac{ELBO} and was first derived in \cite[Thm. 3]{Song2021Maximum}. The ELBO surrogate as used in \cite[p. 5]{Feng2024Variational} is defined as follows,
\begin{align}\label{eq:bsde-full}
b_\theta^{\text{SDE}}(x_0) = 
\mathbb{E}_{x\sim p_{T}(x|x_0)} \left[
\log \pi(x)\right] - \frac{1}{2}\int_{0}^{1} g(t)^2 h(t, x_0) \mathrm d t 
\end{align}
where 
\begin{align}\label{eq:bsde-h}
h(t,x_0) = 
\mathbb{E}_{x\sim p_{t}(x|x_0)} \left[
\| s_\theta(x,t) - \nabla_x \log p(x|x_0)\|_2^2 - \|\nabla_x \log p_{t}(x|x_0)\|_2^2 - 
\frac{2}{g(t)^2} \nabla \cdot f(x, t)
\right].
\end{align}
Under assumptions specified in \cite[App. A]{Song2021Maximum}, the authors show the following inequality \cite[Thm. 3]{Song2021Maximum},
\begin{align}\label{eq:bsdebound}
b_\theta^{\text{SDE}}(x)\leq
\log \hat{p}_\theta(x) \qquad\forall x\in\mathcal{X}.
\end{align}
In practice, $\bSDE$ is approximated by replacing the integral \eqref{eq:bsde-full} and the expectation inside the integrand term $h$ \eqref{eq:bsde-h} with empirical Monte Carlo estimators using a finite number of samples, here reproduced from \cite[Sec.~4.2]{Feng2024Variational}: 
\begin{equation}\label{eq:bsde-empirical}
\begin{aligned}
\bSDE
\approx 
\tilde{b}_\theta^{\mathrm{SDE}}(x) :=\ &  \frac{1}{N_{z}} \sum_{j=1}^{N_{z}} \log \pi\left(x_j^{\prime}\right)
- \frac{1}{2 N_t N_{z}} \sum_{i=1}^{N_t} Z \beta(t)^2 \sum_{j=1}^{N_{z}}
\Bigg[\\
& \quad
    \left\|\mathbf{s}_\theta(x_{i j}^{\prime}, t_i) + \frac{z_{i j}}{\beta(t_i)}\right\|_2^2
    - \left\|\frac{z_{i j}}{\beta(t_i)}\right\|_2^2
    - \frac{2}{g(t_i)^2} \nabla_{x_{i j}^{\prime}} \cdot f(x_{ij}^{\prime}, t_i)
\Bigg]
\\
\text { s.t. } & \quad t_i \sim p(t), z_{i j} \sim \mathcal{N}(\mathbf{0}, \mathbf{I}), x_{i j}^{\prime}=\alpha(t_i) x+\beta(t_i) z_{i j} \\
& x_j^{\prime} \sim \mathcal{N}(\alpha(T) x, \beta(T)^2 \mathbf{I}) \quad \forall i=1, \ldots, N_t, j=1, \ldots, N_{z},
\end{aligned}
\end{equation}
where we set $N_t = 1, N_z = 1$ and also use the re-weighted proposal distribution (c.f. \cite{Song2021Score,Feng2024Variational}) $p(t):=\frac{g(t)^2}{\beta(t)^2 Z}$, where $Z=\int_{0}^T g(t)^2/\beta(t)^2 \mathrm{d} t$. After this modification to \eqref{eq:variational-inference-loss2}, one instead solves
\begin{equation}\label{eq:variational-inference-surrogate3}
    \tilde \phi \coloneqq \argmin_\phi \left\{\bbE_{\img \sim q_\phi} \left[ - \log p(\meas\sbar \img) - \tilde{b}_\theta^{\mathrm{SDE}}(\img)\right] - \calH (q_\phi) \right\},
\end{equation}
where the \ac{ELBO} term $\tilde{b}_\theta^{\mathrm{SDE}}$
%can be estimated efficiently using Monte Carlo integration.
is the Monte Carlo estimator \eqref{eq:bsde-empirical}.
Variational inference naturally allows the parallel evaluation of the likelihood and the prior on image minibatches, consisting of multiple independent image samples, due to the ability to sample from the modeled distribution $q_\phi$. 
We further approximate the expectation in \eqref{eq:variational-inference-surrogate3} using a batch of $B=4$ independent image samples $x_1,\ldots,x_B \sim q_\phi$, used to evaluate both the likelihood term and $\tilde{b}_\theta^{\mathrm{SDE}}$.
%The number of sample time steps per $x_i$ is set to $N_t=1$.
%evaluating both the likelihood term and $\tilde{b}_\theta^{\mathrm{SDE}}$ onthis batch,
%To calculate $\tilde{b}_\theta^{\mathrm{SDE}}$, we pair each image $x_i$ in the batch with a corresponding independent $t_i$ to have $N_t = B$. %% No - if B>1 and we choose one timestep per x, then N_t=1.
\nc
Heuristically, the \ac{ELBO} term induces a \ac{SSP} with density ${\ptheta}^{\mathrm{surr}} \propto \exp(b_\theta^{\mathrm{SDE}})$. The reduced computational time of \eqref{eq:variational-inference-surrogate3} allows to lift the problem dimension to realistic imaging sizes.
\!r%
\begin{remark}
In general, one cannot expect the bound in \labelcref{eq:bsdebound} to be tight and an analytic characterization of the gap is an open problem. %Nevertheless, \cite[sec. 5]{Feng2024Variational} offers some empirical observations on the effect of replacing $\log \hat{p}_\theta$ by the empirical estimator $\tilde{b}_\theta^{\mathrm{SDE}}$.
%and shows differences between solutions to \cref{eq:variational-inference-surrogate3} and \cref{eq:varinf}. 
Empirically, it was shown that this surrogate prior is very effective, in the sense that for small dimensional examples $\tilde \phi \approx \phi^\ast$, i.e., the fitted distribution shows very good agreement with the variational approximation to the posterior \cite[sec. 5]{Feng2024Variational}. 

In the case where the true posterior is Gaussian, \cite[Fig. 3]{Feng2024Variational} shows that solutions to \labelcref{eq:variational-inference-surrogate3} coincide closely with solutions to the original problem \labelcref{eq:variational-inference-loss2}. While \cite[Fig. 4]{Feng2024Variational} highlights differences between using $\log \hat{p}_\theta$ and $\tilde{b}_\theta^{\text{SDE}}$ when applied for more complicated posterior distributions, the experiments still validate that the approach leads to a good posterior approximation.
\end{remark}
\@n

We now describe how \rev this method can be extended \nc to the case of blind inverse problems \eqref{eq:bayes-blind}, using a standard variational Bayes approach for the joint posterior of image $x$ and latent parameter $\bfr$. While we are specifically interested in solving the position-blind ptychography problem \eqref{eq:forward-model}, the method we
\rev
derive
\nc
can be used for general blind imaging inverse problems. 

\subsection{Sampling from the Position-Blind Ptychography Posterior}
Consider now a semi-blind
%\todo{we could add an actual semiblind results section + rough theoretical analysis here. Tim mentioned the PALM paper for this}
or blind setting, where we need to reconstruct the posterior $p(\img,\bfr\sbar \meas)$ in the joint variable $(x,\bfr)$, where $x$ is an image and $\bfr$ the latent parameter. In our position-blind ptychographic setup, $\bfr$ will be a vector containing the measurement positions. 

\subsubsection{Generalized VI Approach for Blind Problems}\label{sec:generalized-vi-for-blind-problems}
Using \eqref{eq:bayes-blind}, we generalize the \ac{VI} approach to the joint posterior $p(x,\bfr\sbar y)$ in the blind case. Under the non-restrictive assumption of independent priors $\log p(x,\bfr) = \log {\ptheta}(x) + \log p(\bfr)$, the derivation of the \ac{VI} objective \eqref{eq:variational-inference-loss2} easily generalizes to
\begin{equation}\label{eq:variational-inference-blind4}
    \phi^\ast = \argmin_\phi \left\{\bbE_{(x,\bfr) \sim q_\phi} \left[ - \log p(y\sbar x,\bfr) - \log {\ptheta}(x) - \log p(\bfr) \right] - \calH (q_\phi) \right\} = \argmin_{\phi} \calL(\phi).
\end{equation}
Note that the variational distribution $q_\phi$ is now the joint distribution of image and parameters. Depending on the forward model and dimensionality, the joint optimization problem can exhibit strong non-convexity and be computationally demanding.
\rev
Empirically, in preliminary experiments with optimizing $\chi,\rho$ jointly according to \eqref{eq:variational-inference-blind4}, we did not achieve any successful reconstructions. We therefore split the joint problem into sub-problems as follows.
\nc
The problem can be provided with more structure by a standard mean-field assumption on the variational Bayesian posterior
\begin{equation}\label{eq:variational-bayes-mean-field-assum}
    q_\phi (x,\bfr) = q_{\chi}(x) q_{\rho}(\bfr),
\end{equation}
where $\phi = (\chi, \rho)$, with $\chi$ encoding the image posterior and $\rho$ the latent parameter posterior. Note that this comes at the cost of less accurately representing the correlation of uncertainties in image and parameters in $q_\phi$, but is necessary to make the optimization feasible at scale.
\rev
Doing so also allows for higher flexibility; for instance, it lets us decouple the number of sequential optimization steps on $\rho$ and on $\chi$, a fact which we will practically make use of.
\nc
We note that a detailed analysis of this approximation is an important open question for future research, but is outside of the scope of this work.
By inspecting the optimality conditions of $\min_\phi L(\phi) = \min_{\chi,\rho} L(\chi,\rho)$ with respect to $\chi$ and $\rho$ separately, under the mean-field assumption, the posterior marginals obey the optimality conditions \cite{Blei2017Variational} 
\begin{align*}
    q_{\chi} \propto \exp\left(\bbE_{\bfr \sim q_{\rho}} \left[ \log p(x \sbar \bfr, y) \right]\right),\qquad
    q_{\rho} \propto \exp\left(\bbE_{x \sim q_{\chi}} \left[ \log p(\bfr \sbar x, y) \right]\right).
\end{align*}
This motivates the following %iterative scheme, 
alternating updates of the image and parameter distributions
\begin{subequations}\label{eq:variational-bayes-scheme}
\begin{align}
    \chi_{k+1} &= \argmin_\chi \calL(\chi, \rho_k) = \argmin_{\chi} \left\{ \bbE_{x\sim q_{\chi}} \bbE_{\bfr \sim q_{\rho_k}} \left[ - \log p(y\sbar x,\bfr) - \log {\ptheta}(x) \right] - \calH (q_{\chi}) \right\} ,\label{eq:variational-bayes-image-step}\\
    \rho_{k+1} &= \argmin_{\rev\rho\nc} \calL(\chi_{k+1}, \rho) = \argmin_{\rho} \left\{\bbE_{x\sim q_{\chi_{k+1}}} \bbE_{\bfr \sim q_{\rho}} \left[ - \log p(y\sbar x,\bfr) - \log p(\bfr) \right] - \calH (q_{\rho}) \right\}.\label{eq:variational-bayes-par-step}
\end{align}
\end{subequations}
The technique of learning a score model using training data applies only to the image prior ${\ptheta}(x)$, while the parameter prior $p(\bfr)$ depends on knowledge in the specific application. For instance, in our ptychography setup, $p(\bfr)$ could be chosen as a unimodal density (if we have an initial estimate of the measurement position, leading to a ``semi-blind'' problem) or constant (reflecting a uniform prior on the two-dimensional cell that the ptychographic measurements are restricted to, with no prior information imposed at all).

\subsubsection{Blind RED-Diff}
The RED-Diff method has been modified for tackling blind imaging inverse problems on the example of MRI with unknown off-resonance field map \cite{Alkan2023Variational}. The authors there employed the same mean-field assumption \eqref{eq:variational-bayes-mean-field-assum} in order to separate optimization steps for the reconstructed image and parameters.
\rev We recall the framework established in \cite{Alkan2023Variational} in the following, written in the notation we use here.
\nc
As before, the objective for the image posterior can be rewritten using
%\todo{We recall their framework} 
\cite[Prop. 1]{Mardani2024Variational} to arrive at a loss similar to \eqref{eq:red-diff-loss-unchanged}, but with the likelihood conditioned on the current parameter estimate, i.e., \eqref{eq:variational-bayes-image-step} becomes
\begin{align*}
    \chi_{k+1} = \argmin_{\chi} \bigg\{ & -\bbE_{x \sim q_\chi} \bbE_{\bfr \sim q_{\rho_k}}\left[ \log p(y\sbar x,\bfr) \right]
    \\ &+ \int_0^T \omega(t) \bbE_{x\sim q_t(\cdot\sbar y)} \left[ \norm{\nabla_x \log q_t(x\sbar y) - \nabla_{x} \log p_t (x)}_2^2 \right]\rmd t \bigg\}
\end{align*}
Like \cite{Mardani2024Variational}, the work \cite{Alkan2023Variational} then introduces time reweighting by replacing $\omega(t)$ by a different $\tilde \omega(t)$. The variational posterior $q_\chi$ is replaced by a point mass and the entropy term removed from the objective in order to arrive at an implementable scheme. We mention again that this essentially replaces the Bayesian character of the reconstructed quantity with an optimization scheme that rather resembles classical \ac{MAP} computation. In our notation, the resulting image optimization step is
\begin{align}
    x_{k+1} &= \argmin_x \left\{ - \log p(y\sbar x,\bfr_k) + \int_0^T \tilde \omega(t) \bbE_{x_t \sim q_t(\cdot\sbar y)} \left[ \norm{\scoremodel(x_t;t) - z}_2^2 \right] \rmd t \right\}\notag\\
    &\eqqcolon \argmin_x \calL^{\mathrm{REDdiff}}(x,\bfr_k),\label{eq:blind-red-diff-image-step}
\end{align}
where now $q_t = \mathcal{N}(\alpha(t) x, \sigma^2(t)I)$ and we abbreviated $z = -\frac{x_t - \alpha(t)x}{\sigma^2(t)}$. The time reweighting $ \tilde \omega(t)$ is chosen such that $\tilde  \omega(0) = 0$, since the gradients of the objective in \eqref{eq:blind-red-diff-image-step} then allow the following %, particularly simple 
form that does not require backpropagation through the score network
\begin{equation}\label{eq:blind-red-diff-image-gradient}
    \nabla_x \calL^{\mathrm{REDdiff}}(x,\bfr_k) = - \nabla_x \log p(y\sbar x,\bfr_k) - \int_0^T \tilde \omega(t) \bbE_{x_t \sim q_t(\cdot\sbar y)} \left[ \scoremodel(x_t;t) - z \right] \rmd t,
\end{equation}
with a suitable weight $\tilde \omega(t)$; see \cite[Prop. 2]{Mardani2024Variational} for details. For the parameter update step \eqref{eq:variational-bayes-par-step}, the authors of \cite{Alkan2023Variational} showed that assuming $p(\bfr)$ and $q_\rho(\bfr)$ are both Laplace distributions allows to obtain closed form representations of the relevant terms $p(\bfr)$ and $\calH(q_\rho)$. Upon replacing $q_\rho$ with a point mass on a single $\bfr$, the entropy is, however, dropped again in \cite{Alkan2023Variational} and the parameter update becomes
\begin{equation}\label{eq:blind-red-diff-par-step}
    \bfr_{k+1} = \argmin_{\bfr} \left\{ - \log p(y\sbar x_{k+1},\bfr) - \log p(\bfr) \right\} \eqqcolon \argmin_{\bfr} \calL^{\mathrm{REDdiff}}(x_{k+1},\bfr).
\end{equation}
We outline the resulting algorithm with our notation in \cref{algo:blind-red-diff}.

\subsubsection{Blind \acf{SSP} method} \label{sec:methods:blind-ssp}
The scheme of \cite{Feng2023Score,Feng2024Variational} can be generalized to the blind setting in a similar way. As before, the image step \eqref{eq:variational-bayes-image-step} could be solved directly by deriving the density via the probability flow \ac{ODE}. However, this becomes computationally infeasible in high dimensions, so the \ac{ELBO} term $b_\theta^\mathrm{SDE} \leq \log {\ptheta}^\mathrm{ODE}$ can be used instead, effectively approximating the learned image prior by the same surrogate prior as in \eqref{eq:variational-inference-surrogate3}. The image optimization step \eqref{eq:variational-bayes-image-step} hence reads
\begin{equation}\label{eq:variational-bayes-surrogate-prior-image-step}
    \chi_{k+1} = \argmin_\chi \calL(\chi, \rho_k) = \argmin_{\chi} \left\{ \bbE_{x\sim q_{\chi}} \bbE_{\bfr \sim q_{\rho_k}} \left[ - \log p(y\sbar x,\bfr) (x) - b_\theta^{\mathrm{SDE}}(x) \right] - \calH (q_{\chi}) \right\}\rev . \nc
\end{equation}
For medium- to large-scale imaging problems, $q_\chi$ can be a Gaussian distribution with diagonal covariance -- more complex models were computationally prohibitive in our setup; see also \cite{Feng2024Variational} for scaling experiments. We solve \eqref{eq:variational-bayes-scheme} using inner loops of stochastic gradient descent, where the partial derivatives $\nabla_\chi \calL$ and $\nabla_\rho \calL$ are approximated using Monte Carlo estimators of the expectations and the typical reparametrization trick \cite{kingma2014variationalbayes} whenever $q_\chi, q_\rho$ are Gaussian.
This sampling-based approach admits the use of a batch size $\batchsize \geq 1$ for each gradient evaluation, which can reduce the variance of estimated gradients. The surrogate prior term $\bSDE(x)$
\rev
is numerically approximated with a Monte-Carlo estimator using a single forward-pass through $\scoremodel$; see \cref{sec:vi-principled-priors} and \cite[sec. 4.2]{Feng2024Variational}.
\nc
This leads to the variational Bayes approach summarized in \cref{algo:var-bayes-principled-prior}. \rev
Note that while the algorithm is presented generically, for simplicity, in practice we do not employ a particular stopping criterion but instead use a fixed number of iterations $N=l_{\mathrm{max}}$.
\nc 
For a clearer presentation, we formulate the updates as single gradient descent steps. However, in the practical implementation, lines 5 and 10 in \cref{algo:var-bayes-principled-prior} are replaced by the Adam optimization scheme \cite{kingma2014adam}.

\begin{algorithm}[t]
\caption{Blind Variational Bayes reconstruction with surrogate prior}\label{algo:var-bayes-principled-prior}
\begin{algorithmic}[1]
\State Initialize parameter distribution $\rho^{(0)} = (\mu^{(0)}_\bfr,\Sigma^{(0)}_\bfr)$, image distribution $\chi^{(0)} = (\mu^{(0)}_x,\Sigma^{(0)}_x)$, $\itc = 0$, data $\meas \in \calY$, forward model $\fwd: \calR \times \calX \to \calY$, $\itc = 0$, maximum number of iterations $\itc_{\max}$, step size sequences $(\tau^{(\itc,i)}), (\eta^{(\itc,i)})$
\While{$\itc < \itc_{\max}$ and stopping criterion on $\rho^{(\itc)} = (\mu^{(\itc)}_\bfr,\Sigma^{(\itc)}_\bfr),\chi^{(\itc)} = (\mu^{(\itc)}_x,\Sigma^{(\itc)}_x)$ is not satisfied}
    \State $ \chi^{(\itc,0)} = \chi^{(\itc)} $
    \For{$i \gets 0, \dots, N_{\mathrm{img}}-1$}\Comment{Solve for image \eqref{eq:variational-bayes-surrogate-prior-image-step}}
        \State $\chi^{(\itc,i + 1)} = \chi^{(\itc,i)} - \tau^{(\itc,i)} \nabla_\chi \calL(\chi^{(\itc,i)},\rho^{(\itc)}) $
    \EndFor
    \State $\chi^{(\itc+1)} = \chi^{(\itc,N_{\mathrm{img}})}$
    \State $ \rho^{(\itc,0)} = \rho^{(\itc)} $
    \For{$i \gets 0, \dots, N_{\mathrm{par}}-1$}\Comment{Solve for parameters \eqref{eq:variational-bayes-par-step}}
        \State $\rho^{(\itc,i + 1)} = \rho^{(\itc,i)} - \eta^{(\itc,i)} \nabla_\rho \calL(\chi^{(\itc+1)},\rho^{(\itc,i)}) $
    \EndFor
    \State $\rho^{(\itc+1)} = \rho^{(\itc,N_{\mathrm{par}})} $
    \State $\itc \gets \itc+1$
\EndWhile
\State \textbf{return} image distribution $q_{\chi^{(\itc)}}$, parameter distribution $q_{\rho^{(\itc)}}$
\end{algorithmic}
\end{algorithm}

We also mention a connection to other previous work on blind inverse problems \cite{Gao2021Deep}, which proposed an \ac{EM} scheme to alternatingly optimize the image and a latent parameter. The \ac{EM} algorithm naturally arises as a special case of the variational Bayes approach derived here, if we slightly abandon the Bayesian perspective and regard the latent parameter as having a true value $\bfr^\ast$. Equivalently, we can assume that $q_{\rho}$ is a point mass in \eqref{eq:variational-bayes-scheme} and remove the entropy regularization. The \ac{EM} iteration then reads as the alternating scheme:
\begin{subequations}\label{eq:em-scheme}
\begin{align}
    \chi_{k+1} &= \argmin_{\chi} \left\{ \bbE_{x\sim q_{\chi}} \left[ - \log p(y\sbar x,\bfr_k) - \bSDE(x) \right] - \calH (q_{\chi}) \right\},\label{eq:em-scheme-e-step}\\
    \bfr_{k+1} &= \argmin_{\bfr} \left\{\bbE_{x\sim q_{\chi_{k+1}}} \left[ - \log p(y\sbar x,\bfr) - \log p(\bfr) \right] \right\}.\label{eq:em-scheme-m-step}
\end{align}
\end{subequations}
To solve the E-step \eqref{eq:em-scheme-e-step}, we use a first-order optimization scheme and the learned surrogate prior.
A closely related work \cite{Laroche2024} proposed a similar scheme with the same M-step for blind parameters, but used a likelihood-approximating sampling method like \ac{DPS} to estimate the image posterior for a given parameter instance in the E-step.

\subsubsection{Variational Minimization as Baseline} While Bayesian techniques like \ac{VI} provide a powerful framework to recover not only a single image, but a whole distribution of images, they are also more computationally demanding. Classical, variational optimization algorithms are generally more economic and form a central methodology in the ptychographic imaging literature \cite{Melnyk2025Convergence}. As already noted, (blind) RED-Diff can be located between the two classes: It is motivated from a \ac{VI} lens, but solves the imaging problem by fitting a single image---or, in distribution terms, a point mass, reflected by the fact that $\calL^{\mathrm{REDdiff}}$ contains no entropy terms. The iterative update of blind RED-Diff is ultimately an alternating first order descent method on the variational minimization (or \ac{MAP} estimation) problem
\begin{equation}\label{eq:general-variational-opt-formulation}
    \min_{x,\bfr} \left\{ - \log p(y\sbar x,\bfr) - \log p(x) - \log p(\bfr) \right\},
\end{equation}
where the image prior is of the specific form $p(x)$ seen in \cref{eq:blind-red-diff-image-step}. By replacing the image prior $p(x)$, we obtain other standard variational regularization formulations. In experiments, we will therefore compare the proposed \ac{SSP} approach to RED-Diff, but also to simpler forms of \eqref{eq:general-variational-opt-formulation} with cheaper, model-based regularization (as opposed to a data-driven score model), or no prior information at all. For the latter, completely omitting $p(x)$ leads to a maximum likelihood estimate of $x$, which we compute as before using first-order optimization methods. In the former case, one may use a typical hand-crafted image prior $p(x)$ that promotes features like sparsity in the image gradients or in a dictionary or frame like a wavelet basis.

\!r
\subsection{On the convergence of \cref{algo:var-bayes-principled-prior}}
\label{sec:convergence-alg1}

We briefly turn towards the convergence analysis of the scheme in \cref{algo:var-bayes-principled-prior}. This section is to be understood as a discussion of the challenges one faces in our particular setting. An in-depth study is left for future work. Moreover, we refer to \cite{hesse2015proximal,filbir2023image,Melnyk2025Convergence} for works on the convergence analysis of other algorithms for ptychography. From optimization literature, \eqref{eq:variational-bayes-scheme} is known as an \ac{AM}, block coordinate descent method or Gauss--Seidel iteration, which has been studied for example in \cite{beck2015convergence,beck2017first,tseng2001convergence}. We state the following standard assumption, which ensures well-definedness of the scheme:
\begin{ass}\label{ass:Lstandard}
The function $\calL$ is proper, continuous and has compact sub-level sets $\{(\chi,\rho): \calL(\chi,\rho)\leq \alpha\}$. 
\end{ass}
By definition we have that $\calL(\chi_{k+1},\rho_{k+1})\leq \calL(\chi_k,\rho_k)$ and the compactness of the sub-level sets then implies that the sequence $(\chi_{k}, \rho_{k})$ is bounded. A further assumption used in \cite{tseng2001convergence} to ensure convergence towards stationary points is that $\calL$ admits coordinate-wise unique minimizers.
\begin{ass}\label{ass:uminc}
The functions $\chi\mapsto \calL(\chi,\tilde{\rho}), \rho\mapsto\calL(\tilde{\chi},\rho)$ admit unique minimizers for any $(\tilde{\chi},\tilde{\rho})$.
\end{ass}
Under this assumption \cite[Thm. 4.1]{tseng2001convergence} yields the following:
\begin{lemma}
Under \cref{ass:uminc,ass:Lstandard}, any cluster point $(\chi^*,\rho^*)$ of $\{(\chi_{k}, \rho_{k})\}$ satisfies
\begin{align*}
\chi^* = \argmin_\chi \calL(\chi, \rho^*), \qquad 
\rho^* = \argmin_\rho \calL(\chi^*, \rho).
\end{align*}
If $\calL \in C^1$ this means $(\chi^*,\rho^*)$ is a stationary point, i.e.,
\begin{align*}
\nabla_{(\chi,\rho)} \calL(\chi^*,\rho^*) = 0.
\end{align*}
\end{lemma}
Regarding the uniqueness in the position-component, we will observe numerically in \cref{fig:probe-comp-shift-difference-norm} that with a randomized probe one can hope to obtain this property. However, formulating the necessary assumptions to prove this statement is not trivial and outside the scope of this work. Regarding uniqueness in the object component of $\calL$ we remark that the same ambiguities as discussed in \cref{rem:wellposed} appear. Therefore, in general, one can only hope that additional prior terms help to ensure \cref{ass:uminc}.

In the absence of this property, the \ac{AM} in \cref{algo:var-bayes-principled-prior} is not guaranteed to converge to a stationary point and counter examples exist, see e.g. \cite[Ex. 14.4]{beck2017first}. Nevertheless, we can exploit the special structure of $\calL$, by splitting it in the parts that depend on both $\chi$ and $\rho$ and the respectively independent parts,
\begin{align*}
\underbrace{\bbE_{x\sim q_{\chi}} \bbE_{\bfr \sim q_{\rho}} \left[ - \log p(y\sbar x,\bfr)\right]}_{=:\mathcal{F}_1(\chi, \rho)}
+
\underbrace{\bbE_{x\sim q_\chi}[- \log {\ptheta}(x)]- \calH (q_\chi)}_{=:\mathcal{F}_2(\chi)}
+
\underbrace{\bbE_{\bfr\sim q_\rho}[- \log p(\bfr)]- \calH (q_\rho)}_{=:\mathcal{F}_3(\rho)}.
\end{align*}
In this situation the arguments in \cite{beck2015convergence} allow us to derive a convergence statement under the following assumption.
\begin{ass}
The functions $\mathcal{F}_2, \mathcal{F}_3$ are proper, convex functions with closed sub-level sets.    
\end{ass}
\noindent We first observe that in the non-parametric setting for any positive distribution $\tilde{p}$ the function 
\begin{align*}
q\mapsto \bbE_{x\sim q}[-\log \tilde{p}(x)] - \calH(q)
\end{align*}
is convex in $q$ since the first part is linear in $q$ and the entropy $\calH$ is concave. However, when we choose a concrete parametrization, this convexity typically does not transfer to the convexity in $\chi, \rho$. Namely, in \labelcref{eq:parachoice},  we choose a Gaussian parametrization as follows:
\begin{align*}
\begin{alignedat}{6}
q_\chi(x|y) := \mathcal{N}_{\mathbb{C}}(x; \mu_\chi, \Sigma_\chi I), &\ \mu_\chi\in\mathbb{C}^d, &\ \Sigma_\chi = \diag(\sigma_\chi^2), &\ \sigma_\chi \in \mathbb{R}_+^d\,, \\
q_\rho(r|y)  := \mathcal{N}(r; \mu_\rho, \Sigma_\rho I), &\ \mu_\rho \in \mathbb{R}^{K\times 2}, &\ \Sigma_\rho = \diag(\sigma_\rho^2), &\ \sigma_\rho \in \mathbb{R}_+^{K}\,.
\end{alignedat}
\end{align*}
The following lemma shows that this choice preserves convexity of the entropy functional.
\begin{lemma}
For the parametrization in \labelcref{eq:parachoice}, the functionals $\chi\mapsto \calH(q_\chi)$ and $\rho\mapsto\calH(\rho)$ are convex.
\end{lemma}
\begin{proof}
The entropy of the Gaussian density $q_\chi$ is given by 
\begin{align*}
\calH(\chi) = C + \frac{1}{2} \log \det(\diag(\sigma_\chi^2)) = C + \frac{1}{2}\sum_{i=1}^d  \log\, (\sigma_\chi)_i^2
\end{align*}
for a constant $C$ independent of $\chi$. In particular $\calH(\chi)$ is constant in $\mu_\chi$ and concave in $\sigma_\chi$. Thus $\chi\mapsto \calH(q_\chi)$ is convex, the convexity of $\rho\mapsto\calH(\rho)$ follows analogously.
\end{proof}
In general we cannot expect convexity of the first term in $\mathcal{F}_2,\mathcal{F}_3$. In order to establish this property, we would need to assume that both $x\mapsto\hat{p}_\theta(x)$ and $\bfr\mapsto p(\bfr)$ are log-concave. We summarize this in the following lemma, which is a direct consequence from \cite{challis2011concave}.
\begin{lemma}\label{lem:prior-log-concavity}
Assume that both $x\mapsto\hat{p}_\theta(x)$ and $\bfr\mapsto p(\bfr)$ are log-concave functions, then $\mathcal{F}_2,\mathcal{F}_3$ are convex.
\end{lemma}
\begin{proof}
This fact is established in \cite[Sec. 3]{challis2011concave}.
\end{proof}
This raises the question whether log-concavity of the priors can be assumed in our case. We first consider the position prior.
\begin{lemma}
The function $\bfr\mapsto p(\bfr)$ defined as in \cref{sec:log-barrier}, i.e., the log-barrier, is log-concave.
\end{lemma}
\begin{proof}
We have that $-\log p(\bfr) = \lambda_{\text{pos}} B(T(\bfr))$, where $T$ is affine and the logarithmic barrier $B$ is convex. Thus $\bfr\mapsto-\log p(\bfr)$ is convex as the composition of an affine and a convex function.
\end{proof}
The data-driven image prior $\hat{p}_\theta$ is usually complex and multimodal, hence convexity of $\mathcal{F}_2$ cannot be guaranteed. A simpler setting, in which the assumptions of \cref{lem:prior-log-concavity} are met, is reached when we replace $\hat{p}_\theta$ by a classical model-based prior. An example of a log-concave prior is the TV baseline in \cref{sec:TV}. In such a setting, we obtain the following result from \cite[Lem. 3.3]{beck2015convergence}.
\begin{lemma}
Assume that $\mathcal{F}_1\in C^1$ with Lipschitz continuous gradients and that both $x\mapsto\hat{p}_\theta(x)$ and $\bfr\mapsto p(\bfr)$ are log-concave. Then, any accumulation point $(\chi^*,\rho^*)$ of $\{(\chi_{k}, \rho_{k})\}$ is a stationary point of $\calL$.
\end{lemma}
Note that a stationary point of $\calL$ does not have to coincide with its minimizer. For this, $\mathcal{F}_1$ would need to be convex, which seems unrealistic considering \cref{rem:wellposed}. While the previous notes offer only limited insights into the convergence behavior, they still highlight the main difficulties connected to the analysis. A more in-depth study of the convergence of the gradient-based schemes under priors that are not log-concave is left for future work.

% NOTE Simon: we could also refer to Fig. 5 here that visually shows the nonconvexity of even the tiny subproblem of recovering two positions correctly i.e. rbold = {r1, r2} (and assuming that x is already recovered correctly). just as a supporting empirical reference
% TR: but this non-conexity in Fig. 5 comes from the likelihood
% SW: that is true but don't you also mention F1 here which is the likelihood..?
% Sorry at this point I always meant F_2, F_3
% My hope was that with the semi-blind setting it might be reasonable to assume convex F_1
% In the non-convex setting we simply have convergence towards stationary points. This I would also write down

\@n

\section{Experiments}
In the following, we describe the computational experiments we perform to arrive at our results and conclusions.
%Relaxations of these assumptions should be possible within our framework through extensions of our formalism,   \todo{SW: re R1: stress more that it is a simplified toy problem...?}

\subsection{Problem Setup}\label{sec:problem-setup}
\rev
First, we describe our simulated problem setup including all relevant parameters for the algorithms and the simulated measurement models. In all experiments, we use an object array of size $\img \in \mathbb{C}^{\imgsizey \times \imgsizex}:= \mathbb{C}^{256\times256}$ and a probe array of size $\probe \in \mathbb{C}^{\probesizey \times \probesizex} := \mathbb{C}^{512\times 512}$.
\nc

\subsubsection{Scan Positions}\label{sec:scan-positions}
We model the ground-truth positions to represent the center of each probe relative to the object coordinates, so if the probe is centered on some pixel of the object, it holds that $0 \leq \shifty_\imeas \leq \imgsizey, 0 \leq \shiftx_\imeas \leq \imgsizex$\revtwo{}, see \cref{ass:probe-centered-on-object}, i.e., $r_\imeas=(\shifty_\imeas,\shiftx_\imeas)\in\shiftdomain$\nc{}
%\todo{TR@SW: I refer back to the Assumption to show that everything fits together}.  % SW: great, thanks :)
Under this assumption, we sample the ground-truth positions from a uniform distribution in the horizontal and vertical directions,
\begin{equation}\label{eq:position-distribution}
\pos_\imeas \sim \mathcal{U}(0,\imgsizey)\times\mathcal{U}(0,\imgsizex)\,.
\end{equation}
%The \ac{DFT} we use in $\shifter$ implicitly treats the input signal as periodic. We use this fact to make it impossible for estimated positions to escape into empty space, as $\pos_\imeas + [a\imgsizey,b\imgsizex]$ is equivalent to $\pos_\imeas$ for all $a,b\in\mathbb{Z}$ under our DFT-based definition of $\shifter$. Since the target image itself is however a single particle and thus aperiodic, we use the padding operator $\padder$, which pads $\img$ with one full probe array size of free space ($1+0\mathrm{i}$) entries. $\shifter(\pos_\imeas, \cdot)$ then effectively models shifting of an aperiodic object inside an indefinitely repeating reconstruction box.
%Furthermore,
To avoid ringing artifacts in the image, we always round $\pos_\imeas$ to the nearest whole integer when evaluating the forward operator but keep the gradients as if we were using fractional shift positions.

\subsubsection{Probe Functions}\label{sec:problem-setup:probe-functions}
We simulate all probe functions $\probe \in \mathbb{C}^{512\times512}$ by taking the \ac{DFT} of the aperture array $\aperture \in \mathbb{C}^{512 \times 512}$ to propagate the wavefront to the focus. The aperture array $\aperture$ contains a centered circular aperture with fractional diameter $\apdiam \in (0, \sfrac{1}{2}]$. In the aperture plane, we assume a wavefront of constant magnitude and a phase profile determined by a random Zernike polynomial \cite{zernike1934} of order 4, with piston and tilt terms set to 0. We sample the random Zernike polynomial only once, shown in the leftmost column of \cref{fig:probe-comparison}. We optionally apply an additional block-wise random phase mask in the aperture plane through pixel-wise complex multiplication:
\begin{equation}
    \aperture_{\text{masked}}[\coordy,\coordx] = \aperture \odot e^{\mathrm{i}\Phi_{\text{mask}}},\quad
    \Phi_\text{mask}[\coordy,\coordx] = M\left[ \left\lfloor \frac{\coordy}{\maskblocksize} \right\rfloor,\left\lfloor \frac{w\vphantom{\coordy}}{\maskblocksize}\right\rfloor \right],\quad
    M \sim \mathcal{U}(0,2\pi)^{\lceil \coordy/\maskblocksize \rceil \times \lceil \coordx/\maskblocksize\rceil}
\end{equation}
where $\maskblocksize$ determines the block size of the random mask in pixels. This construction is inspired by other works on randomized illumination \cite{guizar2012illuminationfreq,levitan2020randomillum}, which show more reliable reconstructions when using structured probes with high-frequency content. We will show in \cref{sec:results:probe-comparison} that the additional structure from these phase masks is also helpful for position recovery in our position-blind setting. By default, we set $\apdiam=\frac{1}{2}, \maskblocksize=4$ in our experiments.

Since the probe is band-limited from the finite extent of the aperture and/or lens, it is in principle space-unlimited. Thus, to avoid the unrealistic case of the probe energy abruptly falling off to zero on some part of the object, we choose the probe array to be twice as large as the object array and set all ground-truth probe positions to be centered on some point on the object. Note that even though the probe array is larger in pixels than the object array, the measurements are ptychographic since the probe energy is concentrated in a smaller region, typically roughly the size of the object or smaller; see \cref{fig:probe-comparison}.

\subsubsection{Noise Model}\label{sec:noise-model}
As the noise model in most of our experiments, we assume that the observation noise $\noise_k$ is independent and identically distributed Gaussian noise with mean zero and known variance $\noisestd^2$ for all $\imeas=1,\dots,\nmeas$. In later experiments, we also simulate more experimentally accurate noise by scaling the signal power of the probe function to match an assumed number of photons diffracted from an ideal non-absorbing object $\nphot$, and drawing $\meas_\imeas$ from a simulated Poisson distribution with mean $\abs{\revtwo\fwdprop\nc(\pos_\imeas,\img)}^2$.

\subsubsection{Simulated Measurements}
To generate a simulated set of measurements $\{\meas_\imeas\}_{\imeas=1}^{\nmeas}$, we take $\img$ to be some simulated test object transmission function from our test dataset (see \cref{sec:dataset}) and $\pos_\imeas$ to be 100 randomly sampled positions according to \cref{eq:position-distribution}, compute each $\meas_\imeas$ according to \cref{eq:forward-model}, and then apply the chosen type of measurement noise. We use $\nmeas=100$ measurement positions in all of our experiments.

%The propagator $A(r_k)$ is a Fourier transform of a shifted and masked image piece and depends on the scan position. It can be written as $A(r_k) = \mathcal{F}PS(r_k)$, with
%\todo{LK:  Revise these three}
% \begin{align}
%     \mathcal{F}\ &\text{being a 2-D Discrete Fourier Transform} \,,\\
%     P &= \diag(p_{11}, \ldots, p_{hw}) \,,\\     % Probe multiplication for cutout (diagonal matrix)
%     S(r_k) &= \mathcal{F}^{-1} \odot \exp\left(\mathrm{i} \left(\Delta{x}_k \frac{x}{2\pi W} + \Delta{y}_k \frac{y}{2\pi H} \right) \right) \odot \mathcal{F} \,, % Cutout operator: in the simplest case, 1 for region centered around $r_k$ where probe is nonzero, 0 everywhere else. Somewhat annoyingly however, this 'centering' in S_{r_k} should probably involve subpixel shifts when r_k is non-integer (in units of object pixels)
%     % TODO replace the $\Delta x$, $\Delta y$ by some function of r
% \end{align}

%\todo[inline]{list all details of the forward model, probe functions, and measurement simulation here}

\begin{figure}
    \centering
    \includegraphics[width=.8\linewidth]{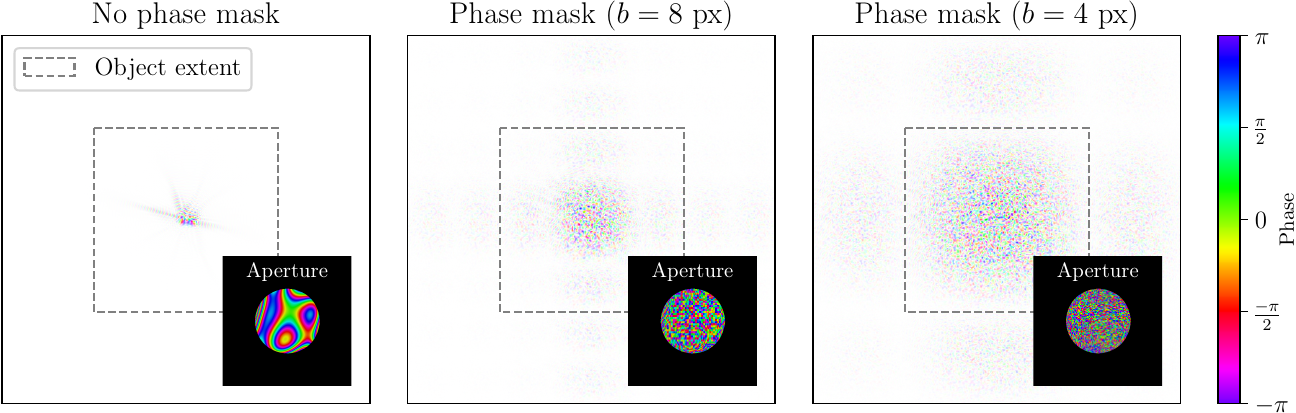}
    \caption{Comparison of several probe functions $\probe$ used in this work. All three are based on the same random Zernike polynomial, with an optional random phase mask of block size $b$ applied as indicated. The dashed white square shows the extent of the imaged object $\img$ for comparison. The diameter of the aperture is half a probe array size here, i.e., 256 pixels.}
    \label{fig:probe-comparison}
\end{figure}

%\todo{SW: Briefly refer back to background section here as a refresher}
Denoting the full set of scan positions by $\bfr = (\pos_1, \dots, \pos_\nmeas)$, the noise model leads to the following measurement likelihood under the Gaussian measurement noise model:
\begin{equation}\label{eq:likelihood-gaussian}
    p(y\,|\,x,\bfr) \propto \exp\left( - \frac{1}{2\sigma_\noise^2} \sum_\imeas \Big\lVert \meas_\imeas - |\revtwo\fwdprop\nc(\pos_\imeas,\img)|^2 \Big\rVert_2^2 \right),
\end{equation}
whereas for Poisson measurement noise, we model the measurement likelihood using a Gaussian approximation to the Poisson distribution, leading to
\begin{equation}\label{eq:likelihood-poisson}
    p(y\,|\,x,\bfr)_{\text{Poisson}} \propto \exp\left( - \sum_\imeas \left\lVert \frac{1}{\sqrt{2 \meas_\imeas}} \odot \left( \meas_\imeas - |\revtwo\fwdprop\nc(\pos_\imeas,\img)|^2 \right) \right\rVert_2^2 \right),
\end{equation}
where each pixel in $\meas_\imeas$ is treated as both the mean and the variance, and the square root is taken element-wise. An alternative would be to use $\abs{\revtwo\fwdprop\nc(\pos_\imeas,\img)}^2$ as the variance of the approximating Gaussian instead of $\meas_\imeas$, but in preliminary experiments we found this option to lead to less stable reconstructions and hence always use \labelcref{eq:likelihood-poisson}. In practice, within the expression $\frac{1}{2\meas_\imeas}$ we clamp $\meas_\imeas$ to have a minimum value of 1 to avoid both division by zero and an overly high weighting of detector pixels with no measured photons.

\subsection{Algorithmic Configuration}\label{sec:algo-configuration}
\rev
In the following, we list all relevant details for the configuration of the tested reconstruction algorithms. A summary of algorithmic parameters is provided in \cref{tab:algorithm-params}.
Note that, due to long reconstruction runtimes, limited computational resources, and a large set of parameters to tune, we did not systematically grid-search any algorithm's set of parameters. Instead, we performed small manual line-searches of parameters using 1-3 example images until successful reconstructions were achieved.
It is not our goal here to establish any particular reconstruction method as a \enquote{state-of-the-art}, but rather to show that decent reconstructions can be attained by each method at all in this difficult problem setting, and to highlight different behaviors of the methods. Therefore, detailed tuning is out of scope for this work, and it is likely that future systematic algorithm parameter optimizations will lead to further improved reconstruction quality.
\nc

\begin{table}[]
    \centering
    \rev
    \begin{tabular}{lccccccccc}
        \toprule
        Method & \#Iterations & $N_{\mathrm{img}}$ & $N_{\mathrm{pos}}$ & $\lambda_{\mathrm{img}}$ & $\lambda_{\mathrm{pos}}$ & $\lambda_{\sigma_{\chi}}$ & $\lambda_{\sigma_\rho}$ & $\lambda_{\mathrm{HTV}}$ & $\lambdaRD$\\
        \midrule
        VI (SSP) & 10,000 & 1 & 10 & 0.1 & 10 & 0.1 & 0.01 & - & - \\
        VI (H-TV prior) & \multicolumn{7}{c}{\ditto} & 5 & - \\
        VI (no prior) & \multicolumn{7}{c}{\ditto} & - & - \\
        RED-Diff \cite{Mardani2024Variational} & \multicolumn{5}{c}{\ditto} & - & - & - & 20 \\
        Opt (H-TV prior) & \multicolumn{5}{c}{\ditto} & - & - & 0.1 & - \\
        Opt (No prior) & \multicolumn{5}{c}{\ditto} & - & - & - & - \\
        \bottomrule
    \end{tabular}
    \caption{\rev Algorithmic hyperparameters used for each reconstruction method in this work, unless otherwise noted. Entries with --\ "\ -- indicate that the same parameter setting is used as in the row above, and entries with - indicate that the parameter is not applicable or unused for the respective method. For easier comparison, $\lambda_{\mathrm{img}}, \lambda_{\mathrm{pos}}$ respectively refer to the step sizes for the image and positions for RED-Diff and optimization-based methods, while for variational approaches they refer to the step sizes for the corresponding distribution's mean, i.e., $\lambda_{\mu_\chi}$ and $\lambda_{\mu_\rho}$, respectively. Note that, for the VI methods, $\lambda_{\mathrm{img}}$ and $\lambda_{\sigma_\chi}$ are scheduled and decay starting from the listed values, see \cref{sec:alg-config-vi} for full details.}
    \label{tab:algorithm-params}
\end{table}
\nc

\subsubsection{Variational Inference} \label{sec:alg-config-vi} For our variational inference methods and all choices of prior distribution, we approximate the posterior by the following variational distributions\rev, which we recall from \cref{sec:convergence-alg1}
\rev
\begin{align}\label{eq:parachoice}
\begin{alignedat}{6}
    q_\chi(x|y) := \mathcal{N}_{\mathbb{C}}(x; \mu_\chi, \Sigma_\chi I), &\ \mu_\chi\in\mathbb{C}^d, &\ \Sigma_\chi = \diag(\sigma_\chi^2), &\ \sigma_\chi \in \mathbb{R}_+^d\,, \\
    q_\rho(r|y)  := \mathcal{N}(r; \mu_\rho, \Sigma_\rho I), &\ \mu_\rho \in \mathbb{R}^{K\times 2}, &\ \Sigma_\rho = \diag(\sigma_\rho^2), &\ \sigma_\rho \in \mathbb{R}_+^{K}\,, \\
    q_{\chi,\rho}(x,r|y) := q_\chi(x|y)q_\rho(r|y)\,.
\end{alignedat}
\end{align}\nc%
We assume complex isotropy of the image uncertainty in that the real and imaginary parts of each image pixel $\img[\coordy,\coordx] \in \mathbb{C}$ share a single variance $\sigma_{\chi,hw}^2 \in \mathbb{R}$, and spatial isotropy of the shift uncertainty in that the horizontal and vertical components of each position $\pos_\imeas \in \mathbb{R}^2$ share a single variance $\sigma_{\rho,\imeas}^2 \in \mathbb{R}$. To enforce stability and positivity of $\sigma_\chi$ and $\sigma_\rho$, we reparametrize and minimize for $\log \sigma_\chi$, $\log \sigma_\rho$. We set the batch size of the Monte Carlo gradient estimators to $\batchsize=4$; see \cref{sec:methods:blind-ssp}.

For both the image and position parameters, we use the Adam optimizer \cite{kingma2014adam}, which has previously been established as a working method for (non-blind) ptychographic image reconstruction \cite{kandel2019autodiffpty,jiang2018tensorflowpty,guzzi2022modularpty}. We set its momentum parameters to $\beta_1 = 0.9, \beta_2 = 0.999$, the default settings in the \texttt{PyTorch} package \cite{paszke2019pytorch}. We run $\Nopt=\text{10,000}$ outer steps unless otherwise noted, each with $N_\mathrm{img}=1$ inner image optimization step and $N_\mathrm{par}=10$ inner position optimization steps. For the position parameters we set the step sizes $\lambda_{\mu_\rho} = 10.0$ and $\lambda_{\sigma_\rho} = 0.01$, where we use a high learning rate of the means ($\lambda_{\mu_\rho}$) due to the difficult position-dependent loss landscape; see \cref{fig:probe-comp-shift-difference-norm}. We keep $\lambda_{\sigma_\rho}$ small to avoid a collapse of the position uncertainty. For the image parameters we set $\lambda_{\mu_\chi} = \lambda_{\sigma_\chi}$ to quickly arrive at an initial image estimate, and employ a falling cosine schedule from $0.1$ to $0.001$ between steps 4,000 and 6,000 to allow the image to stabilize and to improve the effectiveness of the high-variance prior term $\bSDE$ during the later optimization steps.

\subsubsection{TV baseline details}\label{sec:TV}

%As a baseline, we use a total variation prior.
Following the definition in \cite{gao2021complex}, we employ the following variant of isotropic TV for complex valued images $\img \in \mathbb{C}^{H\times W}$, 
\begin{align}
\operatorname{TV}(x) := 
\sum_{\coordy=1}^\imgsizey \sum_{\coordx=1}^\imgsizex \sqrt{\abs{\img[\coordy,\coordx] - \img[\coordy,\coordx+1]}^2 + \abs{\img[\coordy,\coordx] - \img[\coordy+1, \coordx]}^2},
\end{align}
where we set $\img[\imgsizey+1, \coordx] = \img[\imgsizey, \coordx]$ for all $\coordx=1,\ldots,\imgsizex$ and, respectively, $\img[\coordy, \imgsizex+1] = \img[\coordy, \imgsizex]$ for all $\coordy=1,\ldots,\imgsizey$. In order to circumvent the non-differentiability of the TV functional as defined above, we consider a smoothed version
\begin{align}\label{eq:huber-tv}
\operatorname{TV}_\alpha(x) := 
\sum_{\coordy=1}^\imgsizey \sum_{\coordx=1}^\imgsizex
    h_\alpha\left(
        \sqrt{\abs{\img[\coordy,\coordx] - \img[\coordy,\coordx+1]}^2 + \abs{\img[\coordy,\coordx] - \img[\coordy+1, \coordx]}^2}
    \right).
\end{align}
A typical choice for $h_\alpha:[0,\infty)\to[0,\infty)$ is the Huber regularizer (see, e.g., \cite[Example 4.7]{chambolle2016introduction}),
\begin{align*}
h_\alpha(t) := 
\begin{cases}
\frac{t^2}{2\alpha} & \text{if } t\leq \alpha,\\
t - \frac{\alpha}{2} &\text{else},
\end{cases}
\end{align*}
where $\alpha>0$ is chosen small; for our experiments we set $\alpha = 10^{-5}$.

\subsubsection{RED-Diff} 
The algorithm we use for blind RED-Diff in this setup is given in \cref{algo:blind-red-diff}. Following \cite{Mardani2024Variational}, in the notation of \cref{eq:blind-red-diff-image-gradient} we choose the inverse SNR weighting
$
\tilde{\omega}(t) = \lambdaRD\, \frac{\sigma(t)}{\alpha(t)}.
$
In \cite[Prop. 2]{Mardani2024Variational} the weight $\lambdaRD>0$ may also depend on the observation noise $\noisestd$. In our experiments, we choose fixed step sizes $\tau^{(\itc,i)}=\tau=0.1, \eta^{(\itc,i)}=\eta=10$ in \cref{algo:blind-red-diff}
\rev
and $\lambdaRD = 20$.
\nc

\subsubsection{Position log-barrier}\label{sec:log-barrier}
In some of the scenarios we consider, such as phase-only objects, we observe that the loss landscape for the position recovery has an unfavorable structure, which leads to positions moving off the object rather towards reasonable estimates; see \cref{sec:results:phase-only}. To remedy this problem, we add a 2-dimensional log-barrier loss as a hand-crafted prior on the positions, with an empirically chosen weight $\lambda_{\text{pos}}$. Namely, we consider the following barrier function,
$
B(s) := 
-\log(1 - s) - \log(1 + s) + \iota_{\mathbb{R}^2\setminus(-1,1)^2}(s),
$
where for a set $A$, $\iota_A$ denotes the indicator function from convex analysis; see \cite{rockafellar1997convex}. Our prior is then defined by an affine transformation $T$, which first shifts and scales each position $r$, i.e., $p(r)\propto \exp(- \lambda_{\text{pos}} B(T(r))$. We define $T$ such that the object extent within the domain is scaled to a smaller domain $(-a,a)^2\subsetneq (-1,1)^2$. This is done via
%
%\begin{align*}
$T(r) := \frac{r - m}{r_{\max} - r_{\min}}$,
%\end{align*}
%
where we set $r_{\max} = H + l, r_{\min} = H - l$ and $m = (r_{\max} + r_{\min})/2$ with $l=20$, so $l$ grants 20 pixels of leniency for the positions to lie slightly outside of the object extent. To avoid undefined gradients from the log-barrier loss, we clip the positions to lie inside the rectangle spanned by these edges before feeding them to the loss expression.

\subsection{Dataset}\label{sec:dataset}
Since there is not enough realistic data of X-ray wavefront modulation from single proteins or other single nanoscale particles readily available to train a deep generative model, we instead generate artificial complex-valued images from a large public image dataset. We choose the INRIA Aerial Image Labeling Dataset (AILD) \cite{maggiori2017aerialdataset} as a basis, since its images have detailed natural structures and high-frequency content but are nonetheless more predictable and of a simpler distribution than a generic large photograph dataset such as ImageNet \cite{deng2009imagenet}. This allows us to test the reconstruction methods in complex imaging scenarios. We then generate two types of complex-valued images based on random $256 \times 256$ crops of these RGB images, mapped to grayscale, with the following two procedures:
\begin{enumerate}
    \item[(1)] 80\% of the time, to simulate the observation that the absorption image often has less structure than the phase image, we generate the image amplitudes from a Perona--Malik edge-preserving smoothing \cite{perona2002scale} of the grayscale input image. We use a uniformly random number of Perona--Malik iterations in $[30, 100]$ and a random $\kappa$ parameter in $[0.03, 0.075]$. We scale the amplitudes by factors sampled from a log-normal distribution with $\mu=0, \sigma=0.25$. For the phase of the object, we use the grayscale image without any smoothing, mapping all grayscale values $v \in [0,1]$ to a random phase of $4\pi \mathfrak{S} v + \mathfrak{s}$ where $\mathfrak{S} \sim \text{Beta}(3,10), \mathfrak{s} \sim \mathcal{N}(0,\frac{\pi}{2})$ are sampled once for each image.
    \item[(2)] 20\% of the time, to simulate non-absorbing objects, we generate phase-only images by setting all amplitudes to 1 and using the grayscale input image as the phase image. We map to a random phase range of $\mathfrak{S} v + \mathfrak{s}$ with $\mathfrak{S} \sim \mathcal{U}(\frac{\pi}{4}, 3\pi),\mathfrak{s} \sim \mathcal{U}(-\pi, \pi)$ where $v \in [0,1]$ is again the input grayscale value.
\end{enumerate}
In total, we generate 30,000 training images for training our score model from the training subset of AILD \cite{maggiori2017aerialdataset}. For evaluating the reconstruction methods, we generate 10 test images from the test subset of AILD, but here we only follow procedure (1) to generate test objects that exhibit both absorption and phase shifts. For our experiments where we consider phase-only objects (\cref{sec:results:phase-only}) we then generate phase-only variants of these same objects by setting their complex magnitude to 1 everywhere and keeping the phase.
%All 10 complex-valued test objects are shown in \cref{fig:all-test-objects}.
%\todo[inline]{detail the complex-valued dataset generation and motivation}

\subsection{Score model training}
With the training dataset described in the previous section, we train a score model $s_\theta$ using the NCSN++ architecture \cite{Song2021Score} with a channel configuration of $[128, 128, 256, 256, 256, 256, 256]$.
As the diffusion process, we use the VP-SDE \cite{Song2021Score,ho2020ddpm} with $\beta_{\min} = 0.01, \beta_{\max} = 20$ and $t_\varepsilon=0.001$.
For training, we use the Adam optimizer \cite{kingma2014adam} with a learning rate of $10^{-4}$ and an exponential moving average (EMA) weight smoothing with decay $0.999$ \cite{Song2021Score}, and train for 140 epochs.

\subsection{Evaluation}\label{sec:experiments:evaluation}
To evaluate the quality of the reconstructions, we make use of three image metrics (aPSNR, aSSIM, cRMS), as well as one metric for evaluating position recovery which we call \emph{posCorrect}. The metrics aPSNR and aSSIM are evaluated using only the object magnitudes
\begin{equation}
    \aPSNR(\estimg, \img) := \PSNR(\min(\abs{\estimg}, 1), \abs{\img}),
    \quad
    \aSSIM(\estimg, \img) := \SSIM(\min(\abs{\estimg}, 1), \abs{\img}),
\end{equation}
where $\img$ is the ground-truth image and $\estimg$ is a reconstructed image, and we clip the estimated magnitudes to [0, 1] for evaluation to avoid large errors from isolated wrong pixels. \rev For all methods based on variational inference, we treat the final fitted distributional mean $\mu_\chi$ as the image estimate $\hat{x}$ for evaluation, and for producing the images shown in the figures in \cref{sec:results}. \nc We report aPSNR values in dB, while aSSIM takes values in [0, 1]. Since aPSNR and aSSIM ignore errors in the phase structure, we include the complex-valued metric $\cRMS$, introduced as $E_o$ in \cite{maiden2009improvedpty}, which is a normalized root-mean-square error metric with a complex-valued empirical scaling factor $\cRMScorr$ that corrects for a scale ambiguity and the global phase ambiguity inherent in phase retrieval problems:
\begin{equation}
    \cRMS(\estimg, \img) = \frac{
        \sum_{\coordy,\coordx}\abs{\img[\coordy,\coordx] - \cRMScorr \cdot \estimg[\coordy,\coordx] }
    }{
        \sum_{\coordy,\coordx}\abs{\img[\coordy,\coordx]}^2
    },
    \quad
    \cRMScorr = \frac{
        \sum_{\coordy,\coordx} \img[\coordy,\coordx] \cconj{\estimg[\coordy,\coordx]}
    }{
        \sum_{\coordy,\coordx}\abs{\estimg[\coordy,\coordx]}^2
    }
\end{equation}
where $\cconj{(\cdot)}$ indicates the complex conjugate. Our position recovery metric \emph{posCorrect} is defined as the number of estimated positions that are within a box of 3x3 pixels around their respective ground-truth positions. We argue that this small region of allowed error should be enough for further position refinement with well-established subpixel-capable methods, e.g., \cite{zhang2013translation}. In our case where $\nmeas=100$, posCorrect can directly be read as a percentage.

Our scenario of position-blind ptychography exhibits a global shift ambiguity in the reconstructed positions and image, since there is no absolute reference point for the positions. Therefore, before evaluating any metric, we run a simple greedy image registration procedure of the estimate $\estimg$ relative to the ground-truth $\img$: we evaluate every shift between $\pm\,20$ pixels in both directions and choose the global shift with minimum error of the image magnitudes. We then translate the estimated positions and circularly shift the image according to this optimal shift, and use these shifted estimates for evaluation.

For all experiments, unless otherwise noted, we run every reconstruction method for all 10 test images (\cref{sec:dataset}) with 3 repeats, each run with a different random seed, and for every metric we report the mean and standard deviation across these 30 runs.

\section{Numerical Results}\label{sec:results}
\begin{table}
    \centering\tablefontsize%
    {\large Non-blind scenario (baseline)}\\
    \begin{tabular}{lrrr}
    \toprule
    Method / Metric & aPSNR $\uparrow$ & cRMS $\downarrow$ & aSSIM $\uparrow$ \\
    \midrule
    \small\bfseries Optimization-based\\
    No prior  & 12.24 ± 0.99 & 0.55 ± 0.26 & 0.10 ± 0.05 \\
    H-TV prior ($\lambda = 0.1$) & 23.96 ± 1.19 & 0.04 ± 0.01 & 0.51 ± 0.06 \\
    \midrule
    \midrule
    \small\bfseries Variational Inference\\
    No prior & 12.75 ± 0.87 & 0.48 ± 0.22 & 0.11 ± 0.05 \\
    H-TV prior ($\lambda = 5$) & 23.69 ± 1.16 & \U{0.04 ± 0.01} & 0.49 ± 0.06 \\
    H-TV prior ($\lambda = 10$) & 25.04 ± 2.14 & 0.04 ± 0.02 & 0.67 ± 0.04 \\
    H-TV prior ($\lambda = 20$) & 24.49 ± 2.59 & 0.05 ± 0.03 & 0.78 ± 0.05 \\
    SSP & \B{30.36 ± 2.39} & \B{0.02 ± 0.01} & \B{0.90 ± 0.04} \\
    \midrule
    RED-Diff ($\lambdaRD = 20$) & \U{29.41 ± 2.40} & \B{0.02 ± 0.01} & \U{0.87 ± 0.03} \\
    \bottomrule
    \end{tabular}
    \caption{Metrics of reconstructions in the \B{non-blind} baseline setting, comparing different methods. Best in bold, second best underlined. H-TV represents the Huber-TV prior with weight $\lambda$, and \ac{SSP} refers to the surrogate score-based prior method.
    %\textcolor{red}{Intuitively odd that VI with no prior and HTV prior is worse than blind case here -- TODO SW: add a figure that illustrates that position uncertainty seems to act as an implicit image smoothing prior, explaining this effect}
    }
    \label{tab:nonblind-metrics-comparison}
\end{table}

\begin{table}
    \centering\tablefontsize%
    {\large Position-blind scenario\phantom{)}}\\
    \begin{tabular}{lrrrr}
    \toprule
    Method / Metric & aPSNR $\uparrow$ & cRMS $\downarrow$ & aSSIM $\uparrow$ & posCorrect $\uparrow$ \\
    \midrule
    \small\bfseries Optimization-based\\
    No prior & 14.19 ± 1.04 & 0.35 ± 0.15 & 0.11 ± 0.03 & 70.00 ± 19.06 \\
    H-TV prior ($\lambda=0.1$) & 19.08 ± 2.54 & 0.11 ± 0.04 & 0.37 ± 0.05 & 74.90 ± 12.76 \\
    \midrule
    \midrule
    \small\bfseries Variational Inference\\
    No prior & 15.68 ± 0.96 & 0.23 ± 0.08 & 0.16 ± 0.04 & 90.73 ± 6.14 \\
    H-TV prior ($\lambda = 5$) & \underline{23.35 ± 2.74} & \bfseries 0.05 ± 0.03 & 0.65 ± 0.04 & \bfseries 94.73 ± 4.29 \\
    SSP & \bfseries 25.34 ± 3.33 & \bfseries 0.05 ± 0.03 & \bfseries 0.85 ± 0.05 & \underline{94.03 ± 5.24} \\
    SSP, pos. deltas & 21.41 ± 5.39 & 0.20 ± 0.35 & 0.74 ± 0.17 & 65.07 ± 22.92 \\
    \midrule
    RED-Diff ($\lambdaRD = 20$) & 24.17 ± 4.40 & \underline{0.09 ± 0.13} & \underline{0.81 ± 0.09} & 52.67 ± 23.55 \\
    \bottomrule
    \end{tabular}
    \caption{Metrics of reconstructions in the \B{position-blind} setting, comparing different methods. Best in bold, second best underlined. H-TV represents the Huber-TV prior with weight $\lambda$, and \ac{SSP} refers to the surrogate score-based prior method.}
    \label{tab:blind-metrics-comparison}
\end{table}
%\todo{TR: the caption for Table 2 needs to be adapted. Maybe we can also use headline for the tables: with "Blind", "Non-Blind" to easier distinguish them}

In this section, we present the numerical results of the reconstructions achieved by all evaluated methods and compare them.
We evaluate \B{(1)} an optimization-based method with no prior or a Huber-TV prior (see \cref{eq:huber-tv}), \B{(2)} the proposed variational inference approach (\cref{algo:var-bayes-principled-prior}) under three choices of prior (no prior, Huber-TV prior, \ac{SSP} using a score-based model $\scoremodel$) and \B{(3)} the blind RED-Diff method (\cref{algo:blind-red-diff})
%, \cite{Alkan2023Variational,Mardani2024Variational})
using the same score-based model $\scoremodel$.
In each table, \enquote{$a \pm b$} indicates the empirical mean $a$ and empirical standard deviation $b$ of the respective metric, estimated using all 10 test objects, each with 3 independent reconstruction runs of the respective method.

\subsection{Non-blind Baseline}\label{sec:results:nonblind}
We first evaluate the described algorithms in the non-blind case. While this is not the main focus of the present work, we use this simpler case as a validation of our basic methodology and to gain an impression of the reconstruction quality that could be achievable under perfect position recovery. Here we use measurement noise of
\rev
standard deviation $\noisestd = 0.005$, 
\nc
which represents an average measurement \ac{SNR} of 4.5\,dB over all test objects.

In \cref{tab:nonblind-metrics-comparison}, we show the metric values of the compared methods. We can see that the variants \rev of VI and optimization-based methods \nc without image priors only reach PSNR values around 12\,dB, which can already be improved to around 25\,dB by using a Huber-TV prior. The variational \acf{SSP} method achieves a PSNR of 30\,dB and SSIM of 0.90, closely followed by RED-Diff. This demonstrates the usefulness of the score-based data-driven priors for image quality in our problem. The optimization-based methods with no prior and a Huber-TV prior perform similarly to their VI counterparts here. For the weight parameter $\lambda$ in the Huber-TV prior, we compare $\lambda \in \{5, 10, 20\}$, and find that larger $\lambda$ achieve better aSSIM, though aPSNR already decreases again after $\lambda = 10$. Since we aim to reconstruct complex-valued images, we use $\lambda=5$ for the VI method in all subsequent experiments, which achieves the best mean cRMS at the lowest standard deviation.

\begin{figure}
    \begin{subfigure}[t]{\textwidth}
        \centering
        \includegraphics[width=\textwidth]{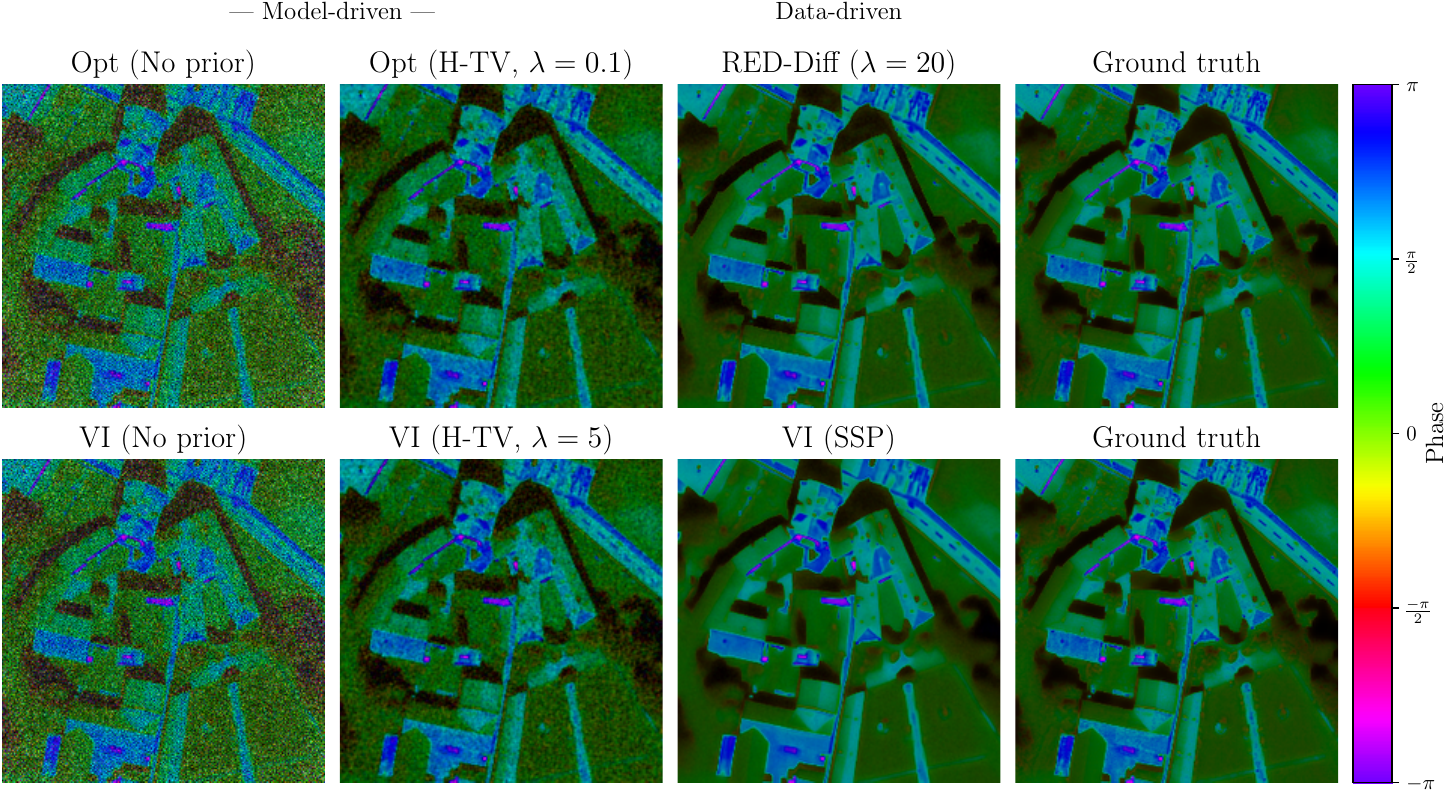}
        \caption{Method and prior comparison in the \B{non-blind} setting}
        \label{fig:example-recons-nonblind}
    \end{subfigure}

    \vspace{\floatsep}
    
    \begin{subfigure}[b]{\textwidth}
        \centering
        \includegraphics[width=\textwidth]{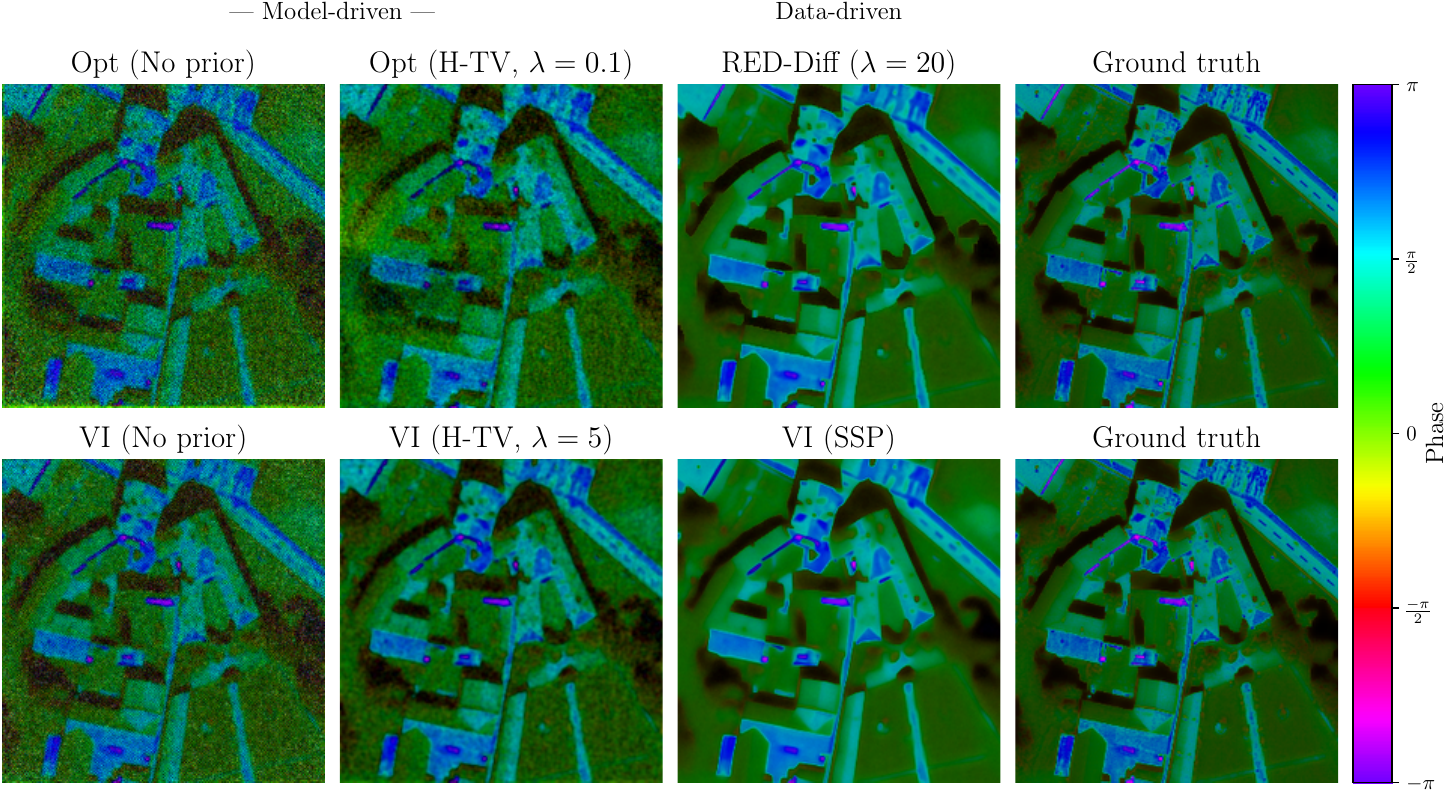}
        \caption{Method comparison in the \B{blind} setting}
        \label{fig:example-recons-blind}
    \end{subfigure}

    \caption{Example image reconstructions of a single test object for the different optimization-based methods (\emph{Opt}) and variational inference methods (\emph{VI}), in \B{(a)} the non-blind baseline setting and \B{(b)} our position-blind setting. The images show complex magnitude as the brightness and complex phase as the hue.}
\end{figure}

To illustrate the qualitative differences between the methods, in \cref{fig:example-recons-nonblind} we show the reconstruction with the best PSNR for each variational method. One can see that the reconstruction with no prior is heavily affected by the measurement noise, and the Huber-TV prior at $\lambda = 5$ only ameliorates this to some extent.
\rev As shown in \cref{suppl:fig:huber-tv-lambda-comparison}, H-TV with $\lambda = 10$ removes more noise but the image edges noticeably begin to blur. \nc
While RED-Diff seems to produce more fine details than \ac{SSP}, these may be partly hallucinated, which is reflected in the slightly lower metric scores in \cref{tab:nonblind-metrics-comparison}.
%\todo{not sure, maybe remove? but seems a good location in the text to point this out again}{
We note here that RED-Diff involves a weighting hyperparameter $\lambdaRD$ for the score-based prior loss%
%}
, the choice of which is ad-hoc and affects the image in a tradeoff between adherence to the measurement and adherence to the prior. RED-Diff therefore does not sample from the actual posterior induced by the likelihood and the prior, similar to issues of the related method DPS \cite{Chung2023Diffusion} as discussed in \cite{Feng2023Score}.

\subsection{Blind case}\label{sec:results:blind}
Having established non-blind baseline performance, we now evaluate the methods in the fully position-blind case. Here, we now also report the \enquote{posCorrect} metric which indicates the percentage of correctly recovered positions; see \cref{sec:experiments:evaluation}. We use the same methods as in the previous \cref{sec:results:nonblind}, and also add a variant of our blind variational \ac{SSP} method where the fitted position distributions are delta distributions, called \enquote{SSP, pos. deltas}. For this method the position estimation effectively turns into a pure optimization method while the image is still fitted as a variational distribution, similar to an \ac{EM} scheme such as DeepGEM \cite{Gao2021Deep}. The goal of evaluating this method is to investigate the usefulness of the variational fitting of the positions in our \ac{SSP} method.

The quantitative results are listed in \cref{tab:blind-metrics-comparison}. The \ac{SSP} methods again achieve the best image metric results of all compared methods and recovers \rev more than 94\% \nc of the positions correctly \rev on average\nc, but regarding position recovery shows no advantage over using only a H-TV prior. This suggests that the score-based prior is helpful for retrieving better images, but not necessarily for easing the overall position-blind reconstruction problem.
RED-Diff and our \ac{SSP} variant with position deltas perform significantly worse at recovering the positions correctly at only \rev 53-65\% \nc correctly recovered. Since RED-Diff also treats the fitted positions as fixed values rather than distributions (see \cref{sec:recmethods}), we conclude that the added randomness from fitting a random distribution on the positions is very helpful for position recovery. Nonetheless, both methods %still
reach fair image metric values, suggesting that using only a subset of the measurements may be enough to retrieve a decent image in this specific simulated measurement setup.

One curious observation, when comparing the values for the non-blind case in \cref{tab:nonblind-metrics-comparison} with the corresponding value for the blind case in \cref{tab:blind-metrics-comparison}, is that the methods with no prior or a Huber-TV prior to some extent reach better metrics in the blind case, in particular a better aSSIM. We found that this effect is due to the variations in the positions during reconstruction -- both from the iterative updates and from the randomness of sampling from the variational distribution -- which seem to act as an implicit smoothing prior on the image, resulting in better metrics in the presence of measurement noise.

\begin{table}
    \centering\tablefontsize%
    \begin{tabular}{lrrrr}
    \toprule
    Probe & aPSNR $\uparrow$ & cRMS $\downarrow$ & aSSIM $\uparrow$ & posCorrect $\uparrow$ \\
    \midrule
    \multicolumn{5}{l}{\bfseries\small No phase mask}\\
    $\apdiam=1/2$ & 10.53 ± 1.04 & 0.97 ± 0.31 & 0.25 ± 0.08 & 0.53 ± 0.68 \\
    $\apdiam=1/4$ & 11.32 ± 1.45 & 0.95 ± 0.37 & 0.23 ± 0.13 & 9.07 ± 10.13 \\
    $\apdiam=1/8$ & \B{21.16 ± 3.51} & \B{0.11 ± 0.07} & \B{0.68 ± 0.14} & \B{89.03 ± 7.85} \\
    \midrule
    \multicolumn{5}{l}{{\bfseries\small Blockwise random phase mask}, $\apdiam=1/2$}\\
    $\maskblocksize=32$ & 12.49 ± 3.19 & 0.81 ± 0.40 & 0.39 ± 0.17 & 15.10 ± 21.21 \\
    $\maskblocksize=16$ & 18.77 ± 5.31 & 0.28 ± 0.38 & 0.65 ± 0.20 & 62.77 ± 31.23 \\
    $\maskblocksize=8$ & 23.75 ± 4.94 & 0.16 ± 0.34 & 0.80 ± 0.14 & 85.97 ± 19.36 \\
    $\maskblocksize=4$ (default) & \B{25.34 ± 3.33} & \B{0.05 ± 0.03} & \B{0.85 ± 0.05} & \B{94.03 ± 5.24} \\
    \bottomrule
    \end{tabular}
    \caption{Metric comparison for the blind \ac{SSP} method using different probe functions. %\textcolor{red}{May be misleading due to different measurement noise floors. But qualitative trend should still be valid. Consider the discussion carefully here}
    }
    \label{tab:metrics-probe-comparison}
\end{table}

\subsubsection{Importance of the probe structure}\label{sec:results:probe-comparison}
\begin{figure}
    \begin{subfigure}{\textwidth}
        \centering
        \caption{Probes without aperture phase mask, compared to our default choice $\apdiam=\sfrac{1}{2}, b=4$.}
        \includegraphics[width=\linewidth]{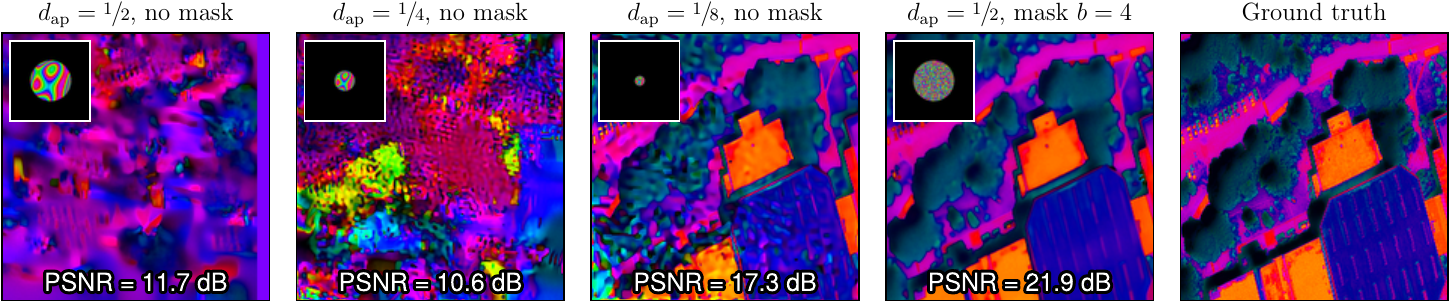}
        \label{fig:example-recons-blind-probe-nomask}
    \end{subfigure}
    
    \begin{subfigure}{\textwidth}
        \centering
        \caption{Probes with aperture phase mask, at different mask block sizes $b$.}
        \includegraphics[width=\linewidth]{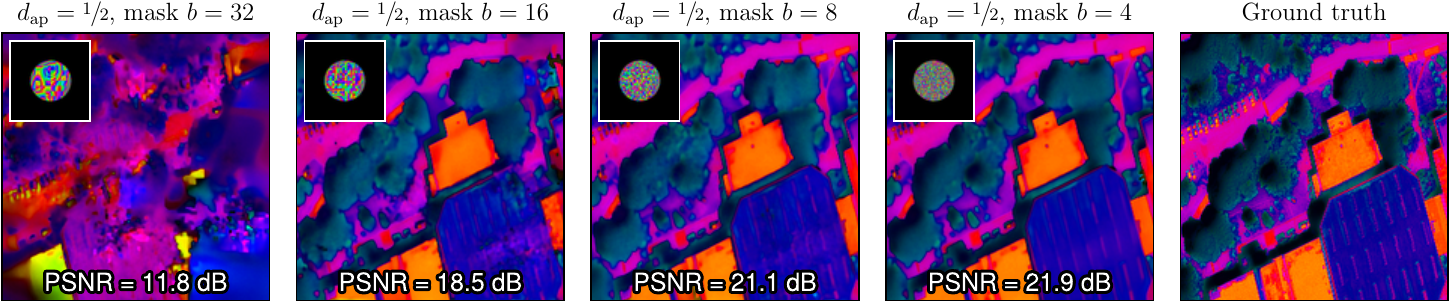}
        \label{fig:example-recons-blind-probe-mask}
    \end{subfigure}

    \caption{Reconstructed images for different probe functions with the \ac{SSP} method. We compare \B{(a)} probes with different aperture diameters $\apdiam$ and \B{(b)} optional random aperture phase masks of different block sizes $b$. The insets show the aperture-plane wavefront generating each respective probe.}
    \label{fig:example-recons-blind-probe-comparison}
\end{figure}

Next we investigate the effect of the choice of probe function. First, we do not use a blockwise-random phase mask and only vary the aperture diameter $\apdiam$, which inversely scales the probe size and therefore the measurement overlap.
Then, for comparison, we add the blockwise-random phase mask to the largest aperture (smallest probe size) at $\apdiam = 1/2$, with block sizes of $b \in \{4, 8, 16, 32\}$. We show the qualitative results visually in \cref{fig:example-recons-blind-probe-comparison} and the quantitative results in \cref{tab:metrics-probe-comparison}. For completeness, we show all evaluated apertures and probe functions in the supplementary material; see \cref{suppl:fig:all-probes-and-apertures}.

Without a phase mask and $\apdiam \in \{\sfrac{1}{2}, \sfrac{1}{4}\}$, we recover almost no positions correctly and fail to produce usable images. While $\apdiam = \sfrac{1}{8}$ recovers 89\% of positions correctly, its final images (\cref{fig:example-recons-blind-probe-nomask}, center) are contaminated with low-frequency artifacts and the image metrics are subpar, which may be explained by the lack of high-frequency structure in the probe due to the low-pass from the small aperture.
In comparison, adding block-wise random phase masks to the previously unusable aperture with $\apdiam=\sfrac{1}{2}$ leads to the best image quality we observed and also improves the position recovery, with a 25.3$\,$dB aPSNR and \rev 94\% \nc of positions recovered correctly for the block size $\maskblocksize=4$\rev, see \cref{tab:metrics-probe-comparison}\nc. We further note that $b=8$ still performs rather well compared to $b=4$ despite illuminating a significantly smaller region of the object (see \cref{fig:probe-comparison}), again suggesting that probe structure is more helpful than probe size for our task.

\begin{figure}
    \centering
    \includegraphics[width=0.75\linewidth]{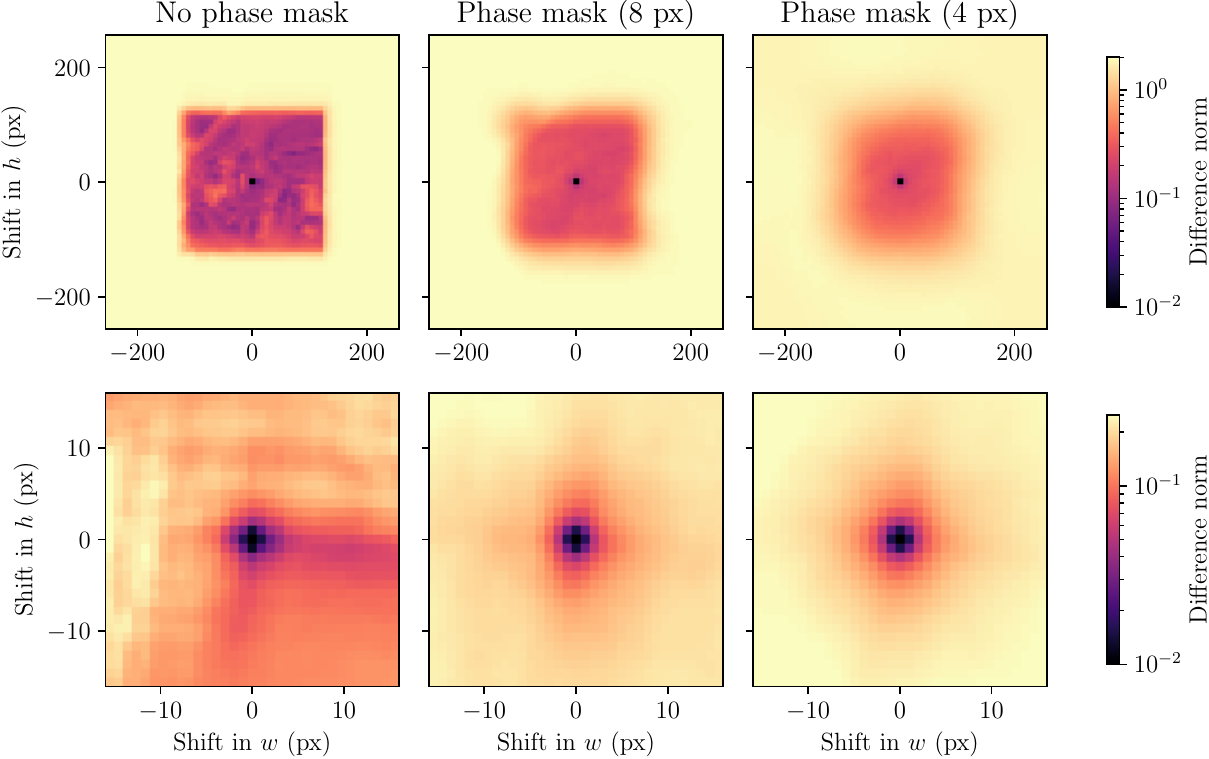}
    \caption{The loss landscape for an idealized position recovery problem. We plot summed squared errors between a noiseless measurement at the central position and simulated noiseless measurements at possible $(\shiftx, \shifty)$ shifts relative to the center. We compare the three probe functions shown in \cref{fig:probe-comparison} with $\apdiam=\sfrac{1}{2}$. The bottom row is zoomed in around (0,0) and of higher resolution.}
    \label{fig:probe-comp-shift-difference-norm}
\end{figure}

These empirical observations are further corroborated by the loss landscapes shown in \cref{fig:probe-comp-shift-difference-norm}. In this figure, we compare three probe functions in the simplified task of recovering the position of a measurement when the ground-truth object is already known. We plot the sum of the squared errors between a noiseless measurement at the ground-truth position (0,0) and noiseless measurements at all other positions, showing the loss landscape of the likelihood term at all possible estimated positions. We can observe that the probe without the phase mask (\cref{fig:probe-comp-shift-difference-norm}, leftmost column) shows strong local minima separated by strong local maxima, and the loss landscape is much less convex than for the other two probes even around the true position $(0, 0)$.

\rev
\subsubsection{Convergence behavior and runtime efficiency}
\begin{figure}
    \centering
    \includegraphics[width=\linewidth]{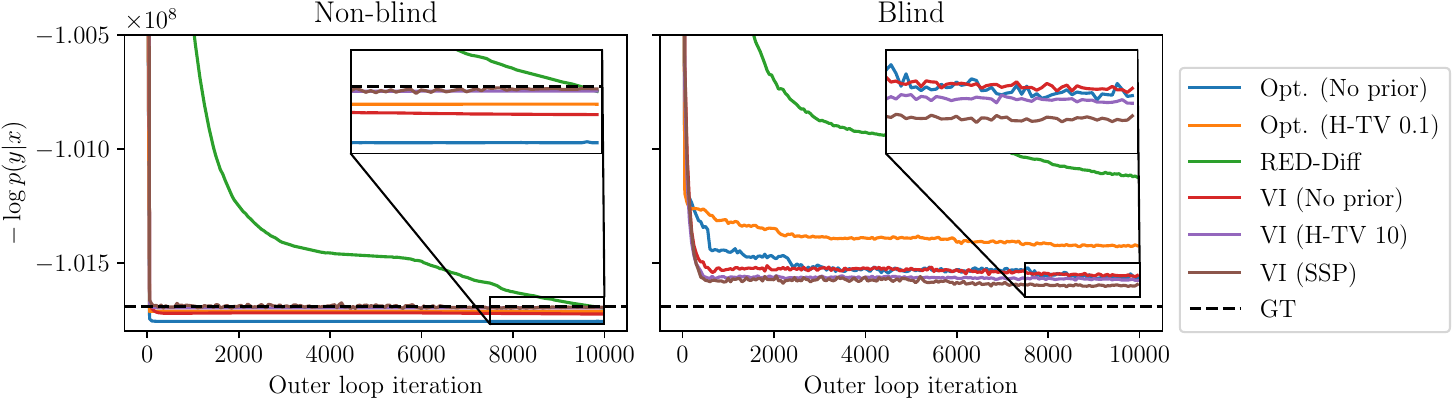}
    \caption{\rev Evolution of the negative log-likelihood along the optimization trajectory for different methods. The dashed line indicates $-\log p(y|x^*)$, where $x^*$ denotes the ground truth image.}
    \label{fig:convergence-curves}
\end{figure}

We now turn our attention to the numerical convergence behavior of the compared methods in the blind and non-blind settings. We treat the \acf{NLL}, $-\log p(y|x)$,
%, which is proportional to the measurement error $\sum_k \lVert y_k - \mathcal{A}_k(\hat{x},\hat{r}) \rVert_2^2$,
as the quantity that should ideally converge.
We show convergence curves, i.e., the \ac{NLL} plotted against the iteration number, for each compared method on a single example image and reconstruction run in \cref{fig:convergence-curves}. We compare against the \ac{NLL} of the ground-truth image and ground-truth positions relative to the simulated noisy measurements $\{y_k\}$, which we call \emph{ground-truth \ac{NLL}}.

\paragraph{Non-blind setting}
In the non-blind setting, see \cref{fig:convergence-curves} on the left, it can be seen that most methods converge below the ground-truth \ac{NLL}, indicating overfitting to the measurement noise. The only methods that avoid this behavior and converge with values close to the ground-truth \ac{NLL} are VI (SSP), VI with the Huber-TV prior, and RED-Diff, indicating that the priors and choice of prior weighting used in these three methods successfully combat the overfitting.
Interestingly, RED-Diff only reaches values close to the ground-truth \ac{NLL} close to the end of the reconstruction. We suspect that this is due to the fixed reverse diffusion schedule that RED-Diff follows \cite{Mardani2024Variational}, which makes the generative prior focus only on the low image frequencies at first and only uses the full image frequency content during the final iterations. In contrast, \ac{SSP} uses all diffusion timesteps through weighted random sampling throughout the reconstruction, see \eqref{eq:bsde-empirical}.
For all methods besides RED-Diff, the convergence plots suggest that the full 10,000 iterations are not needed in this non-blind setting, and one could terminate at much fewer iterations.

\paragraph{Blind setting}
In the blind setting, see \cref{fig:convergence-curves} on the right, none of the methods converge to a value close to the ground-truth \ac{NLL}. This is expected, as the blind problem is of higher difficulty and not all positions are recovered correctly by any method (cf. \cref{tab:blind-metrics-comparison}). In contrast to the non-blind setting, the plotted curves demonstrate that all methods profit from the full 10,000 iterations here, and may even improve the \ac{NLL} further if given more iterations.

The \ac{SSP} method reaches the lowest values, followed by VI with a Huber-TV prior, VI with no prior, and the non-variational optimization method with no prior. For the two methods without a prior, the unfavorable aPSNR and cRMS values (cf. \cref{tab:blind-metrics-comparison}) suggest that the low \ac{NLL} values are due to overfitting to measurement noise. 
% recall that their metrics show rather bad aPSNR and cRMS values , hence their low \ac{NLL} values are likely misleading and again suggest overfitting to measurement noise.
%
%
% \todo{TR: reminder for me to slightly reformulate. Can we interpret this as seeing ambiguities in the positions? SW: do you mean in the direction of recalling that modeling the positions via a distribution is empirically very helpful (our posdeltas experiments)? that would seem sensible, but maybe you mean something different}
%
This is also reflected in the reconstructed images, see \cref{fig:example-recons-blind}. Furthermore, the plot for the prior-free optimization-based method during early iterations suggests a more unstable and complex behavior than for the other methods.
RED-Diff and the optimization method with a Huber-TV prior exhibit a rather large gap in \ac{NLL} to the other methods. We attribute this also to the comparably worse position recovery observed in \cref{fig:example-recons-blind}.

% This suggests that their regularization performs suboptimally in this position-blind setting. 

% This is somewhat concerning, as both methods seem to have been reasonably well-tuned in the non-blind setting according to the metrics and convergence plots, and warrants future investigations.

%\todo{SW@TR: could you read over this? TR: Sounds good to me thanks!}

Related to the convergence behavior, the runtime and memory efficiency of the compared methods is also of practical interest. In \cref{tab:timings-and-memory}, we show and compare the empirical runtime and peak memory usage of each used method variant. It can be seen that the VI-based methods all take more than 4 hours for the reconstruction, while the optimization-based methods and RED-Diff take around one hour. The memory usage of the VI-based methods is also substantially higher.
The reason for this lies mostly in the use of a batch size $B>1$ to perform parallel evaluation of the likelihood and prior terms on multiple image and position samples from the variational distributions, see \cref{sec:vi-principled-priors}. We support this idea by an additional line in \cref{tab:timings-and-memory} when using $B=1$ instead. Evidently, the GPU does not effectively parallelize over the batch, since the runtime increases by a factor very close to $B=4$. The peak GPU memory is increased by a factor of around 3.2 compared to $B=1$.
Interestingly, the use of a diffusion model as in \emph{VI (SSP)} and \emph{RED-Diff} does not significantly increase the runtime nor memory usage, which suggests that both may be dominated by the costly repeated evaluation of the likelihood on $N=100$ high-dimensional diffraction pattern measurements.
Note that a runtime- or memory-optimal implementation is out of scope for this work, and we expect that improvements to the runtime and GPU memory usage can be made in future work.

\begin{table}
    \centering
    \scalebox{0.9}{
    \begin{tabular}{lrr}
        \toprule
        Method & Runtime & GPU Mem. \\
        \midrule
        VI (SSP) & 4:49 h & 19082 MiB \\  % N=10: 13894; N=1: 13496
        \textcolor{gray}{VI (SSP) with $B=1$} & \textcolor{gray}{1:11 h} & \textcolor{gray}{5986 MiB} \\
        VI (H-TV prior) & 4:15 h & 17124 MiB \\
        VI (No prior) & 4:15 h & 17122 MiB \\  % N=10: 3582; N=1: 2079
        RED-Diff & 1:11 h & 6712 MiB \\
        Opt. (H-TV prior) & 1:04 h & 6479 MiB \\
        Opt. (No prior) & 0:55 h & 6472 MiB \\
        \bottomrule
    \end{tabular}
    }
    \caption{\rev Overall runtime in hours and peak GPU memory usage for one image reconstruction, using each method. Determined on an NVIDIA A100 80GB GPU. \nc}
    \label{tab:timings-and-memory}
\end{table}

\nc

\subsubsection{Different levels of measurement noise}
\begin{figure}
\begin{subfigure}[t]{\textwidth}
%\centering
\includegraphics[width=.65\linewidth]{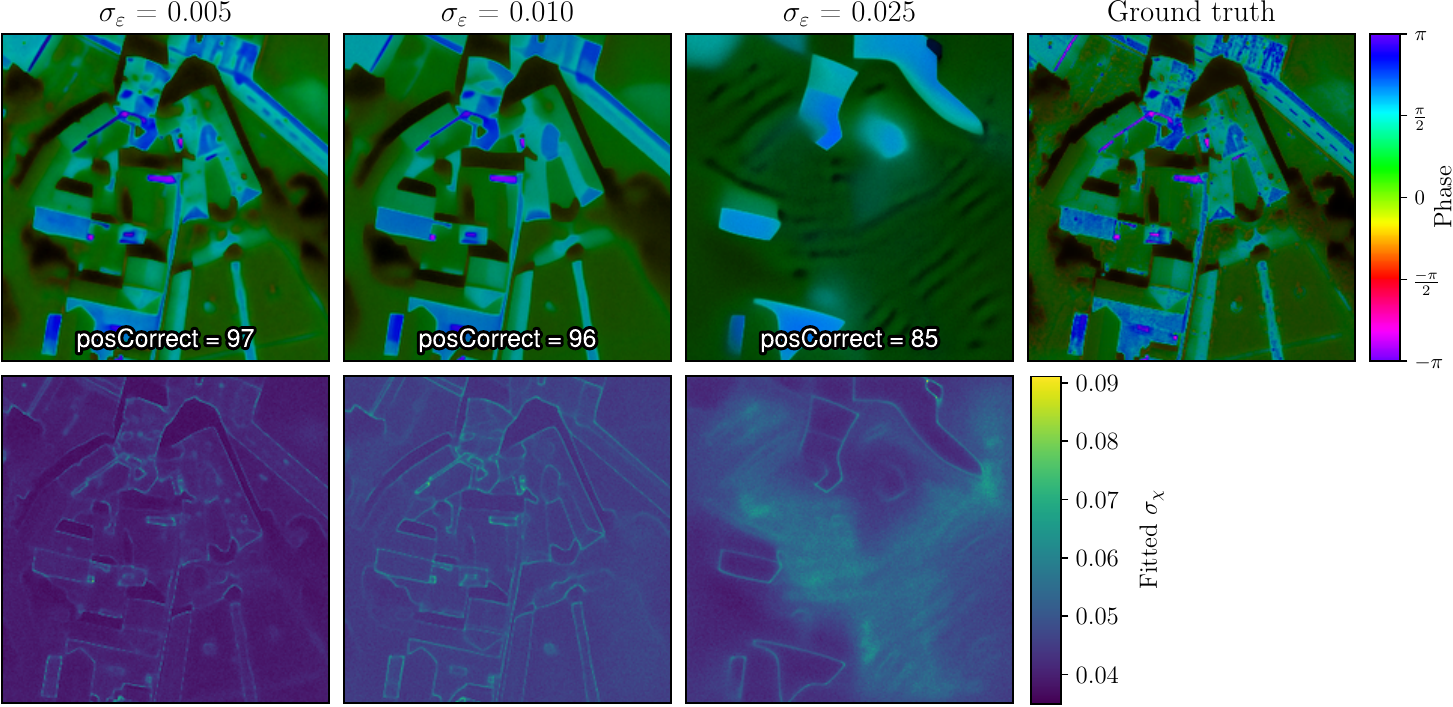}
\caption{With \B{Gaussian} measurement noise at different noise levels $\noisestd$.}
\label{fig:example-recons-gaussian-noise}
\end{subfigure}
\begin{subfigure}[b]{0.65\textwidth}
    \centering
    \includegraphics[width=\linewidth]{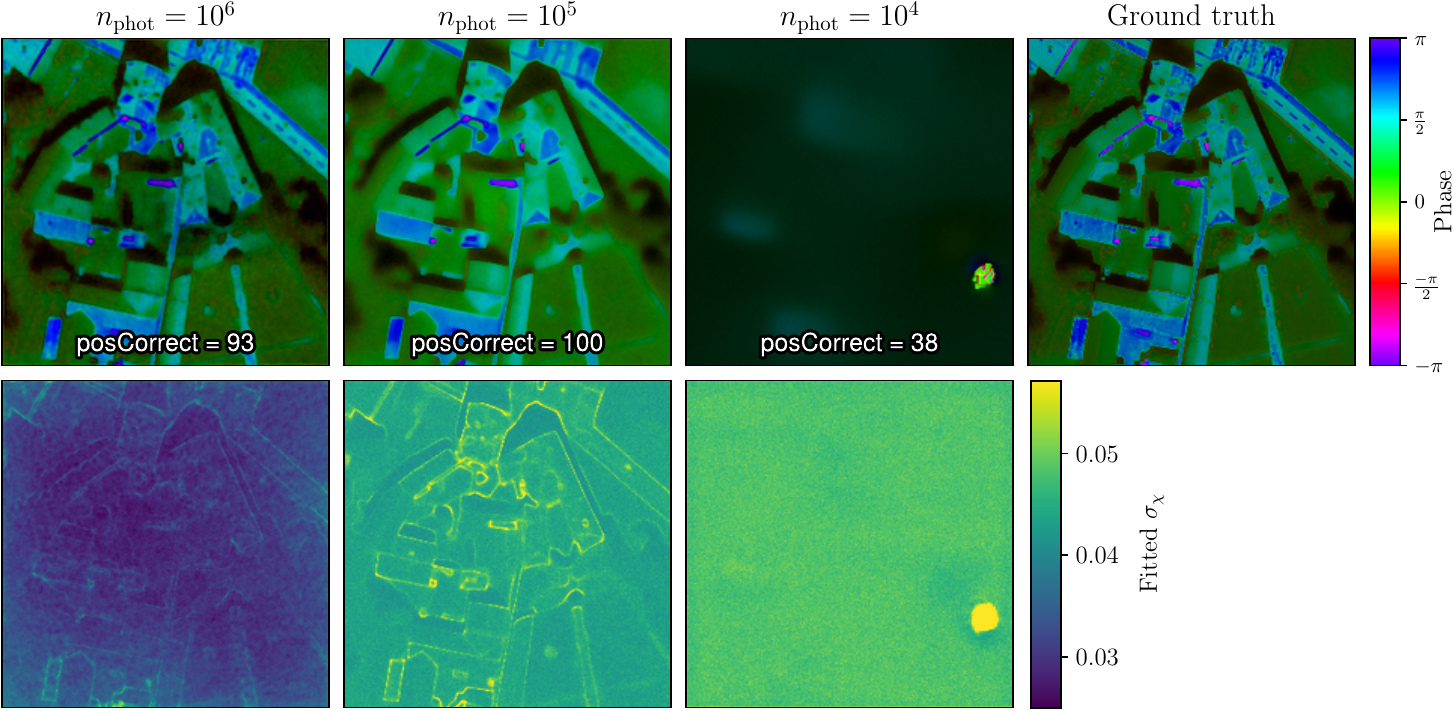}
    \caption{With \B{Poissonian} measurement noise at different expected numbers of photons $\nphot$; see \cref{sec:noise-model}.}
    \label{fig:example-recons-poisson-noise}
\end{subfigure}
\hfill
\begin{subtable}[b]{0.32\textwidth}
\centering\tablefontsize%
\begin{tabular}{lll}
\toprule
$\nphot$ & cRMS $\downarrow$ & posCorrect $\uparrow$ \\
\midrule
$10^4$ & 0.66±0.13 & 37.33±11.89 \\
$10^5$ & 0.13±0.04 & 96.17±6.79 \\
$10^6$ & 0.08±0.05 & 94.20±7.86 \\
\bottomrule
\end{tabular}
\caption{Metrics for SSP under different levels of \B{Poissonian} measurement noise with an expected number of photons $\nphot$.}
\label{tab:ssp-metrics-poisson-noise}
\end{subtable}
    
    \caption{Example reconstructions under different levels of \B{(a)} Gaussian and \B{(b)} Poissonian measurement noise, using the blind SSP method, and \B{(c)} corresponding metrics for Poisson noise. See \cref{tab:metrics-measurement-noise} for the corresponding measurement \acp{SNR}. The bottom image rows show the fitted per-pixel standard deviation $\sigma_\chi$ of the Gaussian variational image distribution $q_\chi$.}
    \label{fig:example-recons-both-noises}
\end{figure}

Here we analyze and compare the behavior of the blind SSP method, blind RED-Diff, and blind VI without an image prior when increasing the level (scale) of the Gaussian measurement noise $\noisestd$ from the default $\noisestd = 0.005$. We show quantitative results in \cref{tab:metrics-measurement-noise}, where we also list the corresponding measurement \ac{SNR}, and qualitative results for the SSP method in \cref{fig:example-recons-gaussian-noise}, where we also show the fitted per-pixel variance of the variational Gaussian that can inform uncertainty estimation. We can observe that \B{(1)} the data-driven priors allow for at least some image and position recovery even under -9.5\,dB measurement noise, where the VI method without an image prior completely fails; \B{(2)} the SSP method performs best and most reliably under all noise levels and, surprisingly, recovers 69\% of positions correctly \rev on average \nc even at the highest noise level; \B{(3)} the per-pixel uncertainty from SSP (\cref{fig:example-recons-both-noises}) is, at least at the highest noise level, informative about regions that are heavily affected by artifacts. %\rev
A complete visual comparison of the methods under all measurement noise levels is shown in \cref{suppl:fig:reconstructions-meas-noise-allmethods}.
%In particular, we observe that at the highest noise level with $\sigma_\varepsilon = 0.025$, the VI method with no prior fully fails. In contrast, both RED-Diff and VI with a \acf{SSP} succeed at reconstructing something resembling the ground-truth object, and VI (SSP) even recovers 85\%\todo{TR: where does the 85\% come from? SW: top row of figure 8, rightmost column, posCorrect=85} of the positions correctly. \todo{SW: revise}
%\nc
We note here that the $\lambdaRD$ parameter of RED-Diff could be tuned further in dependence of $\noisestd$ in order to potentially improve the results, but we do not follow this here.
\begin{table}[]
    \centering
    \resizebox{\textwidth}{!}{%
    \tablefontsize%
    \begin{tabular}{lccccccc}
        \toprule
        &
        & \multicolumn{2}{c}{VI, SSP} 
        & \multicolumn{2}{c}{RED-Diff ($\lambda_{RD}=20$)} 
        & \multicolumn{2}{c}{VI, no prior} \\
        \cmidrule(lr){3-4} \cmidrule(lr){5-6} \cmidrule(lr){7-8}
        $\sigma_\varepsilon$ & {\footnotesize Meas. SNR}
        & cRMS $\downarrow$ & pC $\uparrow$ 
        & cRMS $\downarrow$ & pC $\uparrow$ 
        & cRMS $\downarrow$ & pC $\uparrow$ \\
        \midrule
        0.005 & 4.47 dB
        & \B{0.05±0.03} & \B{94.0±5.2}
        & \U{0.09±0.13} & 52.7±23.6
        & 0.23±0.08 & \U{90.7±6.1} \\
        0.010 & -1.55 dB
        & \B{0.06±0.04} & \B{91.5±6.7}
        & \U{0.21±0.38} & 43.5±26.0 
        & 0.24±0.13 & \U{74.4±12.5} \\
        0.025 & -9.51 dB
        & \B{0.23±0.11} & \B{69.2±15.4} 
        & \U{0.30±0.20} & \U{16.2±11.9}
        & 0.64±0.12 & 0.0±0.0 \\
        \bottomrule
    \end{tabular}%
    }
    \caption{Comparison of methods across Gaussian measurement noise levels in terms of the cRMS and posCorrect (here called \emph{pC} for brevity) metrics. The best value in each row, i.e., for each noise level, is highlighted in bold; the second best is underlined. \emph{Meas.~SNR} indicates the average measurement signal-to-noise ratio corresponding to the respective noise level $\noisestd$.}
    \label{tab:metrics-measurement-noise}
\end{table}

\subsubsection{Poisson noise with SSP}
Here we evaluate the \ac{SSP} method under different levels of Poissonian measurement noise, using the Gaussian approximation detailed in \cref{sec:noise-model} and evaluating for different noise levels via different numbers of expected photons $\nphot$. We list the reconstruction metrics in \cref{tab:ssp-metrics-poisson-noise} and show example images in \cref{fig:example-recons-poisson-noise}. We find that for $\nphot=10^6$ and $\nphot=10^5$ the method performs very well for position recovery and only somewhat worse for image recovery than in the Gaussian noise case with $\noisestd=0.005$. For $\nphot=10^4$, the results degrade substantially, with the corresponding image only containing very coarse features of the ground truth and having a strong local artifact. Interestingly, at the same time, 37\% of positions are still recovered correctly.

\subsubsection{Phase-only objects}\label{sec:results:phase-only}
As a more difficult case, we now turn our attention to phase-only objects, that is, objects that are non-absorbing and only affect the phase of the incoming wavefront.  We argue that this case is closer to real-world biological materials such as proteins, which typically consist only of light elements and thus have complex refractive indices $n = 1 - \delta + i\beta$ with $\delta \gg \beta, \delta \ll 1$ in the X-ray regime \cite{henke1993xray}.

\begin{figure}
    \centering
    \includegraphics[width=0.8\linewidth]{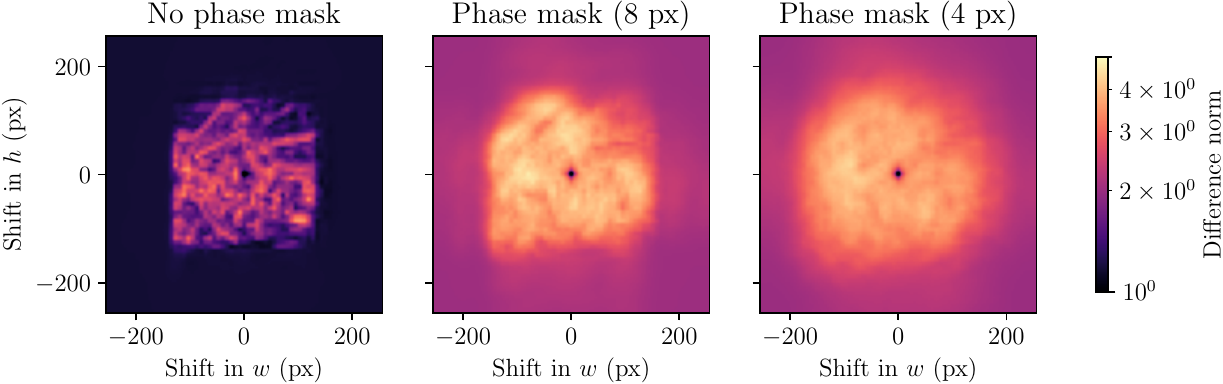}
    \caption{A position recovery loss landscape as in \cref{fig:probe-comp-shift-difference-norm}, but for a phase-only object (no absorption). The increased difficulty of this problem is evident from the broad regions of high loss around the global minimum and the broad regions of misleadingly low loss for positions lying outside of the object.}
    \label{fig:probe-comp-shift-difference-norm-phaseonlyobj}
\end{figure}

One problem we have to contend with in this scenario is the changed position recovery loss landscape, as illustrated in \cref{fig:probe-comp-shift-difference-norm-phaseonlyobj}.
This plot shows that, even with the added phase mask, the position loss landscape in this setting contains misleading broad regions of low loss for positions \emph{off} the object and in fact has the highest loss values for most of the positions \emph{on} the object, with only a small approximately convex region around the true position at (0,0). This is in contrast to the previous scenario with absorption (see \cref{fig:probe-comp-shift-difference-norm}), where the highest-loss regions were always those off the object.
Without further constraints, we observed during initial reconstruction runs that the positions were frequently erroneously estimated to lie in the corners of the domain, i.e., maximally off the object, and we found successful image and position recovery to be unreliable.

\begin{figure}
    \centering
    \includegraphics[width=0.6\linewidth]{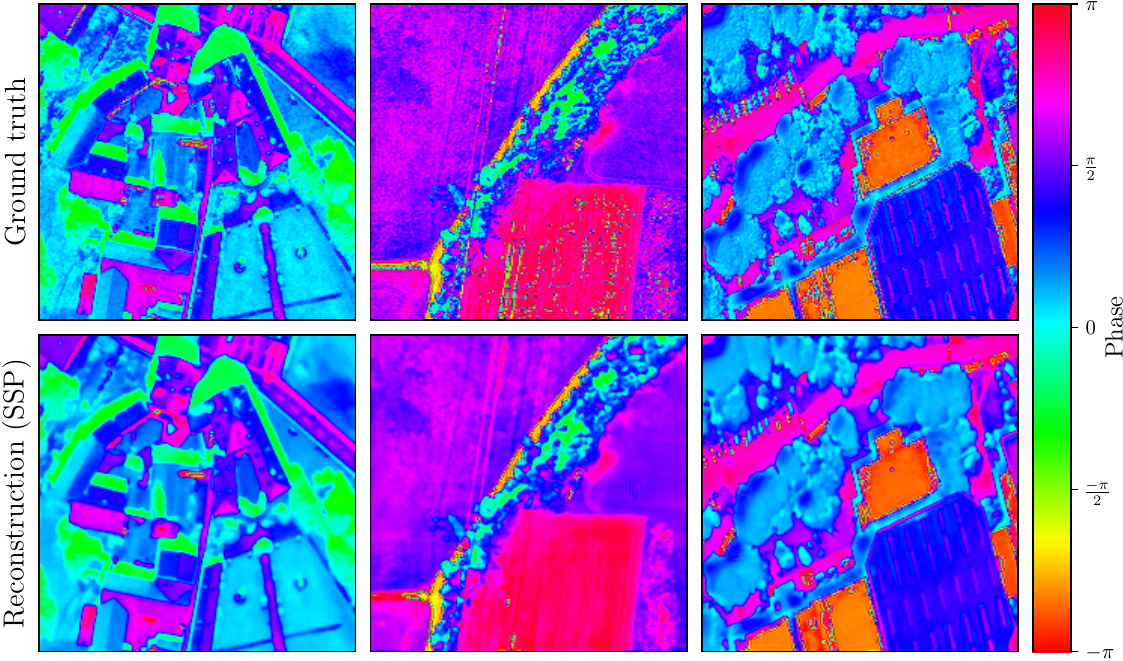}
    \caption{Example reconstructions of three phase-only test objects in the position-blind setting with the variational SSP method, when using a position log-barrier. Only the complex phase is shown as the hue.}
    \label{fig:example-recons-phase-only-objects}
\end{figure}

We therefore added a log-barrier prior loss on the positions as described in \cref{sec:algo-configuration}, with the weight empirically set to $100$. With this, we found a reliable reconstruction ability, with all objects and repeated runs leading to satisfactory image quality. In \cref{fig:example-recons-phase-only-objects}, we show successful example reconstructions for the standard probe $\apdiam=\sfrac{1}{2},\maskblocksize=4$ with the blind SSP method. Nonetheless, in terms of metrics, here we find a worsened cRMS value of 0.06 ± 0.06 and a posCorrect value of only 61.60 ± 14.04,
\rev
which are both worse than the values for blind reconstructions of complex objects that are also absorbing, see \cref{tab:blind-metrics-comparison}. This shows the increased difficulty of this problem variant,
\nc particularly for position recovery.

% \subsubsection{Phase-only weak-phase objects}\label{sec:results:phase-only-weakphase}
% \todo[inline]{SW4SW: results are missing currently, jobs are submitted, write this paragraph after}

\subsubsection{Weak-phase objects with beamstop}\label{sec:results:beamstop}

\begin{figure}
    \centering
    \includegraphics[width=\linewidth]{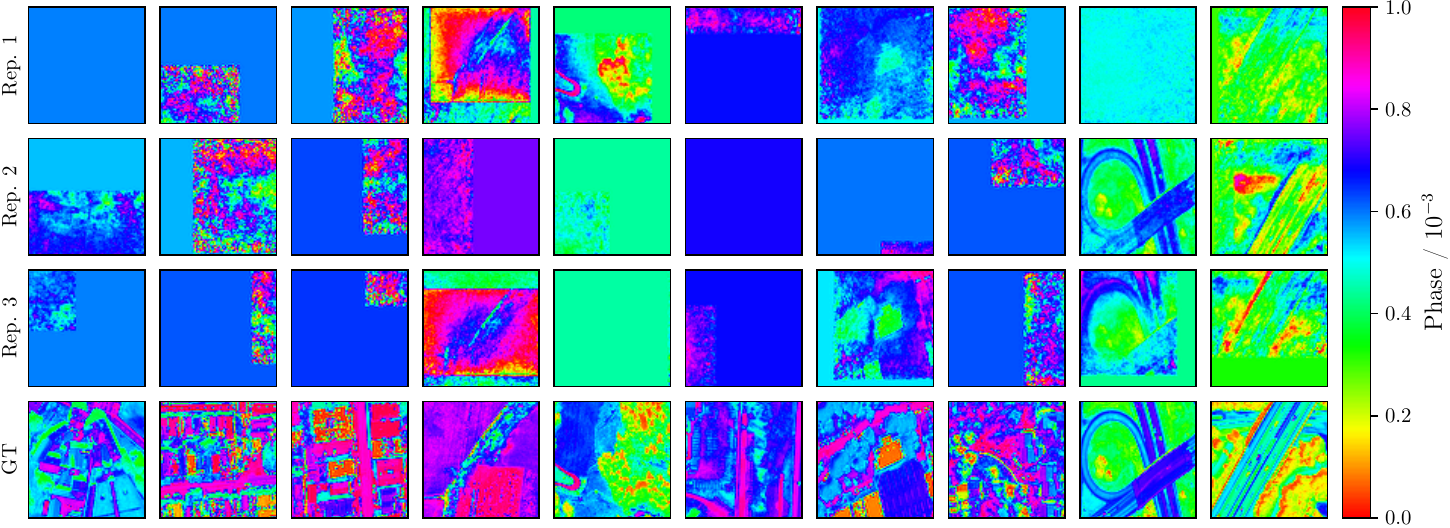}
    \caption{All attempted reconstructions (10 objects, 3 repeats) in the most difficult weak-phase phase-only object scenario with a measurement beamstop, where the maximum phase shift is $10^{-3}$. The ground truth (\emph{GT}) is shown in the bottom row. Only the complex phase is shown as the hue.}
    \label{fig:example-recons-weakphase-beamstopped}
\end{figure}

Finally, we consider the most difficult scenario of position-blind ptychography of phase-only, weakly phase-distorting objects ($|\delta| \ll 1$),  with a circular central beamstop in the measurements. The practical reasoning for using a central beamstop is that the direct portion of the beam has extremely high photon flux in \ac{XFEL} experiments, which would lead to a measurement dynamic range that greatly exceeds what most detectors can handle, and may even damage or destroy the detector. We linearly rescale the phase shift $\delta$ (see \cref{eq:refractive-index}) of all phase-only test objects to the interval $[0, 10^{-3}]$. The beamstop blocks out the zero-order portion (and thus the full imaged aperture) of the diffracted beam on the detector, and is employed when faced with limited detector dynamic range and extreme source brightness such as for \acp{XFEL}. To simulate the beamstop, we mask out all pixels inside the circle that represent the aperture on the detector, i.e., all pixels that are within a circle of radius $(1 + \frac{\apdiam}{2} \cdot 512)$ from the detector center, where we add another pixel to the radius to also block out high intensities that may occur at the edge of the imaged aperture. We inform our reconstruction method about this pixel-wise detector mask by ignoring all masked pixels when calculating the likelihood.

To block out fewer pixels from the measurement than for our default probe with $\apdiam=\sfrac{1}{2}$ where the imaged aperture covers half of each diffraction pattern, here we reduce the size of the aperture to $\apdiam=\sfrac{1}{4}$ but keep the random aperture-plane phase mask with $\maskblocksize=4$. Furthermore, we decrease the measurement noise level to $\noisestd = 10^{-12}$ since the average intensity of pixels that are diffracted outside the region of the imaged aperture is much lower than those inside. We do not use a score-based prior here, since our neural network was not trained for images of weak-phase objects. For the algorithm (VI with no image prior), we set the image optimizer step sizes to $\lambda_{\mu_\chi} = \lambda_{\sigma_\chi} = 10^{-5}$ and use the position log-barrier (\cref{sec:log-barrier}) with a weight of $10^6$.
In this most difficult setting, we find that the probability of successful image recovery is very low, with frequent position misalignment of the entire object, but the method does occasionally achieve at least a partial reconstruction. This suggests that it is possible to further tune and improve our approach towards this scenario in the future. We show all result images in \cref{fig:example-recons-weakphase-beamstopped} for completeness.

\subsection{Towards uncertainty quantification}\label{sec:results:uncertainty}

\begin{figure}
    \centering
    \includegraphics[width=.7\linewidth]{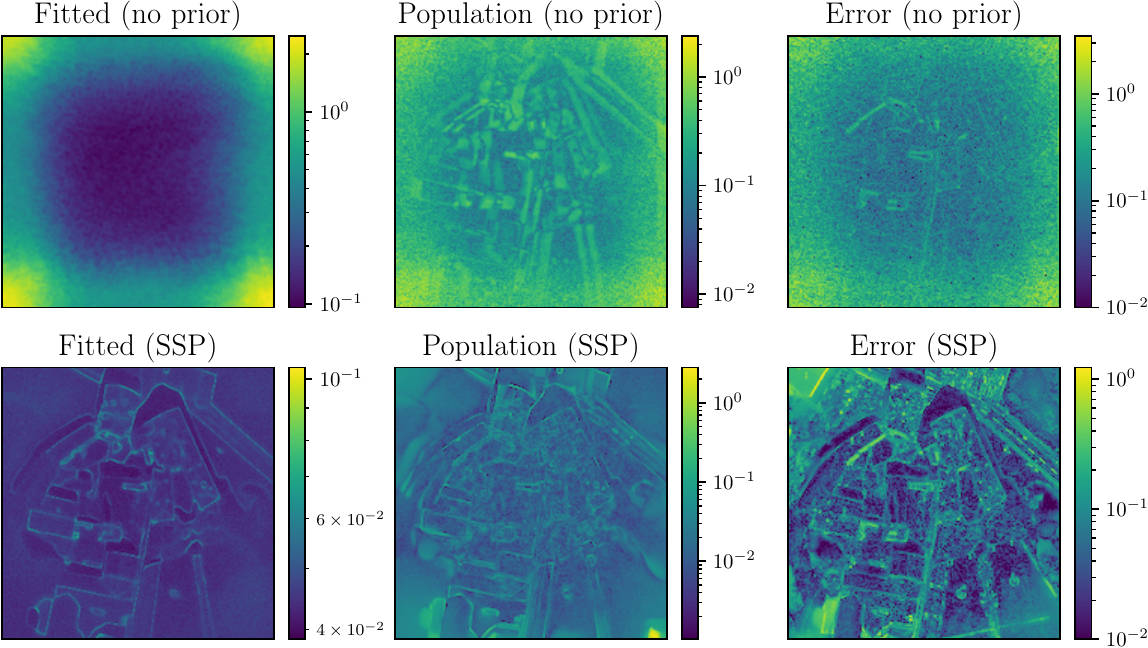}
    \caption{The per-pixel uncertainty without a prior and with a surrogate score-based prior (SSP), either estimated from the fitted diagonal covariance of the Gaussian %variational 
    distribution $q_\chi$ (\emph{Fitted}) or estimated from a population of three independently fitted means (\emph{Population}).\rev{} Note that all reconstructions are obtained by VI.\nc{}  For comparison, we also show the per-pixel to the ground-truth object, i.e., $|x-\hat{x}|$ (\emph{Error}).}
    \label{fig:uncertainty-partial-scan}
\end{figure}

A reliable estimate of uncertainty about each part of the reconstructed images would be helpful to inform experimental practice and scientific results. The variational approach with a Gaussian variational distribution yields a fitted per-pixel standard deviation, which may be used as a coarse approximation of a meaningful uncertainty estimate. Alternatively, the stochastic nature of our VI schemes also allows a straightforward population-based uncertainty estimate, by performing multiple independent reconstructions and constructing a per-pixel standard deviation map from this population.

To test both ideas, here we construct a set of measurements by restricting the simulated positions to the central quarter of the image, which results in low coverage of the outer part of the object image in terms of information encoded in the collection of measurements. One would then hope that this lack of information about the outer part would be reflected in such uncertainty estimates. We compare \ac{SSP} and \ac{VI} with no image prior, and show both per-pixel uncertainty maps (fitted and population-based) in \cref{fig:uncertainty-partial-scan}, in comparison to the actual per-pixel error to the ground truth image. We find that, without an image prior, the fitted uncertainty behaves as expected in the outer part of the image. With a data-driven prior, however, the fitted uncertainty does not reflect the low measurement coverage well, which may suggest an overly high confidence induced by the strong data-driven prior. The population estimate from \ac{SSP} is somewhat more informative, but also does not clearly highlight the expected outer regions as uncertain.
%Nonetheless, we argue that both variants of this simple uncertainty quantification are more informative than fitting a point mass without any associated uncertainty quantity.

\section{Discussion}

\rev

In the following we discuss possible paths towards reconstruction in more realistic measurement setups. Furthermore, we highlight the risks and challenges connected to reconstruction methods based on generative models.

\subsection{Towards a more realistic SPI setting}\label{sec:towards-realistic-spi}

In this paper we treat only a simplified 2-D variant of the full 3-D position- and rotation-blind ptychography problem that would arise in \ac{SPI}. %, we explicitly list our simplifying assumptions for clarity, see \cref{sec:experiments}. W
We expect that it is possible for future works to build upon our Bayesian approach to include scenarios closer to real-world \ac{SPI} experiments, towards 3D images with unknown positions \emph{and} unknown rotations, unstable illumination (unknown or varying probe), and structured measurement noise. In the following we comment on how we hope to eventually unlock this new imaging modality for real-world biological structure investigations.

\subsubsection{On \cref{ass:knownprobe}}
\begin{figure}
    \centering
    \includegraphics[width=0.6\linewidth]{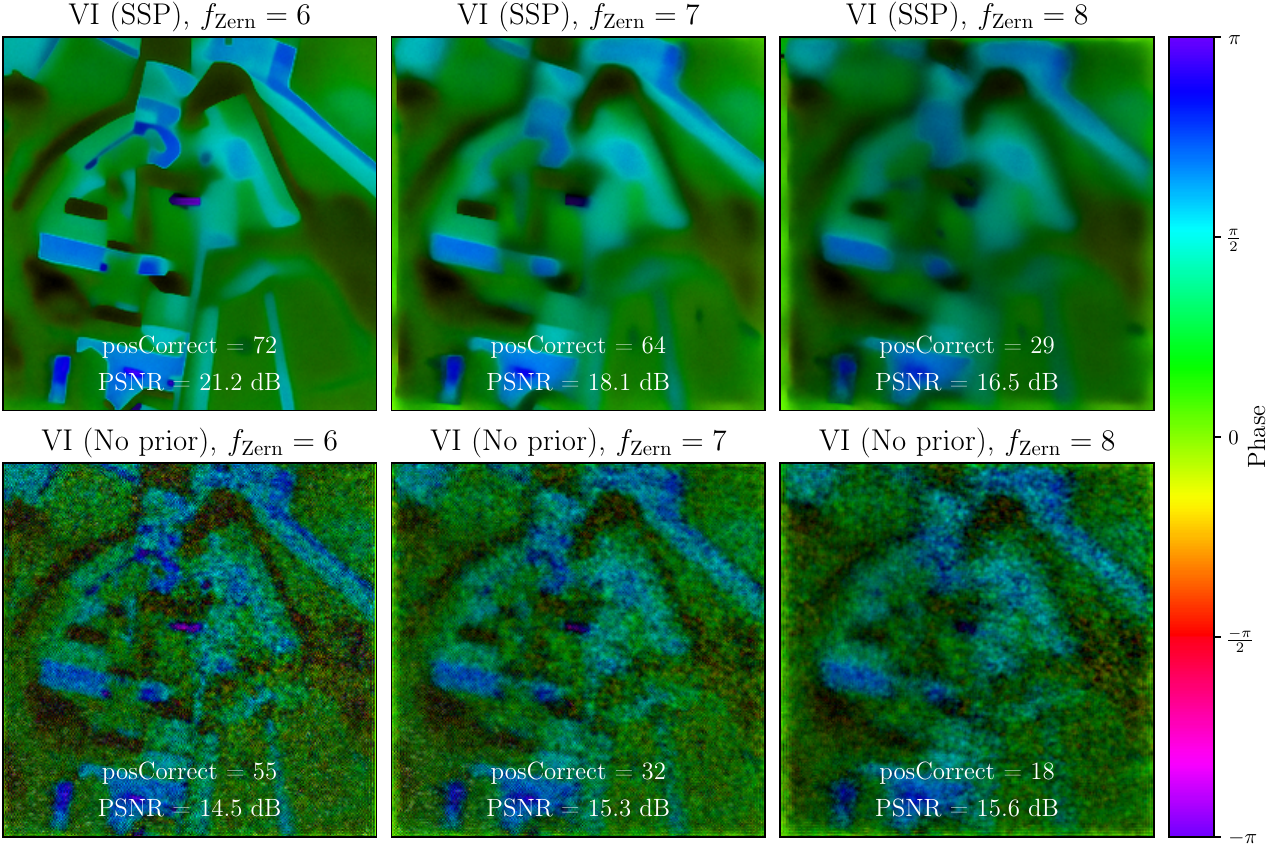}
    \caption{Reconstructions under a mismatched probe function, using variational inference with a score-based prior (top row) or no prior (bottom row). $f_\text{Zern}$ denotes the factor applied to the Zernike polynomial coefficients for the probe used during reconstruction, compared to the original value of 5 used for the simulated measurements.}
    \label{fig:probemismatch}
\end{figure}
Assuming that the probe is known, as in \cref{ass:knownprobe}, is not generally sensible in real ptychographic measurements \cite{Rodenburg2019Ptychography}. A relaxation of this assumption within our framework should be possible by treating the entire probe function as another parameter to recover, given that the scheme remains reasonably stable under an imperfectly estimated probe.

% \todo{TR: mention plethora of works on probe-blind ptychography and that we could incorporate the same. see \cite{Rodenburg2019Ptychography} for possible refs?}

As a motivating example, we briefly investigate the effect of a mismatched probe function on reconstructions by our \ac{VI} scheme. First, we use the same simulated configuration to generate the same measurements as in \cref{sec:results:blind}.
Then, we simulate three mismatched probes by increasing the magnitude of the Zernike aberration coefficients from the \enquote{true} factor of 5 that was used to generate the measurements, to factors 6, 7, and 8. We keep the random phase mask the same, assuming that it can be characterized well and is stable. We then run one reconstruction each for each of these mismatched probes.
We show the results in \cref{fig:probemismatch}. While the scheme recovers a worse and more blurry image, as well as lower PSNR and fewer correct positions -- as one would expect from using the wrong probe function -- it still succeeds to recover a reasonable image in the overall position-blind ptychography problem. Furthermore, interestingly, the use of the generative score-based surrogate prior (SSP) results in more reliable position recovery in this probe-mismatched setting than using no prior.
Empirically, therefore, we argue that the scheme \cref{eq:variational-bayes-scheme} is stable enough under a probe-mismatched estimated measurement operator to allow adding the probe optimization as a third alternating (or joint) optimization step in \cref{eq:variational-bayes-scheme}. We leave detailed investigations of this to future work.

\subsubsection{On \cref{ass:2d}}
Imaging three-dimensional objects with ptychographic methods has been successfully achieved in various works, and we refer the reader to \cite[Sec.~17.6]{Rodenburg2019Ptychography} for a detailed overview. 
Rodenburg \& Maiden \cite{Rodenburg2019Ptychography} note two major types of approaches: Ptycho-tomography and multislice ptychography. Multislice ptychography assumes that the sample is not rotated between measurements, and is hence not applicable to the general SPI scenario, see \cref{sec:on-ass-orientation}.

Ptycho-tomography has been successfully realized both computationally and experimentally in several prior works \cite{Dierolf2010,Gursoy2017PtychoTomo,Floystad2015PtychoTomo,Aslan2021PtychoTomo,Rodenburg2019Ptychography}. However, these works generally assume that scan positions and object rotations are known, which likely presents a substantially easier task than the fully position- and rotation-blind one. While ptycho-tomography typically cannot account for multiple scattering \cite{Rodenburg2019Ptychography}, this should not present a problem for X-ray \ac{SPI} of biomolecules, due to the very small overall sample thickness of even large proteins, where the first-order Born approximation (cf. \cite{BornWolf7ed}) should hold.

Performing ptycho-tomography with generative priors has also previously been successfully realized \cite{Aslan2021PtychoTomo}. In this work, the authors also briefly suggest that future work can add a position correction to their ADMM-based scheme, by introducing additional auxiliary variables. Similarly, we expect that by using a forward model similar to \cite{Aslan2021PtychoTomo}, introducing learnable rotational parameters to our \ac{VI} scheme as described in the next \cref{sec:on-ass-orientation}, treating the imaged target object $\img$ as a 3-dimensional array, and training appropriate three-dimensional generative priors, a future extension to three-dimensional objects is possible. Using three-dimensional diffusion-based priors for multi-slice ptychographic reconstructions has already been successfully performed in \cite{lee2025threedimpty-diff}. Nonetheless, the difficulty of the fully blind problem variant, and any possibility of convergence, is unknown and remains a highly interesting future direction.

%\todo[inline]{Mention something about ptycho tomography literature. Mention that this is a non-trivial extension. Slicing vs 3D input for diffusion models. We expect slicing to not work as well as in MRI settings for example. For SW: Multi-slice}

\subsubsection{On \cref{ass:orientation}} \label{sec:on-ass-orientation}
In \cref{ass:orientation}, we assumed that the object is always oriented the same in every measurement, and no rotational parameters must be recovered. This assumption generally does not hold in real \ac{SPI} settings, where the macromolecules under investigation enter the interaction region with random orientations (and translations, which we however already treat).
To some extent, this problem can potentially be addressed experimentally by physically aligning macromolecules such as proteins, e.g., with lasers \cite{Amin2025Alignment}. However, research on this is still ongoing, and as \cite{Amin2025Alignment} notes, the unknown molecule orientation is so far treated \emph{in silico} instead.
Our variational approach can be extended in this direction, by also modeling and optimizing the distribution of all per-image object rotation angles $\{\varphi_k\}_{k=1,\ldots,K}$ (or in 3D, e.g., quaternions), though care must be taken to use distributions that can correctly model the involved circular statistics. The distribution parameters would then also be optimized through backpropagation with respect to the loss \eqref{eq:variational-inference-blind4}. This is viable with existing software, e.g., the \texttt{kornia} Python package offers a differentiable \texttt{rotate} function. Whether it is better to treat this optimization as part of a single update step for the positions and rotations, or to perform further splitting of the alternating minimization \eqref{eq:variational-bayes-scheme}, is a topic reserved for future work.

\subsubsection{On \cref{ass:probe-centered-on-object}}
In \cref{ass:probe-centered-on-object}, we assume that all diffraction patterns used for reconstruction correspond to probe positions where the probe illuminated the object significantly (no measurements of empty space). This assumption is in line with prior literature on \ac{SPI}, see for instance \cite{Sobolev2020} where patterns were classified as \emph{hits} or \emph{non-hits} by a simple thresholding of the total measurement intensity. Such a classification can also be straightforwardly used as a data preprocessing step for our methods.

\subsection{Risk of generative models for scientific applications}
\begin{figure}
    \centering
    \includegraphics[width=0.8\linewidth]{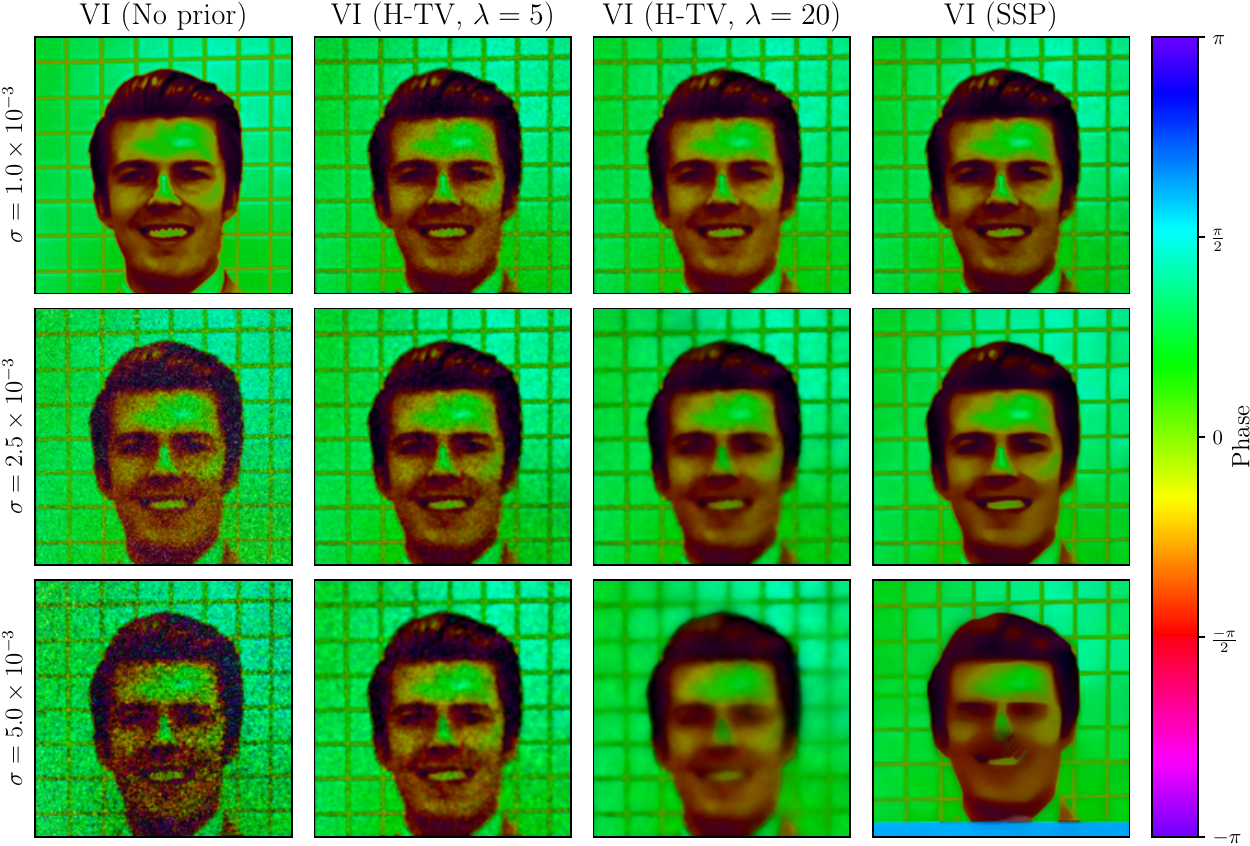}
    \caption{
        \rev
        Position-blind ptychography reconstructions from our variational inference methods %of an image from a human face %dataset
        \revtwo using a human face image\protect\footnotemark{},
        which is mismatched to the aerial photograph training dataset 
        \nc of the score-based prior, see \cref{sec:dataset}.
        The columns indicate different choices of prior, the rows indicate increasing levels of measurement noise. In the most noisy setting, the learned \acf{SSP} hallucinates building-like \revtwo rectangular \nc structures and overly smooths other regions such as the eyes -- both features that are likely under the learned prior, rather than reflect\revtwo{}ing\nc the actual ground-truth image.
        \nc
    }
    \label{fig:datamismatch}
\end{figure}
\footnotetext{\revtwo The displayed individual is one of the authors and has provided explicit consent for the use of their image. \nc} 

Generative models are promising new tools in the field of inverse problems and, as also shown in the results presented herein, can aid with successful reconstructions, especially under strong measurement noise. However, as discussed in various prior works on data-driven reconstruction methods \cite{burger2024learning,antun2020instabilities,muckley2021results,gottschling2025troublesome,kruger2026generativemodelssolveinverse}, they also carry risks with regards to the trustworthiness expected of scientific practice.
For instance, in image regions that are not well-described by the measurement data, generative priors tend to dominate the reconstructed image, which can result in image features that are not meaningfully supported by the measurements and may thus not correspond to physical reality -- a phenomenon which, in an anthropomorphizing manner, is often referred to as \emph{hallucination} \cite{gottschling2025troublesome,muckley2021results}. Such hallucinations are clearly a problem for scientific inquiry and the reliability of results, as they can lead to false and unreliable interpretations. As an extreme example, when used for protein structure investigations, a generative model could produce an entire protein substructure that the real protein does not have, leading to entirely false conclusions and catastrophic downstream decisions, e.g. for drug development.

We note, however, that even simple model-based priors such as \ac{TV}, in principle suffer from errors introduced by said priors, which is inherent to the Bayesian formulation. In an image region that is not captured well by the measurements, the TV prior can dominate, which would result in a uniformly flat region in the reconstruction even if this does not correspond to the imaged object. Nonetheless, the much higher simplicity and interpretability of such model-based priors suggests a higher level of trust since it is considered to be easier to spot and analyze when and why such faulty reconstructions arise.

The problem of so-called hallucinations becomes particularly clear when the generative prior is trained on image data that has distinctly different features than the images captured of the desired real physical objects, which is often referred to as \emph{data mismatch}. To empirically demonstrate that this problem does arise in our context, we run a reconstruction on an image\revtwo{} of a human face with different statistics (256x256 resolution, converted to grayscale and used directly for both amplitude and phase without any smoothing)\nc{} while using the score model trained on our simulated and smoothed INRIA data, see \cref{sec:dataset}. We show the results in \cref{fig:datamismatch}. We can observe the different types of artifacts introduced by different prior choices. In the no prior case the reconstruction tends to overfit to the noise. As expected the \ac{TV} creates flat regions in the image. For the \ac{SSP} prior, we do not observe similar artifacts. Instead the generative model places features into the reconstruction that are comparable to the ones from the INRIA dataset.

%Another problem is that of the weighting between likelihood and prior, ... [note advantage of SSP here compared to various other methods: at least we get the correct automatic weighting under a Bayesian framework] [refer back to \cref{sec:results:uncertainty}] \todo{TR, LK: could you formulate something here? TR: I would be ok with leaving this outr.}

\nc

\subsection{Summary and Outlook}
In this work, we investigate the novel blind inverse problem of position-blind ptychography for the first time, with possible applications in biological \acf{SPI} and ptychographic imaging under extreme movement of the sample. To attack this problem, %
\rev we derive \nc  
and demonstrate a Bayesian variational inference approach that can employ \rev both \nc modern data-driven image priors and classic model-driven priors, and compare this method against non-variational optimization-based approaches and another method from recent literature for data-driven solutions to blind inverse problems, RED-Diff \cite{Mardani2024Variational,Alkan2023Variational}.

We evaluate our approach for increasingly difficult scenarios including phase-only objects, measurements with Poisson noise, and weak-phase phase-only objects under the presence of a beamstop, and show that we can reliably achieve successful reconstructions in all but the most difficult scenario. We investigate the underlying reasons for the difficulty of variants of our position-blind ptychography imaging problem, and propose remedies in the form of structured illumination and additional prior terms on the positions.
\rev
We highlight all assumptions underlying our simplified position-blind ptychography problem, and offer guidance for future extensions towards realistic \acf{SPI} settings. Overall, this work contributes a new interesting inverse problem to the literature and provides a solid foundation for future investigations into this new imaging problem. We hope that the ideas and observations presented here can spark future investigations in this field.
\nc

%\todo[inline]{TR: more opinion / takeaway messages here}

% Limitations:
% \begin{itemize}
%     \item Assumed 2D (thin-sheet) object
%     \item Assumed no rotation
%     \item Assumed known and constant probe
%     \item Assumed Gaussian measurement noise of known level
%     \item Requires a large training dataset (but maybe we can find low-data diffusion model works to cite here)
% \end{itemize}

% Future work:
% \begin{itemize}
%     \item Experimental measurements (currently unavailable)
%     \item Other image reconstruction methods (e.g. an EMC adaptation)
%     \item Alternative generative reconstruction methods (w/ guarantees) -- do we have citations?
% \end{itemize}

\appendix

% \sloppy%
% \printbibliography
\section*{Acknowledgments}
MB, LK and TR acknowledge support from DESY (Hamburg, Germany), a member of the Helmholtz Association HGF.
This research was supported in part through the Maxwell computational resources operated at Deutsches Elektronen-Synchrotron DESY, Hamburg, Germany. 
This work was carried out while LK and TR were a member of DESY. 
MB, TG, LK, TR and SW acknowledge funding by the German Federal Ministry of Research, Technology and Space (BMFTR) under grant agreement No. 16IS24072B and 16IS24072D (COMFORT). SW acknowledges partial funding by DASHH (Data Science in Hamburg - HELMHOLTZ Graduate School for the Structure of Matter) with the Grant-No. HIDSS-0002. MB, LK and TR acknowledge partial funding by the DAAD project 57698811 "Bayesian Computations for Large-scale (Nonlinear) Inverse Problems in Imaging". TR acknowledges the support of the Munich Center for Machine Learning and the ERC Advanced Grant NEITALG, grant agreement No. 101198055.

\newcommand{\FundingLogos}{%
  \raisebox{0pt}{\includegraphics[height=1.5cm]{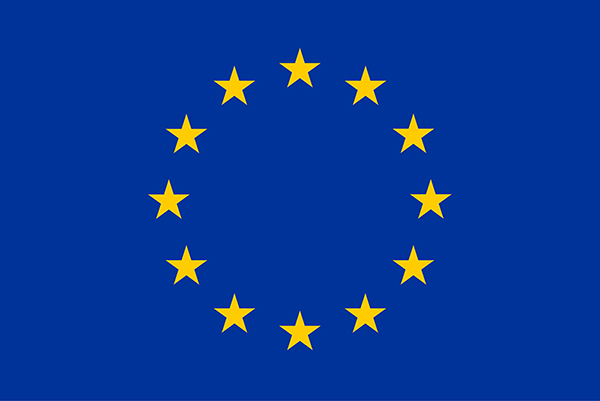}}%
  \hspace{1em}%
  \raisebox{0pt}{\includegraphics[height=1.5cm]{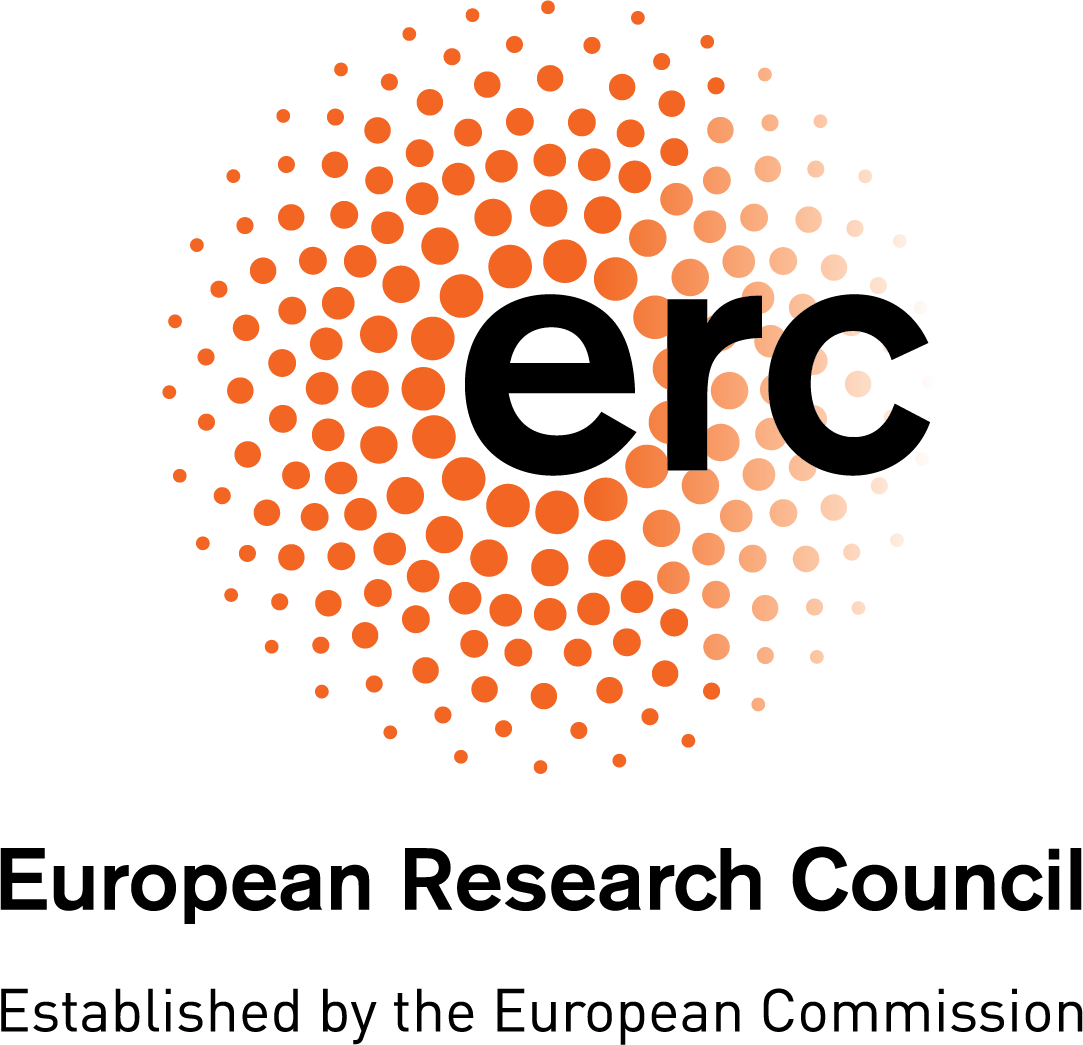}}%
}
\begin{center}
  \FundingLogos
  \vspace{0.5em}
  \begin{tcolorbox}\centering\small
    Funded by the European Union. Views and opinions expressed are however those of the author(s) only and do not necessarily reflect those of the European Union or the European Research Council Executive Agency. Neither the European Union nor the granting authority can be held responsible for them. This project has received funding from the European Research Council (ERC) under the European Union’s Horizon Europe research and innovation programme (grant agreement No. 101198055, project acronym NEITALG).
  \end{tcolorbox}
\end{center}

\printbibliography

\appendix
\clearpage

\section{Additional figures}

\begin{figure}[h]
    \centering
    \includegraphics[width=.9\textwidth]{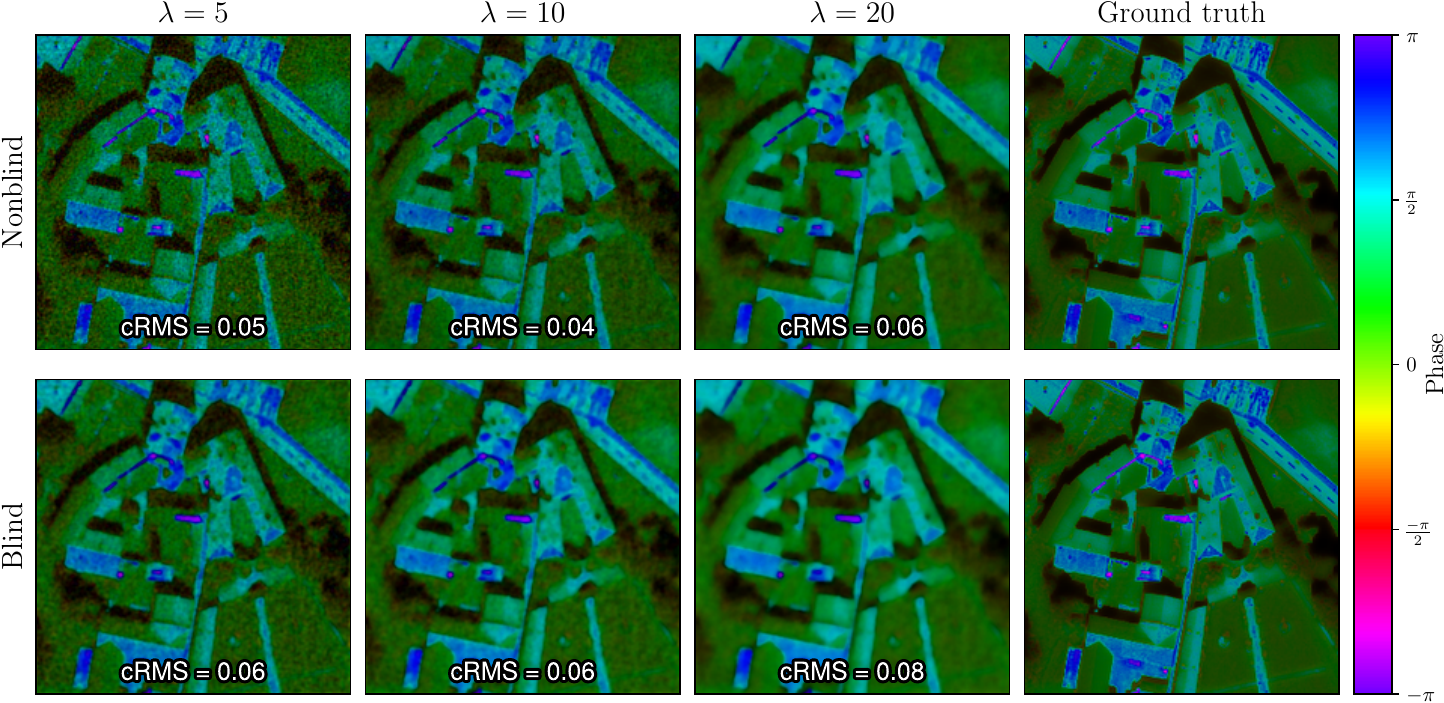}
    \caption{Comparison of VI reconstruction with the Huber-TV prior under different prior weights $\lambda \in {5, 10, 20}$.}
    \label{suppl:fig:huber-tv-lambda-comparison}
\end{figure}

\begin{figure}[h]
    \centering
    \includegraphics[width=\textwidth]{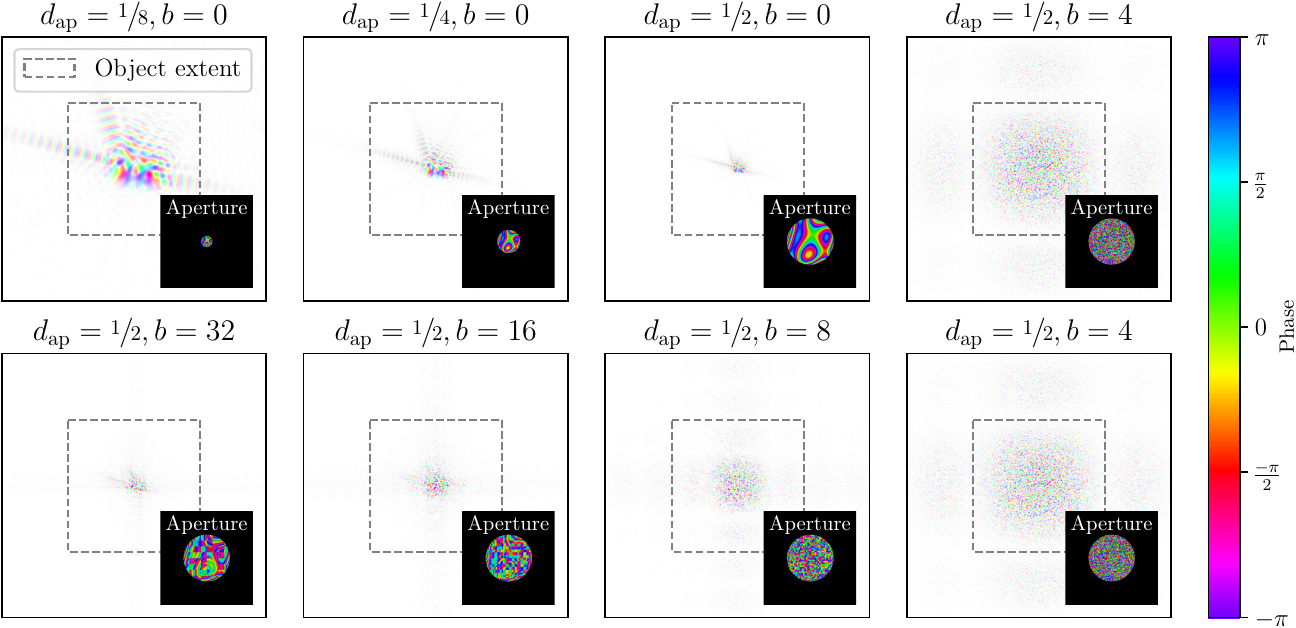}
    \caption{All probes and corresponding apertures used for evaluation in this work, in particular when comparing the effect of different probes. Our default choice throughout the work is the probe with $\apdiam=\sfrac{1}{2}, b=4$, which is visually duplicated here for easier visual comparison and a cleaner layout. $b=0$ indicates no phase mask was used for the respective probe.}
    \label{suppl:fig:all-probes-and-apertures}
\end{figure}

\begin{figure}[h]
    \centering
    \includegraphics[width=0.6\textwidth]{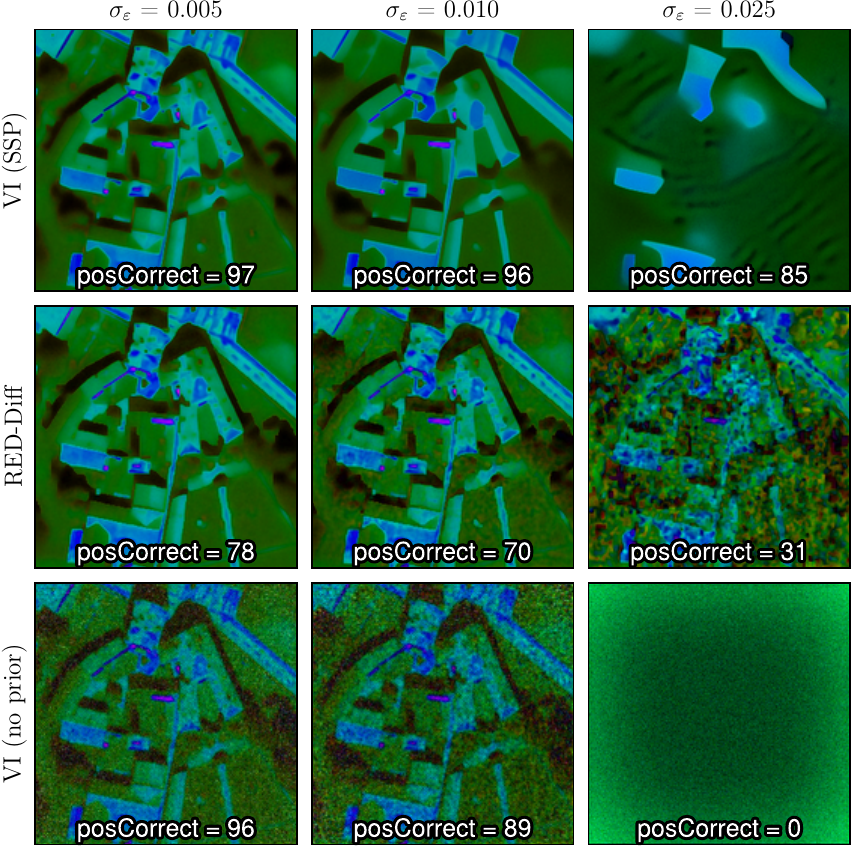}
    \caption{Reconstructed images of a test object from three methods (SSP, RED-Diff, VI without an image prior) under three increasing levels of Gaussian measurement noise $\noisestd$.}
    \label{suppl:fig:reconstructions-meas-noise-allmethods}
\end{figure}

\section{RED-diff Algorithm for Blind Inverse Problems}

\begin{algorithm}[h]
\caption{RED-diff for blind inverse problems}\label{algo:blind-red-diff}
\begin{algorithmic}[1]
\State Initialize parameter estimate $\bfr^{(0)}$, image estimate $x^{(0)}$, $\itc = 0$, data $y \in \calY$, forward model $A: \calR \times \calX \to \calY$, maximum number of iterations $\itc_{\max}$, step size sequences $(\tau^{(\itc,i)}), (\eta^{(\itc,i)})$
\While{$\itc < \itc_{\max}$ and stopping criterion on $x^{(\itc)},\bfr^{(\itc)}$ is not satisfied}
    \State $ x^{(\itc,0)} = x^{(\itc)}$
    \For{$i \gets 0, \dots, N_{\mathrm{img}}-1$}\Comment{Solve for image using gradients% \eqref{eq:blind-red-diff-image-step} using gradients \eqref{eq:blind-red-diff-image-gradient}
    }
        \State $x^{(\itc,i + 1)} = x^{(\itc,i)} - \tau^{(\itc,i)} \nabla_x \calL^{\mathrm{REDdiff}}(x^{(\itc,i)},\bfr^{(\itc)}) $
    \EndFor
    \State $x^{(\itc + 1)} = x^{(\itc,N_{\mathrm{img}})} $
    \State $ \bfr^{(\itc,0)} = \bfr^{(\itc)} $
    \For{$i \gets 0, \dots, N_{\mathrm{par}}-1$}\Comment{Solve for parameters% \eqref{eq:blind-red-diff-par-step}
    }
        \State $\bfr^{(\itc,i + 1)} = \bfr^{(\itc,i)} - \eta^{(\itc,i)} \nabla_\bfr \calL^{\mathrm{REDdiff}}(x^{(\itc+1)},\bfr^{(\itc,i)}) $
    \EndFor
    \State $\bfr^{(\itc+1)} = \bfr^{(\itc,N_{\mathrm{par}})} $
    \State $\itc \gets \itc+1$
\EndWhile
\State \textbf{return} image estimate $x^{(\itc)}$, parameter estimate $r^{(\itc)}$
\end{algorithmic}
\end{algorithm}

\end{document}